\documentclass[final,times,authoryear]{elsarticle}
\usepackage{graphicx}
\usepackage{hyperref}
\usepackage{natbib}
\hypersetup{colorlinks,citecolor=Magenta,linkcolor=Red,urlcolor=Magenta}
\usepackage{amsmath}
\usepackage{amssymb}
\usepackage{empheq}
\usepackage[usenames,dvipsnames]{color}
\usepackage{enumitem}
\setlist{noitemsep}

\allowdisplaybreaks 

\newcommand{\be}{\begin{equation}}
\newcommand{\ee}{\end{equation}}
\newcommand{\simless}{\lower.5ex\hbox{$\; \buildrel < \over \sim\;$}}
\newcommand{\simgreat}{\lower.5ex\hbox{$\; \buildrel > \over \sim\;$}} 
\newcommand{\arad}{{\cal A}} 
\newcommand{\pluto}{{{\rm P}\hskip-4pt{\rm L}}}

\newcommand{\Ham}{\mathcal{H}}

\newcommand{\G}{\mathcal{G}}
\newcommand{\Msun}{M_{\odot}}
\newcommand{\Poincare}{{Poincar$\acute{\rm{e}}$}}
\newcommand{\apj}{{\sl Astrophys. J.}} 
\newcommand{\apjl}{{\sl Astrophys. J. Letters}} 
\newcommand{\apjs}{{\sl Astrophys. J. Suppl.}} 
 
\newcommand{\araa}{{\sl Ann. Rev. Astron. Astrophys.}}
 
\newcommand{\aj}{{\sl Astron. J.}} 
\newcommand{\aap}{{\sl Astron. Astrophys.}} 
\newcommand{\astnach}{{\sl Astron. Nachrichten}}
\newcommand{\annphys}{{\sl Annalen Phys.}} 
 
\newcommand{\celmechda}{{\sl Celest. Mech. Dynam. Astron.}}

\newcommand{\icarus}{{\sl Icarus}} 
\newcommand{\jrasc}{{\sl J. Royal. Astron. Soc. Canada}} 
\newcommand{\mnras}{{\sl Mon. Not. R. Astron. Soc.}} 
\newcommand{\nature}{{\sl Nature}}

\newcommand{\pscript}{{\sl Phys. Scripta}} 
\newcommand{\pss}{{\sl Plan. Space Sci.}} 

\newcommand{\pasp}{{\sl Pub. Astron. Soc. Pacific}} 
\newcommand{\pnas}{{\sl Pub. Nat. Acad. Sci.}}  
 
\newcommand{\science}{{\sl Science}}

\newcommand{\zastro}{{\sl Z. Astrophys.}}

\journal{Physics Reports}

\begin{document}

\begin{frontmatter}

\title{The Planet Nine Hypothesis}

\author{Konstantin Batygin,$^1$ Fred C. Adams,$^{2,3}$ 
Michael E. Brown,$^1$ and Juliette C. Becker$^3$} 

\address{$^1$Division of Geological and Planetary Sciences \\ California Institute of Technology, Pasadena, CA 91125, USA} 
\address{$^2$Physics Department, University of Michigan, Ann Arbor, MI 48109, USA} 
\address{$^3$Astronomy Department, University of Michigan, Ann Arbor, MI 48109, USA} 

\begin{abstract}
Over the course of the past two decades, observational surveys have unveiled the intricate orbital structure of the Kuiper Belt, a field of icy bodies orbiting the Sun beyond Neptune. In addition to a host of readily-predictable orbital behavior, the emerging census of trans-Neptunian objects displays dynamical phenomena that cannot be accounted for by interactions with the known eight-planet solar system alone. Specifically, explanations for the observed physical clustering of orbits with semi-major axes in excess of $\sim250\,$AU, the detachment of perihelia of select Kuiper belt objects from Neptune, as well as the dynamical origin of highly inclined/retrograde long-period orbits remain elusive within the context of the classical view of the solar system. This newly outlined dynamical architecture of the distant solar system points to the existence of a new planet with mass of $m_9\sim 5-10\,M_{\oplus}$, residing on a moderately inclined orbit ($i_9\sim15-25\deg$) with semi-major axis $a_9\sim 400 - 800\,$AU and eccentricity between $e_9 \sim 0.2 - 0.5$. This paper reviews the observational motivation, dynamical constraints, and prospects for detection of this proposed object known as Planet Nine. 
\end{abstract}

\end{frontmatter}

\section{Introduction}\label{sect1}
\label{sec:intro}

Understanding the solar system's large-scale architecture embodies one of humanity's oldest pursuits and ranks among the grand challenges of natural science. Historically, the first attempts to astronomically map the imperceptible structure of the solar system trace back to Galileo himself, and his adoption of the telescope as a scientific instrument some four centuries ago. In terms of sheer numbers, however, the quest to unveil new planets in the solar system has been strikingly inefficient: only two large objects that were not already known to ancient civilizations -- Uranus and Neptune -- have been discovered to date, with no significant updates to the solar system's planetary catalogue since 1930.


While countless astronomical surveys aimed at discovering new solar system planets have consistently resulted in non-detections, the solar system's vast collection of minor bodies has slowly come into sharper focus. Particularly, the past quarter-century witnessed the discovery and characterization of a diverse collection of small icy objects residing in the outer reaches of our solar system, extending from the immediate vicinity of Neptune's orbit to far beyond the heliosphere\footnote{Data collected by the Voyager 1 and 2 spacecraft suggest that the heliosphere nominally extends to a heliocentric distance of approximately 120\,AU.} \citep{jewittlu93}. Intriguingly, rather than the planets themselves, it is this population of scattered debris that holds the key to further illuminating the solar system's intricate dynamical structure and to unraveling its dramatic evolutionary history.

The vast majority of trans-Neptunian small bodies, collectively known as the Kuiper belt, reside on orbits that are consistent with known dynamical properties of the solar system. The most extreme members of this population, however, trace out highly elongated orbits with periods measured in millennia, and display a number of curious orbital patterns. These anomalies include the striking alignment in the orientations of eccentric orbits in physical space, a common tilt of the orbital planes, perihelion distances that stretch far beyond the gravitational reach of Neptune, as well as excursions of trans-Neptunian objects into highly inclined, and even retrograde orbits. All of these otherwise mysterious orbital features can be readily understood if the solar system contains an additional -- as yet undetected -- large planet, residing hundreds of astronomical units away from the sun. 

The existence of this particular solar system object, colloquially known as Planet Nine, has only recently been proposed \citep{phattie}. Nonetheless, the discoveries in the outer solar system that ultimately led to the Planet Nine Hypothesis (\citealt{Brownsedna, trushep2014} and the references therein) represent a veritable revolution in the scientific understanding of our home planetary system. This review will explore these developments, and will provide a status update on the Planet Nine hypothesis. Before delving into specifics, however, we note that despite being characterized by a unique combination of observational evidence and dynamical mechanisms, the Planet Nine hypothesis is by no means the first inference of a new solar system member that is based upon anomalous orbital behavior of known objects. Moreover, it is important to recognize that this general class of scientific proposals has a long and uneven history, with results varying from the stunning success of Neptune \citep{leverriera,galle,adams1846} to the regrettable failure of Nemesis \citep{nemesis}. Accordingly, it is instructive to begin this manuscript with a brief historical review of past claims.

\subsection{The Discovery of Neptune} 
\label{sec:findneptune}

The mathematical determination of the existence of Neptune represents the only successful example of a new planetary discovery motivated by dynamical evidence within the solar system, and epitomizes one of the most sensational stories in the history of astronomy (see \citealt{KrajnovicNeptune} for a recent review). The saga begins with the official discovery of the planet Uranus (then called Georgium Sidus) in 1781 by William Herschel. Although Herschel was the first to record the proper motion of the object and present his results to the {\sl Royal Astronomical Society} (for further details, see the historical accounts of \citealt{alexander,miner}), numerous recorded observations of Uranus already existed, which mistook the planet for a background star.

\begin{figure}[tbp]
\centering
\includegraphics[width=0.65\textwidth]{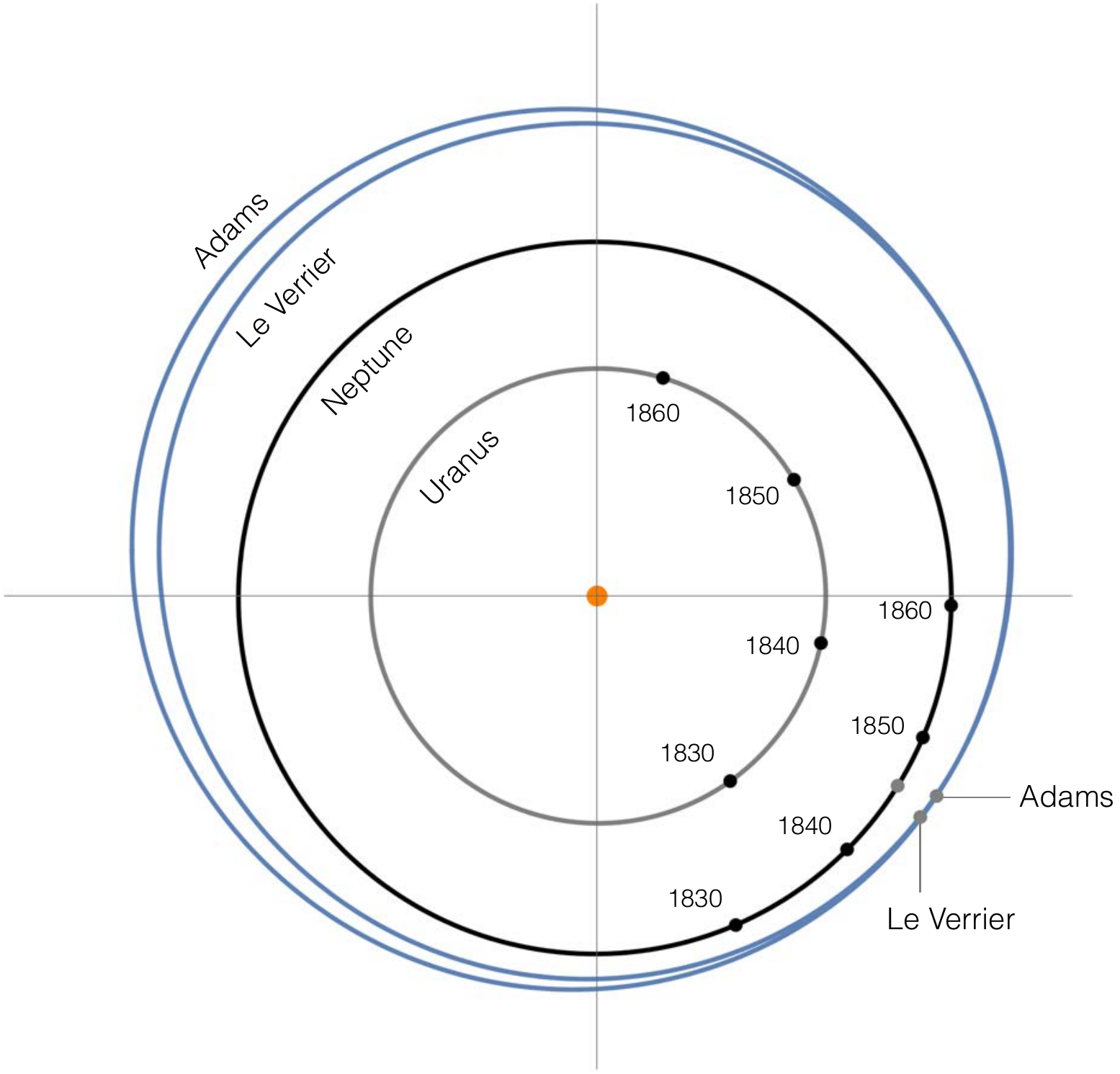}
\caption{Predicted orbit of Neptune. This diagram shows the outer solar system viewed from the north ecliptic pole, including the orbits of Uranus (gray) and Neptune (black), as well as the predicted Neptunian orbits (purple) from \citet{leverriera} and \citet{adams1846}. Note that although the predicted physical location of Neptune was in close agreement with the actual location of Neptune in 1846, the derived orbits are considerably wider and more eccentric than Neptune's true orbit.}
\label{fig:Neptunepredict} 
\end{figure} 

Over the next six decades, astronomers continued to monitor the motion of Uranus along its $84\,$year orbit and computed ephemerides of its position over time based on the then-known properties of the giant planets (e.g., \citealt{bouvard}). The calculations and the observations were not in perfect agreement, with the differences between the theoretical longitudes and the observed longitudes of the orbit growing by approximately $2''$ per year (see page 150 of \citealt{adams1846}). These data led \cite{leverriera,leverrierb} and then \cite{adams1846}\footnote{Although Le Verrier's calculations famously predate Galle \& d'Arrest's astronomical detection of Neptune, Adams's orbital predictions were only published \textit{after} Neptune's location was already known.} to propose an additional planet to account for the differences. It is worth noting that the inferred physical and orbital properties of Le Verrier's and Adams' putative planet differ considerably from those of the real Neptune (Figure \ref{fig:Neptunepredict}). Specifically, instead of the modern value $a_{8}\approx30\,$AU, the orbit of the proposed planet had a semi-major axis (reported as the `assumed mean distance') of $\sim36$ AU \citep{leverrierb} and $\sim37$ AU \citep{adams1846}. Meanwhile, the estimated orbital eccentricity was $e_8\sim0.11$ \citep{leverrierb} and $e_8\sim0.12$ \citep{adams1846}, significantly larger than the modern value of $e_8\approx0.008$. Finally, the mass estimate for the new planet was reported as $m_8\sim36\,M_{\oplus}$ \citep{leverrierb} and $m_8\sim50\,M_{\oplus}$ \citep{adams1846}, two to three times larger than Neptune's actual mass of $m_8\approx17\,M_{\oplus}$. The predicted properties of Neptune were thus somewhat larger in both mass and semi-major axis than the observed body.

We note that resolution of the irregularities found in the orbit of Uranus required rather extensive calculations \citep{leverriera,adams1846}. The analysis had to include perturbations of all the previously known planets, the error budget for the estimated orbital elements of Uranus, and the assumed orbital elements (and mass) of the proposed new planet (an overview of the difficulties is outlined in \citealt{lyttleton}). The discrepancies between Le Verrier's and Adams' theoretical expectations and the observed properties of Neptune therefore constitutes a gold standard for dynamically motivated planetary predictions. Moreover, it is worth pointing out that the most significant quantity in perturbing the orbit of Uranus was the anomalous acceleration in the radial direction produced by the new body, $\arad \propto m_8/r_8^2$ -- a ratio that was predicted with much higher accuracy than the individual values of mass and semi-major axis. As we discuss below, comparable degeneracies between mass and orbital parameters exist within the framework of the Planet Nine hypothesis as well.

\subsection{Planet X and the Discovery of Pluto} 
\label{sec:planetx}
Following Le Verrier's triumphant mathematical discovery of Neptune, unexplained behavior in the motions of objects in the solar system continued to inspire predictions of the existence, and sometimes locations of new planets beyond the boundaries of the known solar system. Some of the early trans-Neptunian planetary proposals include Jacques Babinet's 1848 claim of a $\sim12\,M_{\oplus}$ planet beyond Neptune \citep{grosser64}, David P. Todd's 1877 speculation about a planet at $52\,$AU \citep{hoyt1976}, Camille Flammarion's inference of a planet at $48\,$AU \citep{Flammarion}, as well as the prediction of two planets at $100$ and $300\,$AU by George Forbes, whose calculations were motivated by an apparent grouping of orbital elements of long-period comets \citep{forbes1880}. An expansive set of planetary hypotheses was later put forth by W. H. Pickering, who predicted seven different planets between 1909 and 1932, with masses ranging from $0.045\,M_{\oplus}$ to $20,000\,M_{\oplus}$ (see \citealt{hoyt1976} for an excellent historical overview). Arguably, the most emblematic planetary prediction, however, can be attributed to Percival Lowell, who championed the search for what he called ``Planet X" and founded Lowell Observatory in Arizona in hopes of finding it.

The story of the search for Planet X is well documented in the literature (e.g., \citealt{levybook}). Briefly: despite the addition of Neptune to the solar system's ledger of planets, small apparent discrepancies in the orbits of the giant planets remained. With the observations and analysis available at the time, the inferred orbital anomalies in the orbit of Uranus implied a planetary object with a mass of $m_X\sim7\,M_{\oplus}$, about half the mass of Neptune. Lowell died suddenly in 1916, without having found the elusive planet. Nevertheless, the search continued. The new director, Vesto Slipher (who used the observatory to measure the recession velocities of galaxies; see \ref{appA}), handed off the search for Planet X to Clyde Tombaugh. By 1930, Tombaugh had examined countless photographic plates containing millions of point sources for possible planetary motion and finally discovered a moving object \citep{tombaugh1946,tombaugh1996}. Because it was found in the approximate location on the sky where Planet X was envisioned to be, and because Planet X was the object of the original search, the newly found body was initially considered to be the long-sought-after Planet X \citep{slipher1930}.

Immediately there was a problem. An object with physical size comparable to Neptune should have been resolved with the observational facilities of the era, but the new planet appeared point-like. It was also dim, coming in about six times fainter than the estimates. It soon became clear that the newly found member of the solar system was not {\sl the} Planet X, as it was not massive enough to account for the perceived perturbations in the orbit of Uranus. The new body was subsequently named Pluto, after the Greek god of the underworld\footnote{As a bonus, the first two letters of the planet's name were coincident with the initials of Percival Lowell, although W. H. Pickering was evidently under the impression that it stood for Pickering-Lowell \citep{hoyt1976}.}. In the end, the observed irregularities in Uranus' and Neptune's motion turned out to be spurious, and were fully resolved by a 0.5\% revision of Neptune's mass, following the Voyager 2 flyby \citep{standish1993}.

\begin{figure}
\centering
\includegraphics[width=0.8\textwidth]{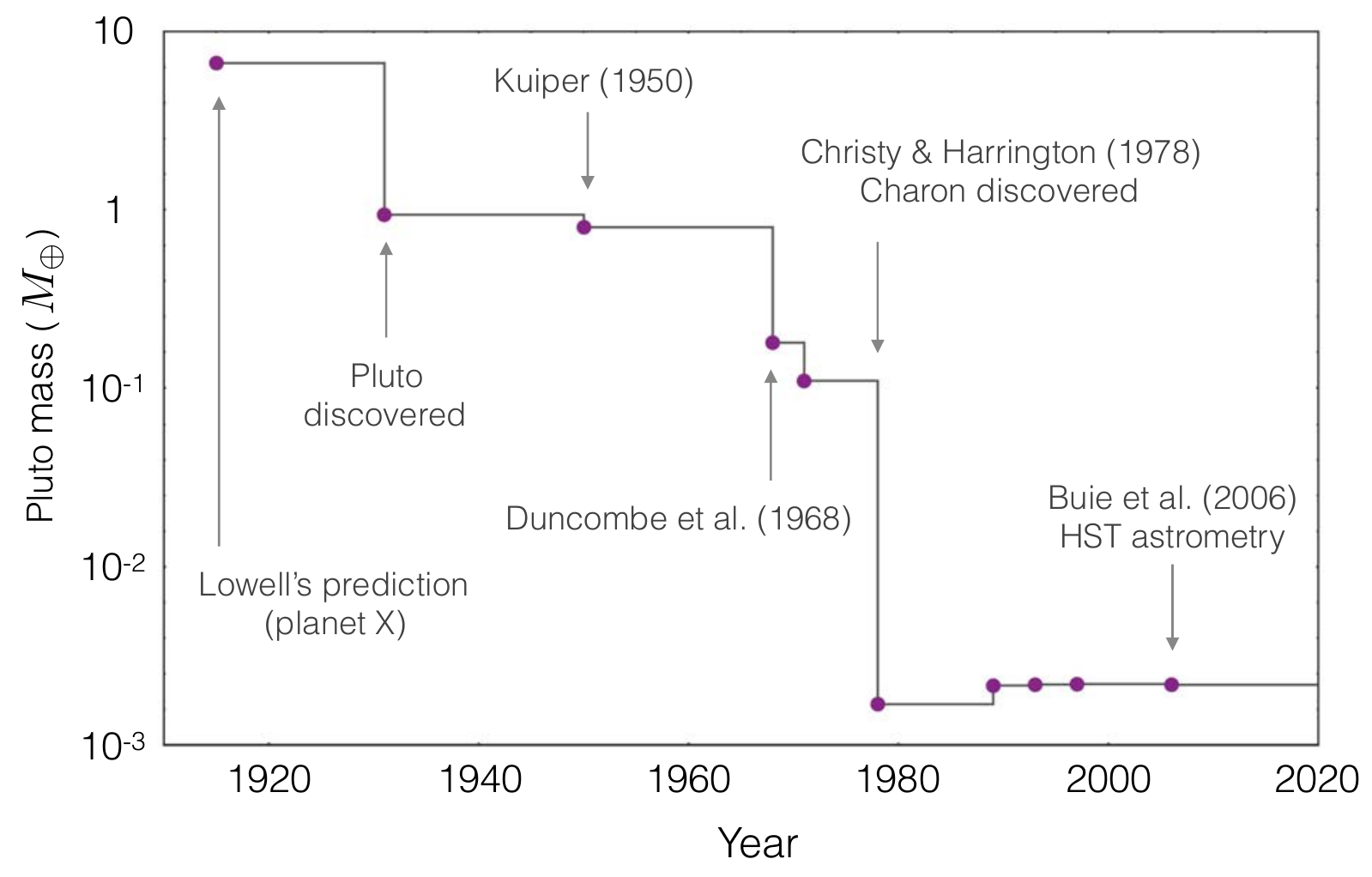}
\caption{Estimated mass of Pluto as a function of time. The first (and largest) estimate is from the dynamical calculations that inpired the initial search. The subsequent estimates are based on observations (as labeled) and show a steady downward trend until relatively recently. The particularly steep decline of Pluto's mass in 1978 is marked by the discovery of its largest moon, Charon.}
\label{fig:plutomass} 
\end{figure} 

Figure \ref{fig:plutomass} shows the estimated mass of Pluto as a function of time. The initial estimate ($m_X\approx7\,M_{\oplus}$) is the mass required to account for the perceived perturbations of the giant planet orbits \citep{tombaugh1946}. The first observational estimate for the mass of Pluto \citep{nicholson} was already down to $M_\pluto$ $\approx1\,M_{\oplus}$, and subsequent observations led to steadily lower values as shown in the Figure. Note that a precipitous drop in the mass estimate that came in 1978, with the discovery of the moon Charon and the first clean measurement of Pluto's mass \citep{christy}. The current mass estimate is only $M_\pluto$ $\approx$ 0.00218 $\,M_{\oplus}$ \citep{buie}, roughly 3200 times smaller than the original mass estimate that inspired the search. In spite of its diminutive size, Pluto was considered as the ninth planet of the solar system until 2006, when it was demoted to the status of a dwarf planet \citep{websiteA}. 

\subsection{The False Alarm of Vulcan} 
\label{sec:vulcan} 


By the mid 19th century, observations of the planet Mercury became precise enough to detect variations in its orbital parameters, lending a handle on the gravitational perturbations exerted by the terrestrial planets upon one-another. By making precise measurements of transits of Mercury across the Sun, 19th century astronomers thus realized that the orbit of Mercury was precessing forward at a rate that could not be fully accounted for by known bodies. This determination inspired \citet{leverriermercury,leverrier1859} to propose that an additional planet interior to Mercury is responsible for the extra perihelion precession necessary to fit the observations. This hypothetical new planet become known as Vulcan, the god of fire, volcanoes, and metal working in Roman mythology.

In order to generate the anomalous $(d\varpi/dt)=43"$/century perihelion advance of Mercury, the parameters of the unseen planet ($m_{\rm{v}},a_{\rm{v}}$) would have been constrained by (see section \ref{sec:data_apsidalconf})
\be
\bigg(\frac{d\varpi}{dt}\bigg)_{\rm{v}}\approx\frac{3}{4}\sqrt{\frac{\G\,\Msun}{a_1^3}}\frac{1}{\big(1-e_1^2\big)^2}\frac{m_{\rm{v}}\,a_{\rm{v}}^2}{\Msun\,a_1^2} \Leftrightarrow \bigg(\frac{d\varpi}{dt}\bigg)_{\rm{gr}} = 3\sqrt{\frac{\G\,\Msun}{a_1^3}}\frac{\G\, \Msun}{a_1\,c^2}\frac{3}{\big(1-e_1^2\big)^2},
\label{delgr} 
\ee
where $\G$ is the gravitational constant and $c$ is the speed of light. Coincidentally, the above expression is satisfied by a $m_{\rm{v}}\sim3\,M_{\oplus}$ Super-Earth type planet with an orbital period of approximately 3 days, analogues of which are now known to be very common around sun-like stars \citep{kepler2010,kepler2013}. In the early 20th century, however, Einstein developed his theory of General Relativity, which self-consistently resolved the ancillary precession problem (yielding a rate of apsidal advance given by the RHS of the above equation; \citealt{einstein}). This alleviated the need for intra-Mercurian planets within the solar system. Accordingly, unlike the case of Uranus (discussed in section \ref{sec:findneptune}), dynamical anomalies in the inner solar system paved the way to the discovery of fundamentally new physics rather than a planet.

\subsection{The Nemesis Affair} 
\label{sec:nemesis} 

The 1980s witnessed the proposal that the Sun was actually part of a binary star system. This time, the motivation for the proposed stellar companion stemmed from paleontology: approximately 65\,Myr ago, a mass extinction event wiped out three quarters of the species then living on Earth. This cataclysm, which famously included the removal of (non-avian) dinosaurs from the biosphere, occurred at what is known as the Cretaceous-Tertiary (K-T) boundary.  At the layer of rock corresponding to this geological age, sediments are observed to contain high levels of iridium, an element that is relatively rare in crustal rocks on Earth, but is much more plentiful in asteroids. The association of the iridium layer with the extinction boundary led to the hypothesis that the event was caused by a collision of a large ($\sim10-15$ km) asteroid with Earth \citep{alvarez}. The resulting impact would have catastrophic consequences on the climate, and would, in turn, explain the mass extinction event. 

Although the K-T extinction event is often seen as being the most dramatic, it is far from unique: the geologic record unequivocally shows that the biosphere of Earth has experienced a series of mass extinction events. Moreover, although the data are sparse, with 12 events distributed over a time span of $\sim250$ Myr, some analyses have suggested that these extinction events are periodic, and recur every $\sim26\,$Myr \citep{raup1984}. One way to achieve a periodic signal in the extinction events is for the Sun to have an eccentric stellar companion, which perturbs comets in the Oort cloud on the necessary time interval \citep{nemesis}. The envisaged red/brown dwarf companion would therefore have an orbital period of about 26\,Myr and hence a semi-major axis of $a_{\rm{nem}}\sim88,000$\,AU. This hypothetical body became known as Nemesis, named after the Greek goddess of retribution. Although the orbital stability of the proposed companion was inconclusively analyzed by \citet{hills1984,hut1984,torbettsmol}, more recent work indicates that the probability of the ejection for Nemesis from the solar system by passing stars is of order unity over the age of the Sun \citep{liadams2016}.

A number of searches for Nemesis have come up empty, starting with a University of California search \citep{perlmutter}, and continuing with infrared surveys carried out by {\sl IRAS} \citep{beichman1987} and {\sl 2MASS} \citep{2mass}. Moreover, the evidence for periodic extinction events itself is tenuous, meaning that the Nemesis hypothesis may simply be unnecessary. A more modest proposal, motivated by the distribution of aphelion directions of comets, has been subsequently put forward, where the Sun has a Jovian mass companion (referred to as Tyche) on a distant orbit that perturbs comets in the Oort cloud \citep{matese1999,matese}. However, the Wide-field Infrared Survey Explorer ({\sl WISE}) has placed stringent limits on the existence of such bodies in the distant solar system. Specifically, objects with the mass of Saturn and Jupiter are ruled out to distances of 28,000 AU and 82,000 AU, respectively \citep{luhman2014}. Taken together, these observational surveys leave little parameter space for any putative binary companion to the Sun.

\subsection{Recent Planetary Proposals} 
\label{sec:recentprops} 

Since the discovery of the Kuiper belt \citep{jewittlu93}, the orbital architecture of this population has often been examined in hope that it can provide insight into massive astrophysical objects that may reside beyond the current observational frontier of the solar system. The first suggestion that the observed distribution of objects in the Kuiper belt pointed to the presence of a perturbing planet came from \citet{brunini2002}, who argued that the dramatic drop-off in the numbers of Kuiper belt objects (KBOs) with semi-major axes beyond $a\gtrsim48\,$AU could be explained by an approximately Mars-sized body with a semi-major axis of $a\sim60\,$AU. \citet{melita2004}, however, quickly determined that such an object was incapable of reproducing the observations. 

The idea of a sub-Earth-sized outer planet was revisited by \citet{lykawka} who suggested that such a body could explain many of the detailed properties of the main region of the Kuiper belt. The next suggestions of a planetary perturber came after the discovery of objects with high eccentricities and with perihelia beyond the immediate reach of strong Neptunian perturbations. Particularly, the KBO 2000\,CR$_{105}$ inspired speculation that its large perihelion distance of $q=44.3\,$AU can most readily be attributed to the current or former presence of an external perturber, although slow chaotic diffusion driven by the known planets also constitutes a viable explanation for this seemingly unusual trajectory \citep{gladman2002, gladmanchan}. The orbit of the distant object Sedna (2003 VB$_{12}$), on the other hand, which has perihelion distance of $q=76\,$AU, can only be explained by perturbations from an external agent. To this end, \citet{Brownsedna} suggested that even though an approximately Earth mass object at $\sim70\,$AU could in principle be responsible, perturbations arising from stars within the birth cluster constitute a more plausible explanation for Sedna's detachment from Neptune \citep{al2003,morbyhal2004}. Nevertheless, \citet{gomes2006} carried out an extensive series of numerical simulations showing that the orbits of both 2000\,CR$_{105}$ and Sedna could be explained by a Neptune-to-Jupiter mass planet with a semi-major axis of $100$s to $1000$s of AU. 

A separate suggestion of a planetary perturber surfaced when \citet{trushep2014} discovered an additional outer solar system object, 2012\,VP$_{113}$, with a perihelion well beyond the planetary region, and noted that all known KBOs in the outer solar system with perihelion distance beyond Neptune and semi-major axis greater than $a>150\,$AU have argument of perihelion\footnote{Not to be confused with the \textit{longitude} of perihelion, the \textit{argument} of perihelion is an angle made by the vectors pointing to the perihelion and the ascending node (see Figure \ref{fig:elements}).}, $\omega$, clustered around zero. Orbits with $\omega = 0$ or $\omega = 180\deg$ come to perihelion at the heavily observed ecliptic, so observational bias could naturally facilitate a clustering around both of those values, particularly for the highest eccentricity objects. However, no conceivable bias could lead to only observing objects clustered around $\omega = 0$, implying that \citet{trushep2014}'s result is both statistically robust and not a product of survey strategy.

To explain the observations, \citet{trushep2014} speculated that a several Earth mass planet at approximately $200\,$AU could maintain the arguments of perihelion alignment through the Kozai-Lidov effect (see section \ref{sec:secularforcing} for a short discussion). In particular, \citet{trushep2014} demonstrated that a $5\,M_{\oplus}$ body on a circular orbit at $a=210\,$AU could cause Kozai-Lidov oscillations of 2012\,VP$_{113}$, thereby maintaining its argument of perihelion in libration around zero. Nevertheless, they also pointed out two difficulties with this scenario. First, in their simulations, they failed to find a single planet that can cause all of the KBOs -- which have semi-major axes between 150 and 500\,AU -- to undergo Kozai-Lidov oscillations. This difficultly is not surprising; objects librate about $\omega = 0$ or $180\deg$ only for an interior perturber with a semi-major axis relatively close to the semi-major axis of the object being perturbed \citep{ThomasMorby}. Thus, Kozai-Lidov libration of all objects with semi-major axes between 150 and 500\,AU would likely require a special configuration of several carefully placed planets (see \citealt{marcosmarcos3} for further discussion). Second, internal perturbers can cause affected objects to have Kozai-Lidov oscillations about either $\omega = 0$ or $\omega = 180\deg$, and as already stated above, the evidence for clustering about only $\omega = 0$ is robust. \citet{trushep2014} suggested that this dilemma might be overcome if a close stellar encounter had previously aligned the arguments of perihelia, a process demonstrated by \citet{Feng2015}. The \citet{trushep2014} planetary proposal thus requires multiple external planets and a close stellar flyby to explain the argument of perihelion clustering, along with some additional mechanism to explain the high perihelia of many of the distant objects.

Even more recently, \citet{volkmalhotra} examined the observational census of long-period KBOs, and argued that the mean orbital plane of the Kuiper belt in the $a\sim 50-80\,$AU domain is inclined with respect to the ecliptic in an unexpected manner. This led the authors to suggest that the observed warping may be facilitated by a small (e.g. $\sim$Mars-mass) planet, residing in the solar system on an appreciably inclined orbit with $a\sim65-80\,$AU. The possibility that a body of this type may indeed exist in the solar system is bolstered by the recent simulations of \citet{silsbee2018}, who find a significant probability that sub-Earth-mass planetary embryos can become trapped on orbits with $a\lesssim200\,$AU and $q\sim40-70\,$AU early in the solar system's lifetime. Given that the expected visual magnitude of such an object would be of order $\sim17$ or less, the most likely location on the sky where this body could have avoided being discovered to date is the galactic plane, which remains relatively unexplored by solar system surveys.
\\

This brief overview of gravitationally motivated planetary proposals illuminates the fact that over the course of the last 170 years, numerous variants of trans-Neptunian perturbers have been considered (and subsequently abandoned), with the aim to explain a broad range of dynamical phenomena at the outskirts of the solar system. In light of this multiplicity, a natural question emerges: what distinguishes the different planetary proposals and the models that accompany them? A simple answer may be that each proposition of a trans-Neptunian planet is characterized by the unique combination of the anomalous data it seeks to explain, and the specific dynamical mechanism through which the putative planet generates its observational signatures. With an eye towards characterizing the Planet Nine hypothesis specifically within this framework, in the text below we will present an up-to-date account of the orbital architecture of the trans-Neptunian region of the solar system, and outline a theoretical description of the dynamical mechanisms through which Planet Nine sculpts the small body population of the distant solar system.

The remainder of this review is structured as follows. In the following section, we sketch out the various sub-categories of objects residing in the trans-Neptunian solar system, and summarize the relevant nomenclature. In section \ref{anomalous}, we provide a detailed account of the anomalous structure of the distant solar system, including a brief discussion of the characteristic timescales and observational biases. Section \ref{sec:analytical} outlines a theoretical description of the dynamical mechanisms through which Planet Nine sculpts the population of small bodies in the distant solar system and thereby accounts for these anomalies. In section \ref{sec:numerical} we present results from a large ensemble of numerical simulations that fully capture the non-linear and chaotic nature of Planet Nine-induced dynamics. These integrations conform to the expectations of the analytical theory from the previous section and constrain the allowed properties of Planet Nine. The prospects for detecting this as-yet-unseen planet are described in section \ref{sec:detection}. Given its requisite large distance from the Sun, the formation of Planet Nine poses a challenging problem, and various scenarios are discussed in section \ref{sec:formation}. The review concludes in section \ref{sec:conclude} with a summary of results, possible alternative explanations, and a brief outline of questions that remain open within the broader framework of the Planet Nine hypothesis.

\section{Inventory and Structure of the Trans-Neptunian Solar System} 
\label{sec:inventory}  

As astronomical surveys have continued to push to ever greater depth and unveil the population of small bodies beyond Neptune, it has become progressively evident that the trans-Neptunian region of the solar system encompasses a rich diversity of objects that exhibit distinct modes of gravitational coupling with Neptune. Much of this orbital structure is a frozen-in relic of the solar system's violent dynamical past \citep{Levisonetal2008, Nesvorny2015a}, and plays virtually no role in the formulation of the Planet Nine hypothesis. Meanwhile, other aspects of the distant solar system's architecture are ceaselessly being sculpted by both known and inferred long-period planets, and are therefore of key importance. Thus, in order to better understand any lines of evidence for external gravitational perturbations, it is of considerable use to first outline the various dynamical classes that make up the Kuiper belt and their respective significance for the analysis that will follow. 

\begin{figure}[tbp]
\centering 
\includegraphics[width=0.75\textwidth]{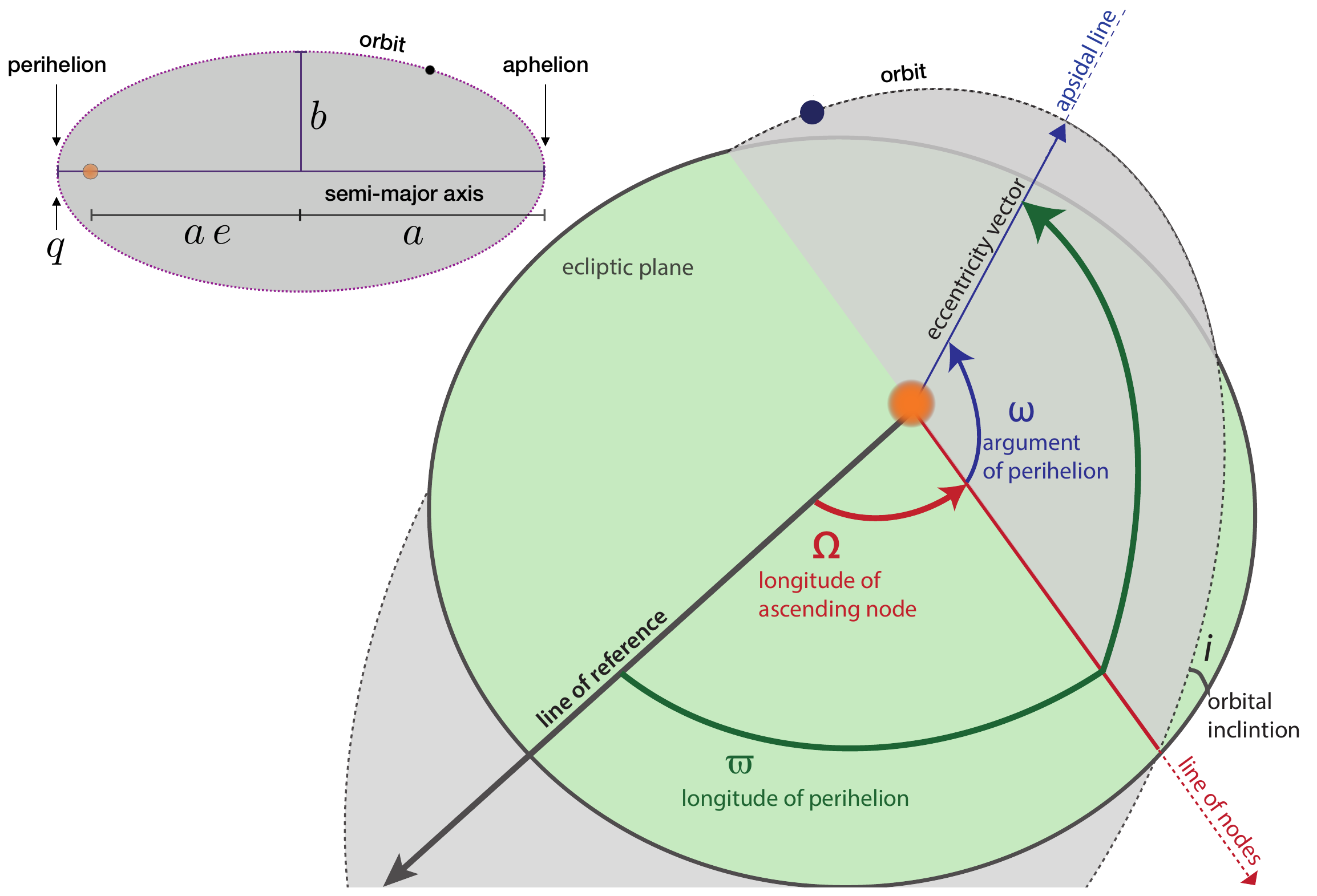}
\caption{Definition of Keplerian orbital elements, illustrated by a schematic of an inclined, eccentric orbit in the solar system. The size and ellipticity of the orbit are parameterized by the semi-major axis, $a$, and the eccentricity, $e$, as shown in the inset. The longitude of ascending node (denoted $\Omega$) informs the direction into which the orbit is tilted, and is measured by location at which the orbit intersects the ecliptic plane from below. The argument of perihelion (denoted $\omega$) describes the angle between the line of nodes and the direction of the planet's closest approach to the sun (also referred to as the apsidal line). The combined angle, $\varpi = \omega + \Omega$, defines the overall direction of the apsidal line and is called the longitude of the perihelion. Finally, the tilt of the orbit with respect to the ecliptic is informed by the inclination, $i$.} 
\label{fig:elements} 
\end{figure}

\begin{figure}[tbp]
\centering 
\includegraphics[width=1.0\textwidth]{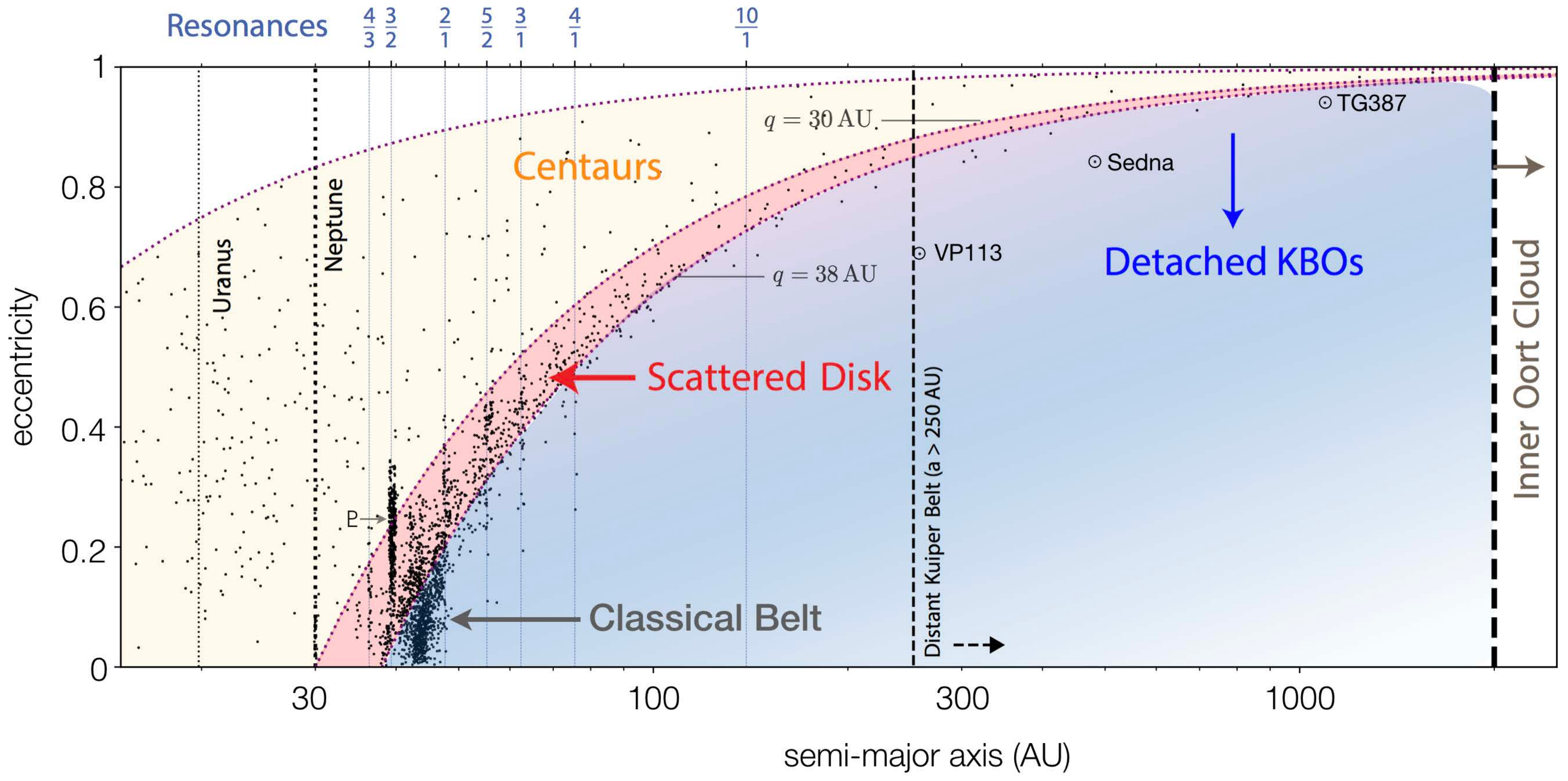}
\caption{Main dynamical classes of trans-Neptunian objects. This diagram shows the current observational census of the Kuiper belt in the $a$-$e$ plane (black points), and outlines the primary sub-populations of Kuiper belt objects. Resonant Kuiper belt objects have orbital periods that are commensurate with that of Neptune. The classical Kuiper belt is primarily composed of comparatively low-eccentricity objects that reside between the 3:2 and the 2:1 mean motion resonances, extending out to $\sim48\,$AU. The scattered disk contains objects with higher eccentricity and perihelion distance in the range $q\sim30-38\,$AU. The few objects with higher perihelia and large semi-major axes represent the detached
population, and are keenly relevant to the Planet Nine hypothesis. Objects with perihelion distance smaller than the semi-major axis of Neptune are referred to as Centaurs.} 
\label{fig:outersys} 
\end{figure}

The current census of trans-Neptunian objects (TNOs) is comprised of a few thousand bodies, the vast majority of which are considerably smaller than 1000\,km in diameter. The outer solar system also contains a relatively large number of dwarf planets -- bodies large enough for gravitational forces to render them quasi-spherical, yet too small to appreciably influence their orbital neighborhoods. In particular, the inventory of currently known objects includes 10 bodies larger than $D>900\,$km and another 17 bodies with estimated sizes in the range $D$ = 600 -- 900 km (see \citealt{brown2008} for a review). The number of such objects will continue to grow as the trans-Neptunian region is surveyed to greater depth. Nevertheless, the \textit{total} mass of the present-day Kuiper belt is estimated to be a small fraction of an Earth-mass \citep{pitjevapitjev}, meaning that Kuiper belt objects can be essentially thought of as tracers of dynamical evolution facilitated by the outer planets. Accordingly, typical classification of KBOs is based upon their orbital, rather than physical properties (Figure \ref{fig:elements}), and we will follow this convention here (Figure \ref{fig:outersys}). For more comprehensive discussion of specific attributes of the trans-Neptunian solar system, we direct the reader to reviews of the observational characteristics of the Kuiper belt (\citealt{luu2002,nomenclature}), planet formation \citep{armitage2010}, early dynamical evolution \citep{nesvorny}, and the birth environment of the solar system \citep{adams2010}, as well as the references therein. 

\subsection{Resonant Kuiper Belt} 

An emblematic sub-population of Kuiper belt objects (that includes Pluto itself) is composed of objects entrained in mean motion resonances with Neptune. Simply put, resonant orbits have periods that can be expressed as rational multiples of Neptune's orbital period, although we note that having a nearly rational period ratio is a necessary but insufficient condition for resonance. In addition to period commensurability, the angular orbital elements must have the proper form so that the relevant resonance angle (a Fourier harmonic of the gravitational potential, also called a ``critical argument") oscillates (or ``librates'') around a particular value, instead of steadily increasing or decreasing its value (a regime known as ``circulation''). KBOs residing on resonant orbits generally experience coherent exchange of orbital energy and angular momentum with Neptune, and can be long-term stable even at high eccentricities, thanks to a phase-protection mechanism that allows KBO orbits to overlap the orbit of Neptune without being rapidly destabilized by close encounters \citep{peale1976, nesvorny2001a, nesvorny2001b}.

The most prominent orbital commensurabilities in the Kuiper belt correspond to the 3:2 and 2:1 mean motion resonances. Drawing on the fact that these resonances are densely populated, \citet{malhotra1995} demonstrated that Neptune likely formed much closer to the sun, and must have experienced long-range outward migration, capturing resonant KBOs along the way. Subsequent characterization of resonant dynamics in the Kuiper belt further revealed that Neptune's migration must have had a stochastic component, and likely occurred during a transient period of dynamical instability experienced by the outer solar system (\citealt{tsiganis2005, Levisonetal2008}; see also \citealt{Nesvorny2015a} for a recent study). Despite the constraints that the structure of the resonant Kuiper belt places on the early evolution of the solar system, it plays no role within the context of the Planet Nine hypothesis, and its existence can be safely ignored for the purpose of the calculations that will follow.

\subsection{Classical Kuiper Belt} 

The Classical Kuiper Belt is primarily comprised of icy bodies that have semi-major axes in between the 3:2 and 2:1 mean motion resonances, corresponding to $a\approx42-48$ AU. By virtue of not being locked into resonances with Neptune, classical KBOs dominantly experience phase-averaged (so-called ``secular") interactions that are considerably more subdued than their resonant counterparts. The broader class of classical KBOs is often divided into dynamically `cold' and `hot' sub-populations, where the distinction is based on their orbital inclinations, which show a bimodal distribution \citep{brown2001}. The dividing point between the cold and hot populations is often taken to be $i\approx5\deg$, but the boundary is not sharp. \citet{caceres2018} recently considered the gravitational constraints upon Planet Nine's orbit and mass that may ensue from the lack of orbital excitation of the cold belt, and demonstrated that no strong limits arise within this framework. Thus, like the resonant Kuiper belt, the classical belt is largely inconsequential for the Planet Nine hypothesis. 

\subsection{Scattered Disk} 

If the resonant and classical populations were the only constituents of the Kuiper belt, gravitational evidence for \textit{any} distant, massive perturber is unlikely to have surfaced. Thankfully, this is not the case, and the Kuiper belt hosts a third dynamical class of icy objects that spans a much more extended range of orbital elements -- the so-called scattered disk. The vast majority of scattered disk objects reside on dynamically metastable orbits, and have perihelion distances in the vicinity of Neptune's orbit i.e., $q\sim30-38\,$AU. While the relatively confined perihelion range exhibited by scattered disk objects necessitates persistent orbital diffusion facilitated by Neptune's chaotic perturbations, the semi-major axes of scattered disk objects range from the neighborhood of Neptune's orbit ($a\sim q$) to thousands of AU, connecting this population to the much more distant Oort cloud \citep{gomes2008review}. Because the orbits of long-period scattered disk objects sample a broad range of heliocentric distances, they are keenly relevant to the Planet Nine hypothesis, and provide some primary lines of evidence for the existence of Planet Nine.

Although comparatively rare, some members of the scattered disk have perihelia well outside of the immediate gravitational reach of Neptune (typically $q\gtrsim40\,$AU). Placed within the framework of the currently known eight-planet solar system, this sub-class of ``detached" Kuiper belt objects exhibits dynamically stable evolution. The existence of such objects is important for two reasons. First, such objects cannot be generated simply through interactions with Neptune, and require an additional source of external gravitational perturbations \citep{morbyhal2004}. Second, because the orbital evolution of such objects is not contaminated by chaotic interactions with Neptune, they provide a particularly intelligible probe of dynamical evolution that unfolds in the far reaches of the solar system. We will return to the characterization of the distant scattered disk in sections \ref{sec:data_apsidalconf} and \ref{sec:data_orbitalplanes}.

\subsection{Centaurs} 

Facilitated by short-periodic interactions with Neptune, some KBOs get scattered inwards, attaining perihelion distances below $q\leqslant30\,$AU. For the purposes of this review, we will refer to the population of objects with $q\leqslant30\,$AU and $a\geqslant30\,$AU as Centaurs, although we note that a more general definition also includes objects with semi-major axes in between the giant planets \citep{nomenclature}. Because the orbits of Centaurs veer into the inter-planetary space, their orbital evolution is chaotic, with typical dynamical lifetimes measured in tens to hundreds of millions of years \citep{Tiscareno2003,Tiscareno2009,lykawka}. Despite having small perihelion distances, however, like the scattered disk, the Centaur population spans a broad range of orbital periods with some known objects having semi-major axes in excess of $2000\,$AU. Therefore, in spite of dynamical contamination from the known giant planets, long-period Centaurs may still be strongly influenced by Planet Nine's gravity.

A point of greater importance is that Centaurs exhibit a very broad dispersion of orbital inclinations, with a significant number of bodies occupying strongly retrograde orbits. Such a wide scatter of inclinations cannot be accounted for by interactions with the known giant planets alone, and requires some additional dynamics to explain the observations. As discussed further below, a radically excited distribution of Centaur inclinations is a natural outcome of Planet Nine induced evolution, and constitutes a tantalizing line of evidence that points to Planet Nine's existence (\citealt{batbrown2016b, batmorby, bp519}; see also \citealt{gomes2015}). In other words, if Planet Nine exists, highly inclined Centaurs almost certainly represent an outcome of dynamical evolution sculpted by an interplay between P9 and the canonical giant planets.

\subsection{The Oort Cloud} 

At heliocentric distances more than one thousand times that of Neptune, resides yet another, nearly spherical reservoir of debris known as the Oort cloud. Distinct from the Kuiper belt, the Oort Cloud is a collection of icy bodies that provides the source for long-period comets \citep{opik1932,oort1950}. The region is thought to have a radial extent from about 20,000 to 200,000\,AU, where the outer distance scale corresponds to the effective gravitational boundary to the solar system and is roughly given by the Hill Sphere of the Sun (as dictated by the potential of the Galaxy). 

It is noteworthy that although its existence is seemingly well established \citep{hills1981}, the only concrete evidence for the Oort Cloud stems from the observations of comets with semi-major axes in excess of $a\geqslant20,000$. The observed cometary flux yields a cumulative mass estimate for the Oort cloud of approximately $\sim1-2\,M_{\oplus}$, although this value is highly uncertain \citep{francis2005}. With its large outer boundary, the Oort Cloud is subject to disruption by passing stars \citep{huttremaine,heislertremaine}. The continued existence of the Oort Cloud thus puts constraints on the severity of such interactions, and also constrains how disruptive such events would be for distant Kuiper belt objects as well as Planet Nine itself (\citealt{duncan2008}, see also section \ref{sec:formation}). 


\subsection{Remaining Parameter Space for New Planets} 

The overview of TNOs outlined above delineates the main dynamical classes of the observable Kuiper belt. Before leaving this section, however, let us remark on the uncharted domain of the distant solar system, with a specific focus on generic constraints that restrict the properties of any putative trans-Neptunian planets. Intriguingly, independent of any particular model, reasonably tight observational and gravitational limits can be placed on the permissible parameters for any as-yet-undiscovered major bodies.

\begin{figure}[tbp]
\centering 
\includegraphics[width=.65\textwidth]{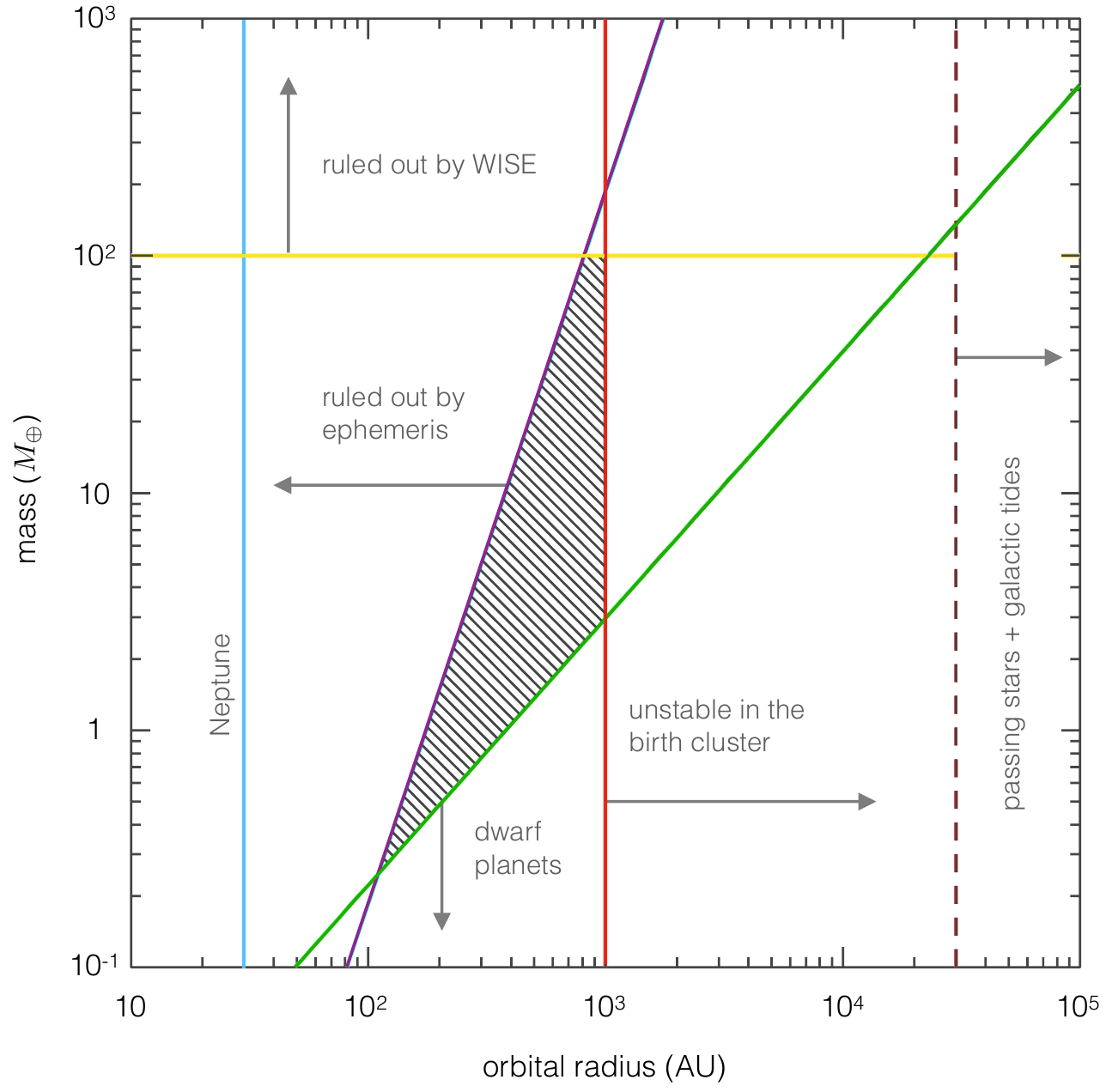}
\caption{Available parameter space for yet-undiscovered planetary members of the solar system. Planets that are massive enough to gravitationally clear their orbits must lie above the green line. In order to survive the stochastic perturbations within the solar birth cluster, planets must have sufficiently tight orbits, with semi-major axes to the left of the red line. To avoid producing anomalously large perturbations of giant planet orbits, new planets must fall below the purple line. Finally, infrared surveys require planets to fall below the yellow line. The resulting admissible portion of parameter space is shown as the hatched region.}
\label{fig:possible} 
\end{figure}

In principle, new planets could orbit with semi-major axes ranging from just outside Neptune (tens of AU) out to the boundary specified by galactic tides ($\sim10^{4-5}$ AU). In terms of mass, new planets could be as small as Mars ($\sim0.1\,M_{\oplus}$; smaller entities are not considered to be major bodies) or as large as a few thousand Earth masses (due to Deuterium fusion that ensues in objects more massive than $\sim13$ Jupiter masses, entities larger than $\sim4100\,M_{\oplus}$ are considered brown dwarfs). This initial parameter space, spanning four decades in both semi-major axis and mass, is strongly restricted by considerations of both dynamics and observational surveys, as outlined in this section. These results are summarized in Figure \ref{fig:possible}, which delineates the available parameter space in the $(a,m)$ plane for possible new planets.

The orbits of the known planets are well determined, so that any additional bodies must be small and far away in order to avoid contradicting the working ephemerides. For example, a hypothetical planet with mass $m=10\,M_{\oplus}$ would produce perturbations to Saturn's orbit that would be detectable via telemetry of the Cassini spacecraft, if it were currently closer to the Sun than about $r\lesssim370\,$AU \citep{fienga2016}. In general, distant planets interact primarily through their stationary tidal effects over timescales of modern observations (which are short compared to the orbital period of such distant orbits). As a result, the bound on additional planets from the orbits of known giant planets scales as $m/r^3$ (see the discussion of \citealt{silsbee2018}). This constraint requires planets to fall below the purple line in Figure \ref{fig:possible}, although we note that such determinations are sensitive to the specifics of the employed dynamical model \citep{folknerdps,pitjevapitjev}.


On the other end of parameter space, solar system bodies cannot have overly large orbits and remain bound to the Sun. Passing stars can disrupt wide orbits over the age of the solar system. This outer boundary is not sharp, and this process must be addressed statistically \citep{liadams2015}. As a working estimate, orbiting bodies with semi-major axis $a\simgreat30,000$ AU are likely to be stripped from the sun by passing stars in the field (or at least have their orbital elements drastically altered), as marked by the dashed line in the Figure.  The corresponding limits from the solar birth cluster are much more restrictive, so that surviving planets must have semi-major axes $a\simless1000$ AU (\citealt{liadams2016}; see also section \ref{sec:formation} for more detail). This limit requires planets to lie to the left of the red line in Figure \ref{fig:possible}.

A lower limit on the mass arises from the definition of a planet. One of the characteristics of planets specified by the International Astronomical Union is that the body must be massive enough to clear its orbit over the age of the solar system. This requirement can be written in several different forms, and implies that the minimum planetary mass is an increasing function of orbital distance. Here we use the constraint advocated by \cite{margot}, which can be written in the form 
\be
\Pi = k {m \over M_\odot\, a^{9/8}} > 1 \qquad {\rm or} \qquad 
m \simgreat 1\,M_{\oplus} \left({a\over380 {\rm AU}}\right)^{9/8}\,.
\label{margot}
\ee
This lower limit on the planetary mass is shown as the green line in Figure \ref{fig:possible} (see \citealt{soter} for an alternate treatment of this criterion). 

An upper limit to the mass of any possible new planets is provided by {\sl WISE} observations, which rule out bodies larger than Saturn out to distances $\sim3\times10^4$ AU (see \citealt{luhman2014} for further details).  This upper limit is shown as the yellow line in Figure \ref{fig:possible}. Note that stronger limits can be derived for smaller distances, but such results are model-specific \citep{meisner2017a, meisner2017b}. This is because infrared emission is dominated by the internal energy sources of the planet, rendering its observational signature (surface emission in the {\sl WISE} bands) dependent on the detailed interior structure of the planet \citep{ginzburg16,linder2016}.

In light of the constraints delineated above, we can speculate that generically, planetary semi-major axes must be measured in hundreds of AU, but cannot exceed $\sim1000\,$AU. Meanwhile, the allowed masses range from the mass of Earth to that of a sub-Saturn\footnote{See e.g., \citet{petiplanets} for an extrasolar characterization of such an object.}. As we will see below, the P9 parameters necessary to explain the dynamical anomalies in the orbits of distant TNOs requires the hypothetical planet to have semi-major axis and mass that reside precisely within the allowed region.

\section{Anomalous Structure of the Distant Kuiper Belt}
\label{anomalous}

Observational characterization of the orbital architecture of the classical ($a<100\,$AU) domain of the Kuiper belt discussed in the previous section has had a profound effect on reshaping our understanding of the outer solar system's evolutionary history \citep{Levisonetal2008,batygin2011}. Indeed, as the structure of the trans-Neptunian region came into sharper focus a little over a decade ago, the hitherto conventional, in-situ formation narrative of the solar system (e.g. \citealt{Cameron1988, pollack1996}) was gradually replaced with a strikingly dynamic evolution model, wherein the giant planets are envisioned to have formed closer to the sun and subsequently scattered onto their current orbits during a transient period of instability \citep{malhotra1995, tsiganis2005, batygin2010, Nesvorny2011,Nesvorny2015a}. As illuminating as mapping of the $a<100\,$AU domain of Kuiper belt may have been, an important aspect of its architecture is that very little of it is \textit{anomalous} -- that is, the vast majority of the observations can be readily understood as being a consequence of gravitational sculpting facilitated by the known giant planets of the solar system. Remarkably, the same statement does not hold true for trans-Neptunian objects with semi-major axes in excess of $a \gtrsim 250\,$AU.

\begin{figure}[tbp]
\centering
\includegraphics[width=\textwidth]{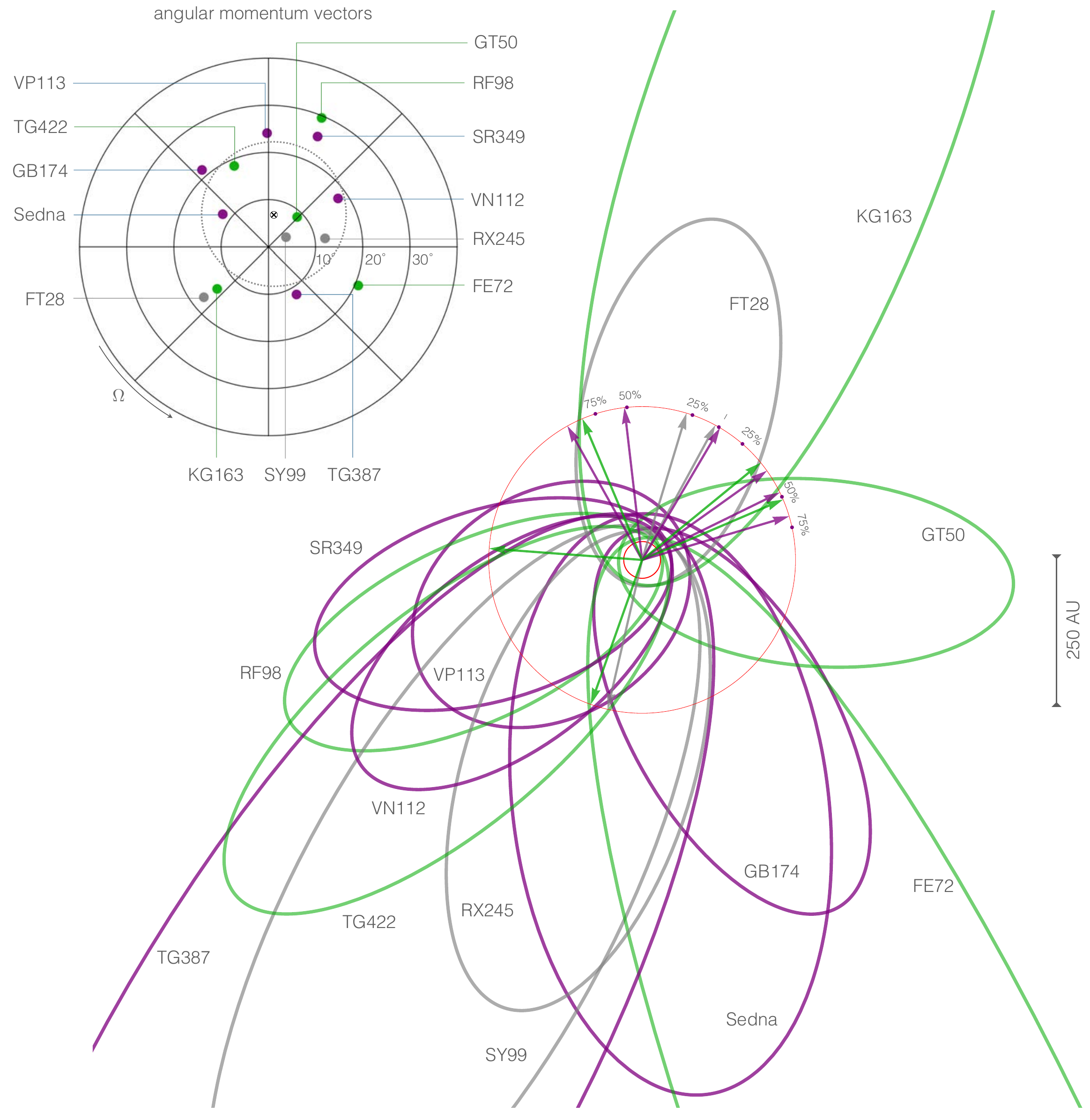}
\caption{Orbits of the distant Kuiper belt objects in physical space. The 14 illustrated objects have semi-major axis $a\ge250$ AU, perihelion $q\ge30$ AU, and inclination $i\le40$ deg. The arrows depict the perihelion directions measured from the position of the Sun, where all of the vectors extend out to 250\,AU to illustrate the non-uniformity of their apsidal orientations. The locations of the first, second, and third quartiles corresponding to the $\varpi-$distribution of (meta)stable objects is marked on the surrounding circle. The polar inset plot shows the positions of the angular momentum vectors of the same 14 KBOs, where the radial coordinate informs the orbital inclination and the azimuthal angle corresponds to the longitude of ascending node. The mean polar coordinates of stable and metastable KBOs are marked by the $\otimes$ sign, and the dispersion of the vectors around the mean is shown with a dotted circle. Each object is color-coded in accordance with its present-day dynamical stability as follows. Orbits depicted in purple correspond to the Neptune-detached population, and have dynamical lifetimes much longer than the age of the solar system. Orbits shown in green experience comparatively rapid dynamical chaos due to interactions with Neptune. An intermediate class of orbits that only experience mild diffusion over the age of the solar system are shown in gray. Note that the dynamically (meta)stable objects exhibit significantly tighter apsidal confinement as well as clustering of the orbital poles than their unstable counterparts.}
\label{fig_orbits}
\end{figure}

A diagram depicting the fourteen presently known\footnote{In this review, we consider the census of trans-Neptunian objects to be comprised of all objects listed in the minor planet center database, as of October 10th, 2018.} long-period KBOs with $a\geqslant250\,$AU, $q\geqslant30\,$AU, and $i\leqslant40\deg$ is presented in Figure \ref{fig_orbits}. The orbits are shown from the perspective of the north ecliptic pole, and have their apsidal lines (vectors pointing into the perihelion direction from the sun) marked by arrows, which are $250\,$AU in length. Additionally, the inset on the top left corner of the diagram shows a polar projection of the angular momentum vectors of the individual KBOs, and thus informs the magnitude, and the direction of the orbital tilts. By and large, these objects have (apparently) randomly distributed semi-major axes ranging from hundreds to thousands of astronomical units, and large eccentricities that typically equate to perihelion distances that approximately cradle Neptune's orbit, with $q\sim35-45\,$AU. Notable exceptions to this rule of thumb include Sedna \citep{Brownsedna}, 2012\,VP$_{113}$ \citep{trushep2014}, 2015\,TG$_{387}$ \citep{goblin} which have $q = 76, 80,$ and $65\,$AU respectively. 

The broad range of perihelion distances exhibited by the population of long-period KBOs translates into a widely varied degree of gravitational coupling between Neptune and the minor bodies, which in turn determines the dynamical stability of their orbits. In particular, distant objects with $q$ somewhat smaller than $40\,$AU are typically embedded within a chaotic region of phase-space, generated by overlapping exterior mean motion resonances with Neptune. Correspondingly, they experience stochastic orbital evolution over timescales that greatly exceed the orbital periods of the objects \citep{MorbyTNOreview}. This point is of considerable importance to understanding the intrinsic architecture of the distant Kuiper belt, since chaotic dynamics inevitably acts to erase any innate orbital structure that the population of bodies may otherwise have had. Moreover, objects that experience comparatively rapid semi-major axis evolution due to interactions with Neptune could plausibly represent (relatively) recent additions to the distant population of KBOs that were scattered out from the $a<250\,$AU region of the solar system and have not yet been strongly affected by P9-induced dynamics\footnote{The strong observational bias that favors the detection of low-perihelion objects leads to a considerable over-representation of dynamically unstable bodies in the observational sample of KBOs.}. Accordingly, in an effort to ascertain which subset of long-period KBOs reside on orbits that are likely to have been substantially altered by chaotic evolution (within the last $\sim$Gyr), we generated ten clones of each member of the distant Kuiper belt, and evolved them for 4\,Gyr under the influence of the known giant planets.

\begin{figure}[tbp]
\centering
\includegraphics[width=\textwidth]{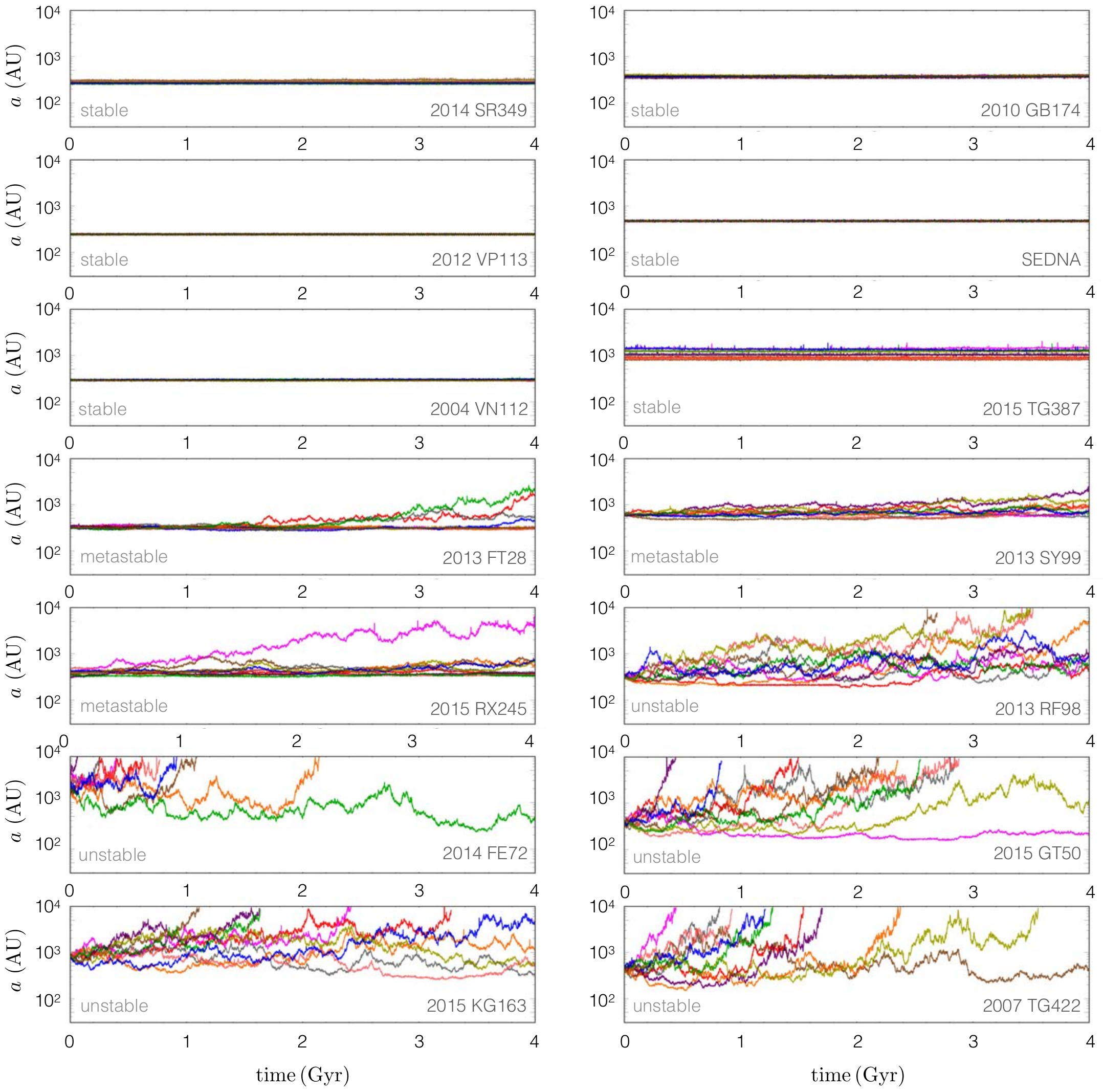}
\caption{Dynamical stability of the distant KBOs. Each Kuiper belt object shown in Figure \ref{fig_orbits} was cloned 10 times, and integrated forward for 4\,Gyr under perturbations from the canonical giant planets. The panels of this figure depict the semi-major axis time-series of the clones, which are individually colored. The perihelion-detached objects 2014\,SR$_{349}$, 2010\,GB$_{174}$, 2012\,VP$_{113}$, Sedna, 2004\,VN$_{112}$, and 2015\,TG$_{387}$ reside on long-term stable orbits and are rendered in Figure \ref{fig_orbits} as purple ellipses. The objects 2013\,FT$_{28}$, 2013\,SY$_{99}$ and 2015\,RX$_{245}$ experience limited orbital diffusion on $\sim$Gyr timescales, but are stable over the lifetime of the solar system. These metastable objects are shown in gray in Figure \ref{fig_orbits}. Finally, the dynamically unstable objects 2013\,RF$_{98}$, 2014\,FE$_{72}$, 2015\,GT$_{50}$, 2015\,KG$_{163}$, and 2007\,TG$_{422}$ are depicted in green on Figure \ref{fig_orbits}.}
\label{fig_stability}
\end{figure}

Figure \ref{fig_stability} shows the semi-major axis time-series of each of the objects depicted in Figure \ref{fig_orbits}. Upon examination, the observational census of distant KBOs can be qualitatively organized into three broad categories, based upon their dynamical stability. The KBOs 2014\,SR$_{349}$, 2012\,VP$_{113}$, 2004\,VN$_{112}$, Sedna, 2010\,GB$_{174}$ and 2015\,TG$_{387}$ experience essentially no orbital diffusion, and are completely stable. The orbits of these bodies are shown in purple on Figure \ref{fig_orbits}. In stark contrast, 2014\,FE$_{72}$, 2015\,GT$_{50}$, 2015\,KG$_{163}$, 2013\,RF$_{98}$, and 2007\,TG$_{422}$ exhibit rapid dynamical chaos, and are irrefutably unstable. These orbits are depicted on Figure \ref{fig_orbits} in green. Finally, the objects 2013\,FT$_{28}$, 2015\,RX$_{245}$ and 2013\,SY$_{99}$ evince only a limited degree of orbital diffusion, and therefore can be thought of as being dynamically metastable. This intermediate class of objects is shown on Figure \ref{fig_orbits} in gray. For uniformity, we will maintain this color-scheme for the remainder of the manuscript, whenever graphically representing the observational data.

Recall that while the semi-major axis and eccentricity define the size and shape of an orbit, its spatial orientation is determined by three Keplerian angles:
(1) longitude of perihelion, $\varpi$, which serves as a proxy for the apsidal orientation of the orbit
(2) the inclination, $i$, which determines the tilt of the orbital plane, and
(3) the longitude of ascending node, $\Omega$, which dictates the azimuthal direction into which the orbit is tilted (see Figure \ref{fig:elements}).
Importantly, all three of these angles show unexpected patterns beyond $a\gtrsim250\,$AU, and we briefly summarize them below. Throughout this review, we will consistently emphasize the stable and metastable subsets of Kuiper belt objects, which exhibit these anomalous patterns more clearly than their unstable counterpart (although we note that simply using the full dataset leads to qualitatively identical, and quantitatively similar conclusions). Accordingly, we will also apply the same demands for long-term dynamical stability to the theoretical calculations that will follow, with the aim of accentuating the closest points of comparison between theory and observations.

\subsection{Apsidal Confinement}
\label{sec:data_apsidalconf}

\begin{figure}[tbp]
\centering
\includegraphics[width=0.85\textwidth]{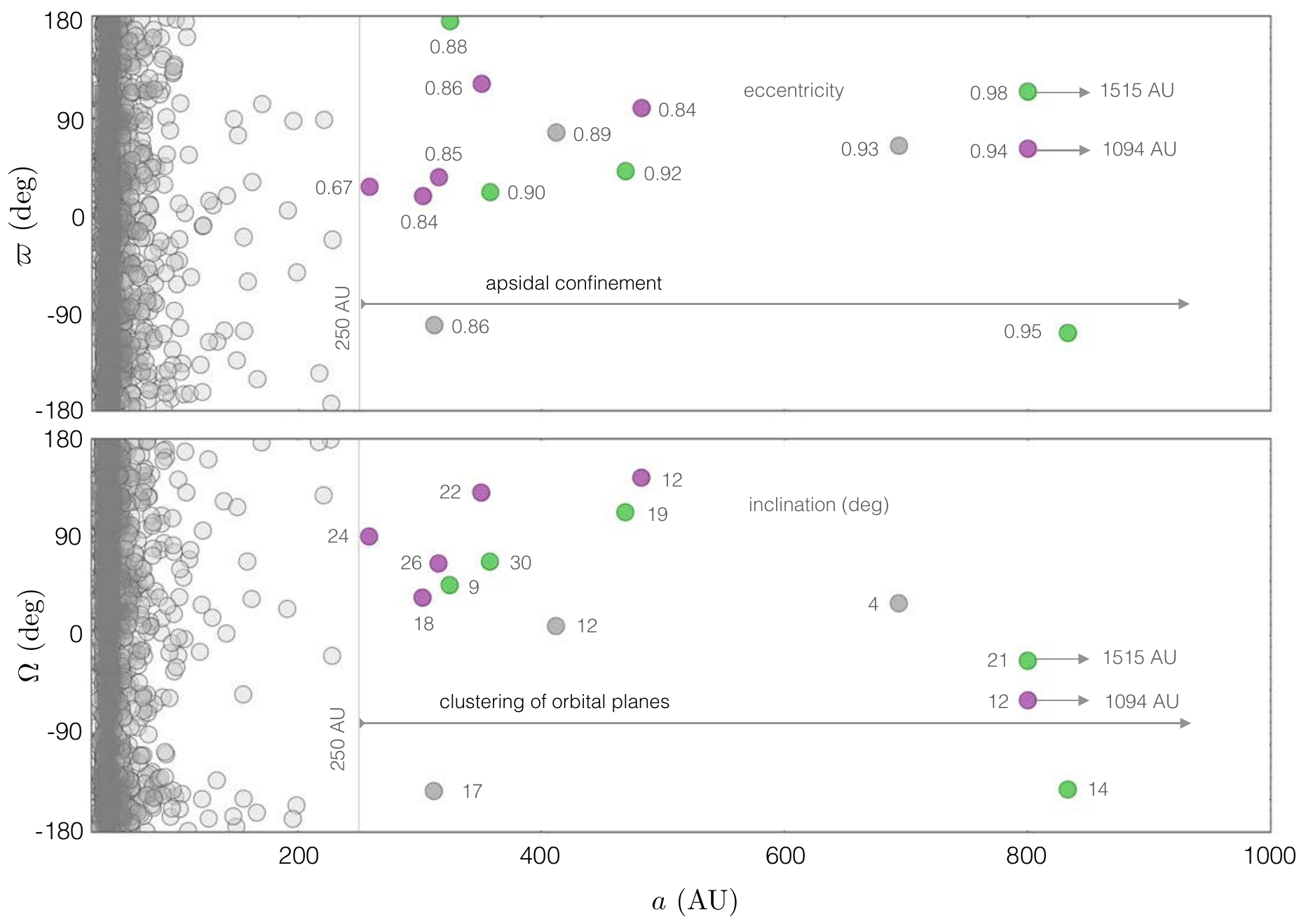}
\caption{Orbital elements of the distant KBOs (for the same 14 objects shown in Figure \ref{fig_orbits}). The top and bottom panels show the longitude of perihelion $\varpi$ and the longitude of ascending node $\Omega$ as a function of the semi-major axis, $a$, respectively. Each object is classified according to its dynamical stability (Figure \ref{fig_stability}) and is color-coded in the same way as in Figure \ref{fig_orbits}. The individual data points are further labeled by their corresponding eccentricity (top panel) and inclination (bottom panel). In both panels, the angular elements show a wide range of scatter and a nearly uniform distribution for small semi-major axes ($a\lesssim250\,$AU). For wider orbits with $a\gtrsim250$ AU, both panels show an emergent pattern of clustering among (meta)stable KBOs.}
\label{fig_data}
\end{figure}

Arguably the most visually striking characteristic of the distant Kuiper belt is the apsidal confinement of the orbits. While clearly evident in the top-down view of the orbits in physical space (Figure \ref{fig_orbits}), the transition between apsidally randomized and clustered population of the Kuiper belt at $a\sim250\,$AU is most readily seen in the top panel of Figure \ref{fig_data}, where the longitude of perihelion is shown as a function of the semi-major axis. A simple way to quantify this confinement is to separate the $a\geqslant250\,$AU data into two $180\deg$ wide $\varpi$ bins, with one bin centered on the mean longitude of perihelion, $\langle\varpi\rangle\approx60\deg$ and the other on $\langle\varpi\rangle-180\deg$. Notably, 8 out of 9 dynamically (meta)stable objects reside within $\langle\varpi\rangle\pm90\deg$, with the third quartile of the data located $Q_3-\langle\varpi\rangle\approx48\deg$ away from the mean.


A point of key importance is that if left to evolve exclusively under perturbations arising from the canonical giant planets, the observed apsidal confinement of long-period orbits would not persist, as a result of \textit{differential precession} \citep{dm1999}. As is well known, orbits in a purely Keplerian potential are perfect ellipses with turning angle of exactly $2\pi$ over a full period \citep{contopoulos}. Any departure from a pure $\mathcal{V}\propto1/r$ potential thus results in orbits that do not close, but rather precess, leading to the slow dispersion of the apsidal lines. To illustrate this quantitatively, consider the orbit-smoothed gravitational potential of the giant planets, averaged over the Keplerian trajectory of a Kuiper belt object. In terms of orbital elements, this expression (which serves as the Hamiltonian for the problem at hand) reads:
\begin{align}
\bar{\mathcal{V}}=-\frac{1}{8}\frac{\G\,\Msun}{a}\frac{3\cos^2(i)-1}{\big(1-e^2\big)^{3/2}} \sum_{j=5}^{8}\frac{m_j\,a_j^2}{\Msun\,a^2},
\label{potentialJSUN}
\end{align}
where the sum runs over the individual contributions from Jupiter, Saturn, Uranus, and Neptune \citep{Mardling2010, Gallardo2012}. 

Application of Lagrange's planetary equations (Hamilton's equations in non-canonical form) to equation (\ref{potentialJSUN}) yields the perihelion precession rate \citep{dm1999}:
\begin{align}
\frac{d\varpi}{dt} =-\frac{\sqrt{1-e^2}}{e\sqrt{\G\,\Msun\,a}}\frac{\partial \bar{\mathcal{V}}}{\partial e}= \frac{3}{4}\sqrt{\frac{\G\,\Msun}{a^3}}\frac{1}{\big(1-e^2\big)^{2}} \sum_{j=5}^{8}\frac{m_j\,a_j^2}{\Msun\,a^2},
\label{varpidot}
\end{align}
where we have assumed that the inclination is small enough to approximate $\tan(i)\approx0$ and $\cos(i)\approx1$. As an example, for SR$_{349}$ and Sedna, this expression yields apsidal precession rates of $\dot{\varpi}_{\mathrm{SR}} \approx0.8\deg/$Myr and $\dot{\varpi}_{\mathrm{Sedna}}\approx0.15\deg /$Myr, respectively. Accordingly, the steep inverse dependence of the apsidal precession rate on the KBO's semi-major axis implies that the presently confined group of objects would become uniformly distributed in $\varpi$ on a timescale of order a few hundred million years -- that is, an order of magnitude shorter than the age of the solar system. Thus, whatever perturbation is responsible for the apsidal clustering of the long-period orbits, it is very likely to operate continuously and have a characteristic timescale not exceeding a $\sim\,$Gyr.

While it is tempting to assume that apsidal confinement must be explained by some gravitational mechanism, the possibility exists that the observed alignment is simply due to random chance. A simplistic manner in which we can gauge the probability of a chance alignment is to assess its statistical significance. The Rayleigh Z-test, for example, which is used to determine if angles on a circle deviate from a uniform distribution, indicates that the 14 KBOs with $a\geqslant250$ AU are clustered at the 94\% confidence level. 

This calculation does not, however, take into account observational biases which could affect the observed distribution of Kuiper belt objects in the solar system. A well-known example of such bias is that objects on highly eccentric orbits -- including those depicted on Figure \ref{fig_orbits} -- are predominantly found near perihelion where they are closest and brightest. If astronomical surveys are biased in where they detect KBOs, these survey biases could directly translate into biases in the distribution of longitude of perihelion. There is another well-known longitude bias in KBO surveys: nearly all surveys avoid the galactic plane, where the density of background stars makes the discovery of faint KBOs challenging. This bias is, however, symmetric and would be unable to cause clustering of $\varpi$ in one direction. Nonetheless, it is important to evaluate the possibility that the combination of choices of survey observing locations, weather, and stellar density could conceivably conspire to cause apparent clustering where none is present.

\citet{brown2017} addresses the issue of observational bias in $\varpi$ by using the entire catalog of KBO discoveries to construct individual bias functions for each of the distant KBOs. The probability distribution function of $\varpi$ for each distant object indeed shows considerable bias, particularly against the galactic plane, but no overall preference for a single direction is found. \citet{brownbatygin2018} updated the probability that the observed clustering is due to random chance by using these calculated bias functions to perform Monte Carlo simulations where they randomly select $\varpi$ for each of the distant objects by assuming a uniform distribution modified by the observational bias of the individual object. The clustering seen in these Monte Carlo samples (as measured by a modified version of the Rayleigh Z-test) exceeds that of the real data only 4\% of the time. Thus, including observational biases, $\varpi$ is clustered at the 96\% confidence level. It is interesting to note that although observational biases exist, they make little difference to the assessment of significance of the clustering. This result is not surprising given the lack of biases that would plausibly cause clustering in a single direction. 


\subsection{Clustering of the Orbital Planes}
\label{sec:data_orbitalplanes} 

A second intriguing feature exhibited by long-term stable, distant KBOs is that they cluster around a common orbital plane, which is appreciably inclined with respect to the ecliptic. This clustering is easily discernible in the polar projection of the angular momentum vectors shown in Figure \ref{fig_orbits}, where the gray and purple points predominantly occupy the upper half of the graph. It is this effect, combined with the perihelion confinement discussed above, that gives rise to the approximate alignment of orbits in physical space \citep{phattie}. An important side-effect of this alignment is the grouping of the argument of perihelion, $\omega=\varpi-\Omega$, first pointed out by \citet{trushep2014}. It is further worth noting that although clustering of $\omega$ remains robust beyond $a\gtrsim250\,$AU, detections of new KBOs have largely eliminated this feature in the $150\,$AU\ $\lesssim a \lesssim 250\,$AU orbital domain.

Unlike apsidal confinement -- which is largely captured by a single parameter -- clustering of the orbital planes requires simultaneous alignment of the longitudes of ascending node as well as the inclinations of long-period KBOs, in order to ensue. For this reason, rather than quoting a single degree of grouping as we did for $\varpi$, we instead consider the mean inclination and node of the distant objects, as well as the rms dispersion of KBO inclinations about this mean plane. Quantitatively, these numbers are $\langle i \rangle \approx 7\deg$, $\langle \Omega \rangle \approx 82\deg$ and $\sigma_{i} \approx 15\deg$ respectively. These quantities are illustrated graphically in the inset of Figure \ref{fig_orbits} by a dashed circle that is centered on a point marked with an $\otimes$ symbol. To further exemplify the $(e,\varpi)$ and $(i,\Omega)$ data in a uniform fashion, in the bottom panel of Figure \ref{fig_data} we show the longitudes of ascending node as a function of the semi-major axis, and label each object by its inclination.

Much in the same way that the apsidal confinement discussed in the previous sub-section is susceptible to differential precession, clustering of the orbital planes is subject to dispersal due to the differential regression of the ascending node, induced by the gravity of the giant planets. Applying Lagrange's equations as before, we obtain the rate of nodal regression from equation (\ref{potentialJSUN}) as follows \citep{dm1999}:
\begin{align}
\frac{d\Omega}{dt} =\frac{-1}{\sqrt{\G\,\Msun\,a\,(1-e^2)}\sin(i)}\frac{\partial \bar{\mathcal{V}}}{\partial i}=-\frac{3}{4}\sqrt{\frac{\G\,\Msun}{a^3}}\frac{\cos(i)}{\big(1-e^2\big)^{2}} \sum_{j=5}^{8}\frac{m_j\,a_j^2}{\Msun\,a^2}.
\label{omegadot}
\end{align}
Noting that the differential regression of the node operates on a comparable timescale to perihelion precession (given by equation \ref{varpidot}), we assert that in absence of a sustained restoring torque that maintains the near-alignment of long-period KBO orbits in physical space, the quadrupolar gravitational field associated with the known giant planets would randomize the orientations of the distant orbits on a timescale that is short relative to the age of the solar system\footnote{Accounting for Neptune scattering (which equations \ref{varpidot} and \ref{omegadot} explicitly neglect) would further reduce the characteristic timescale for randomization. However, by restricting our consideration of the observations to the subset of stable and metastable objects, we largely circumvent this complication.}.

Once again, we need to consider the possibility that the clustering of the orbital planes is due to chance, along with observational biases \citep{shankman2017}. Like $\varpi$, the measured distributions of both $\Omega$ and $i$ are easily biased by where surveys have detected objects. For example, most KBO surveys target at or near the ecliptic, requiring that objects in the survey be near either their ascending or descending node. A strong bias likewise exists for survey inclinations to be approximately equal to the ecliptic latitude at which the survey is undertaken. As before, no obvious set of biases should lead to the sort of clustering seen in the data. Nonetheless a full analysis is clearly needed.

\citet{brownbatygin2018} extend the method previously developed in \cite{brown2017} to calculate simultaneous bias functions for $\Omega$ and $i$. As before, they use the entire collection of KBO discoveries to calculate a probability distribution function of discovered orbital planes for each distant object under the assumption that $\Omega$ is uniform and $i$ is distributed identically to the $a\leqslant250\,$AU scattered KBOs. They find that the 14 KBOs with $a>250$ AU are clustered in their orbital poles at the 96.5\% confidence level. 

Both the apsidal orientation and orbital plane clustering are moderately significant, but the combined probability of both occurring simultaneously is only 0.2\% \citep{brownbatygin2018}. The distant KBOs are thus distinctly clustered at the 99.8\% confidence level. 


\subsection{Highly Inclined TNOs}

\begin{figure}[tbp]
\centering 
\includegraphics[width=0.99\textwidth]{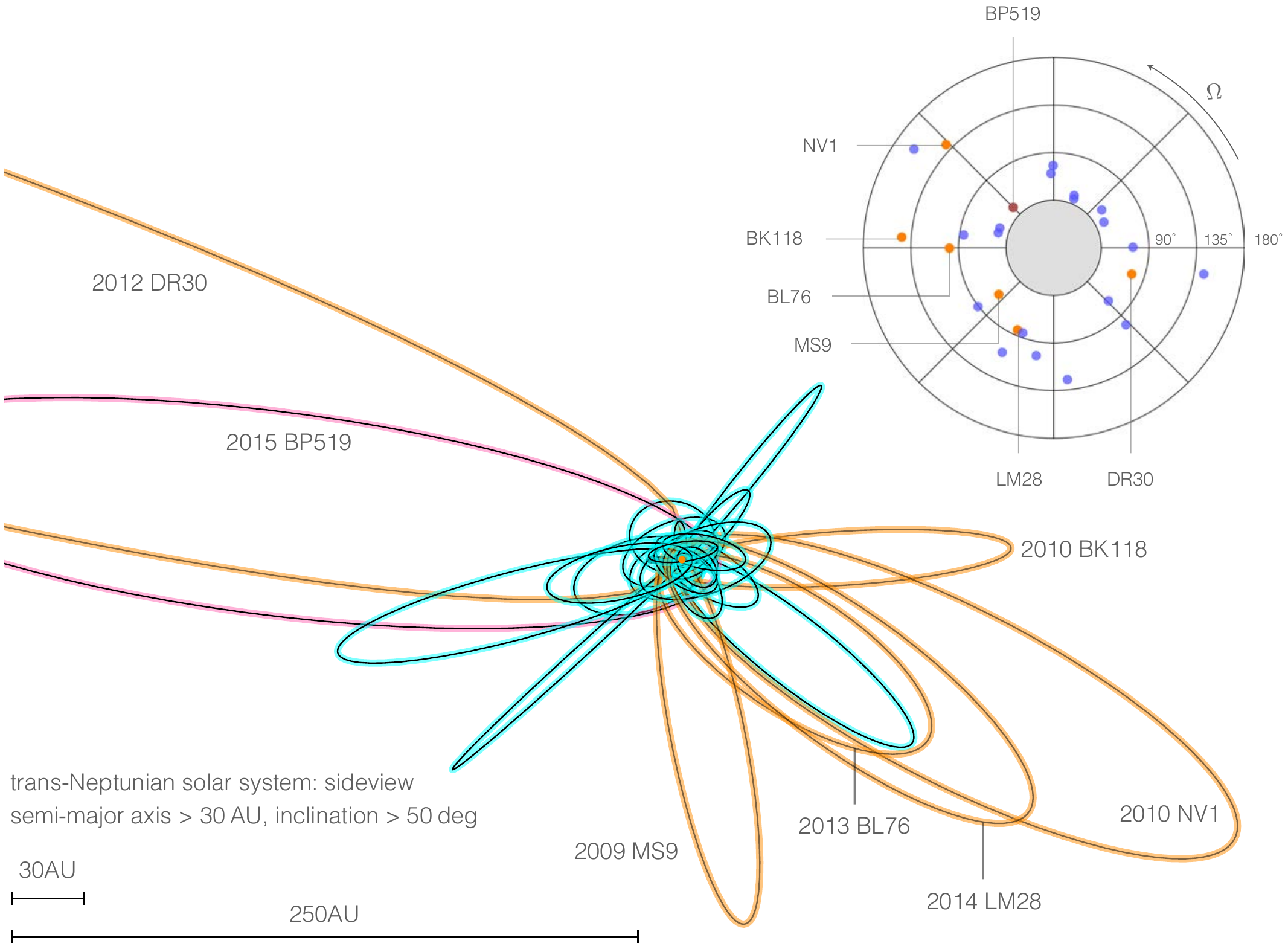}
\caption{Trans-Neptunian objects with high inclinations. This diagram shows the $i>50\deg$ subset of the distant solar system's small body population, as viewed from the ecliptic plane. Long-period Centaurs with $q<30\,$AU and $a\geqslant250\,$AU are shown in orange, while the KBO 2015\,BP$_{519}$ is denoted in pink. The orbits of more proximate high-inclination TNOs are rendered in cyan. Generically speaking, trans-Neptunian objects with $i>50\deg$ cannot be natually explained by the standard model of the solar system's formation and dynamical evolution.} 
\label{fig:highincTNOs} 
\end{figure}

A third puzzling population of trans-Neptunian objects is comprised of minor bodies that reside on orbits that are strongly inclined with respect to the plane of the solar system. Indeed, the current census of known TNOs contains 49 objects with inclinations above $i>40\deg$, with 10 of them occupying orbits with $i>90\deg$. With the exception of the recently discovered KBO 2015\,BP$_{519}$ ($i=54\deg$, $q= 36\,$AU, $a = 450\,$AU; \citealt{bp519}) all presently known high-inclination objects are Centaurs, meaning that they have $q<30\,$AU, and therefore veer into inter-planetary space at perihelion. For consistency with the preceding discussion, we sub-divide the high-inclination population of TNOs into a long-period component with $a>250\,$AU, and a sub-population occupying the more proximate part of the Kuiper belt (Figure \ref{fig:highincTNOs}).

Objects with orbital inclinations in excess of a few tens of degrees are neither a natural result of the solar system formation process (which unfolds in a geometrically thin, dissipative disk of gas and dust; \citealt{armitage2010}), nor an expected outcome of the solar system's post-nebular evolution \citep{MorbyTNOreview}. Indeed, detailed numerical simulations of the solar system's early dynamical relaxation carried out by \cite{Levisonetal2008} generate an inclination dispersion of TNOs that is largely confined to $i\lesssim30\deg$. While the more finely-tuned simulation suite of \citet{Nesvorny2015b} boost the upper envelope of the inclination distribution to $i\approx40\deg$, perpendicular and strongly retrograde objects such as Drac \citep{gladman2009}, Niku \citep{Chen2016} as well other long-period Centaurs \citep{gomes2015} are never produced in these calculations. This picture is further consistent with the recent simulation suite of \citet{bp519}, who demonstrate that even the $i=54\deg$ orbit of 2015\,BP$_{519}$ has a negligible chance of being produced self-consistently through Neptune scattering.

In order to understand why interactions with the canonical giant planets do not routinely yield highly inclined orbits, it is useful to consider a simplified toy-model of the dynamical evolution of scattered disk objects, wherein Neptune's orbit is taken to be circular, and acts as the sole source of perturbations. Within the framework of this circular restricted three body problem, the conservation of the test particle's specific energy is replaced by a near-conservation of the Tisserand parameter \citep{dm1999}:
\begin{align}
T=\left( \frac{a_8}{a} \right)+2\sqrt{\frac{a}{a_8} \left(1-e^2\right)}\,\cos(i),
\label{Tisserand}
\end{align}
where $a_8$ is Neptune's semi-major axis. Notably, this quantity is nothing other than the Jacobi constant (which is an exact integral of motion), in the limit of small planet/star mass ratio.

Employing the conservation of $T$, let us consider the fate of a particle that is initialized on a nearly planar, circular orbit, in the immediate vicinity of Neptune (e.g. at $a=33\,$AU). Facilitated by chaotic diffusion arising from overlap of Neptunian resonances (\citealt{Wisdom1980, murray1986}; see also \citealt{gomes2008review} and the references therein), the test particle will experience a random walk in energy, thereby slowly increasing its semi-major axis, while maintaining its perihelion distance close to its point of origin, at $q\sim a_8$. The characteristic inclination attained by this particle throughout its evolution is then
\begin{align}
i = \arccos\Bigg(\frac{a\,T-a_8}{2}\sqrt{\frac{a_8}{a\,q\,(2a-q)}} \Bigg) \sim10\deg,
\label{lowinc}
\end{align}
where we have set $\cos(i)\sim1$ in evaluating equation (\ref{Tisserand}), due to the assumed near-coplanarity of the initial condition. This simple calculation shows that it is the symmetry entailed by the conservation of $T$ that prevents significant excitation of orbital inclination in the scattered disk by Neptune.

Of course, the real dynamics of the solar system are more complicated than those encompassed by the circular restricted three body problem. Nevertheless, the qualitative picture implied by the above calculation illuminates the relevant limitations on the degree of orbital excitation that can ensue in the scattered disk population of the Kuiper belt. To restate this result simply, an external gravitational influence is required to generate the high-inclination orbits observed in the distant solar system.

One potential solution to this conundrum is to imagine that rather than being native to the Kuiper belt, these highly inclined objects are sourced from the Oort cloud. That is, perturbed by the combined action of the Galactic tide and passing stars (e.g., \citealt{heislertremaine}), very long period objects acquire near-parabolic trajectories, and upon crossing the orbit of Neptune get scattered inwards, becoming exotic members of the scattered disk \citep{levison2001}. Although such a scenario can in principle be invoked for high-$i$ Centaurs (see, however, \citealt{gomes2015}), considerable fine-tuning would likely be required to account for $q>30\,$AU objects like 2015\,BP$_{519}$ in this manner.

As always, we could consider observational bias here, but in this case there is nearly no need. Surveys for KBOs which observe near the ecliptic are biased against high inclination objects by a factor of $1/\sin (i)$ \citep{brown2001}. Thus, the detection of {\it any} KBOs and Centaurs with large inclinations suggests that the true population of high-$i$ TNOs is much more prominent. In other words, independent of observational bias, nearly perpendicular as well as strongly retrograde TNOs exist, and require an explanation beyond that which can be formulated within the framework of an eight-planet solar system.

\section{The Planet Nine Hypothesis: Analytical Theory}
\label{sec:analytical}

The current observational census of long-period Kuiper belt objects indicates that the dynamical origin of the anomalous structure of the distant Kuiper belt requires sustained perturbations beyond those that can be generated by the known giant planets of the solar system. In other words, a separate source of gravitational influence in the trans-Neptunian region is required to explain the anomalous patterns exhibited by the data. A series of recent studies \citep{phattie, batbrown2016b, batmorby, millholland, beckerp9, hadden, tali, caceres2018, li2018} have demonstrated that the existence of an additional, multi-Earth mass planetary member of the solar system -- Planet Nine -- can successfully resolve this theoretical tension. More specifically, (1) the apsidal confinement of distant KBOs, (2) perihelion detachment of long-period orbits (3) clustering of the $a\gtrsim250\,$AU orbital planes, and (4) excitement of extreme TNO inclinations can all be simultaneously explained by Planet Nine (P9)-induced dynamical evolution.


In order to define the Planet Nine hypothesis as a \textit{specific} theoretical prediction, we begin by presenting a purely analytical description of the dynamical mechanisms through which P9 is envisioned to sculpt the distant Kuiper belt. By and large, the following discussion will draw upon ideas of secular perturbation theory of celestial mechanics. Within the framework of this formalism, the Keplerian motion of all objects in the system is averaged over, leaving long-term exchange of angular momentum (but not energy) among the constituent bodies as the only physically active process. In other words, instead of computing the evolution of test-particles under the influence of point masses as done in the direct treatment of the gravitational N-body problem, the aim of secular theory lies in calculating the long-term behavior of test-orbits, subject to perturbations from massive wires that trace out the planetary trajectories\footnote{Formally speaking, the secular normal form is attained by canonically averaging the Hamiltonian over the mean longitudes, which are the fast angles of the system \citep{Morbybook, touma2009}}. Inherently, this mathematical procedure is equivalent to the so-called Gaussian averaging method, wherein the orbiting bodies are replaced by massive wires with line-densities that are inversely proportional to the instantaneous Keplerian velocities \citep{touma2009}.

\subsection{Secular Forcing}
\label{sec:secularforcing}
In order to classify the various flavors of secular interactions that can arise due to a distant planetary perturber, it is of considerable use to examine the approximate form of the P9-KBO coupling function i.e., Planet Nine's orbit-averaged gravitational potential. In terms of Keplerian elements, the octupole-level expansion of this function has the form \citep{Mardling2010}:
\begin{align}
&\bar{\mathcal{V}}_9 = -\frac{1}{16} \frac{\G\Msun}{a_9}\bigg(\frac{a}{a_9} \bigg)^2\frac{1}{\left(\sqrt{1-e_9^2}\right)^3} \Bigg[\underbrace{\bigg(1+\frac{3}{2}e^2\bigg)\big(3\cos^2(i)-1\big)\big(3\cos^2(i_9)-1\big)}_\text{Precession} \nonumber \\
&+\underbrace{15\,e^2\sin^2(i)\cos(2\omega)}_\text{Kozai-Lidov effect} + \underbrace{3\left(\sin(2i)\sin(2i_9)\cos(\Delta\Omega)+\sin^2(i)\sin^2(i_9)\cos(2\Delta\Omega)\right)}_\text{Interactions of the Planes} \nonumber \\
&-\frac{15}{8}\frac{e\,e_9}{1-e_9^2} \bigg(\frac{a}{a_9} \bigg) \bigg[\underbrace{\big(1+\cos(i)\big)\big(15\cos^2(i)-10\cos(i)-1\big)\cos(\Delta\varpi)}_\text{Eccentricity Coupling} \nonumber \\
&+\underbrace{\big(1-\cos(i)\big)\big(15\cos^2(i)-10\cos(i)-1\big)\cos(\varpi+\varpi_9-2\Omega)}_\text{High-Inclination Dynamics}\bigg] \Bigg],
\label{perfart}
\end{align}
where the subscript 9 refers to parameters of P9. Each harmonic in the above expansion governs a particular dynamical effect, and if dominant, entails a specific orbital architecture of the distant Kuiper belt. Let us briefly consider each of these terms and their physical meanings in isolation, as this will help us make sense of simple perturbative models for P9-induced dynamics that will follow.

\paragraph{Precession} The first term in equation (\ref{perfart}) governs apsidal precession and nodal regression of a KBO forced by P9. Akin to the discussion surrounding equations (\ref{varpidot}) and (\ref{omegadot}), on its own this effect only acts to randomize the physical orientation of the orbits, leading to an un-clustered orbital distribution. Thus, irrespective of its magnitude, this component of Planet Nine's potential only accelerates the evolution already enabled by the quadrupolar fields of the known giant planets.

\paragraph{Kozai-Lidov Effect} The second term in the above expansion possesses (two times) the argument of perihelion as its critical angle, and governs the Kozai-Lidov mechanism, which can facilitate a periodic exchange between eccentricity and inclination of a test-particle while conserving the $\hat{z}$-component of the orbital angular momentum vector $h_z=\sqrt{1-e^2}\cos(i)$ \citep{Lidov, Kozai}. If dominant, this resonance can lead to clustering among the arguments of perihelion (as first discussed in the context of the distant Kuiper belt by \citealt{trushep2014}), such that $\omega$ librates about $90$ and $270\deg$ for well-separated orbits, or alternatively about $0$ and $180\deg$ for orbits close to the perturber \citep{ThomasMorby}. It is worth noting, however, that the Kozai-Lidov resonance is easily destroyed by external sources of perihelion precession (such as that arising from the canonical giant planets) and even if active, would yield an orbital distribution in the distant Kuiper belt that is not compatible with the data.

\paragraph{Interactions of the Planes} The third harmonic in equation (\ref{perfart}) has the difference of the longitudes of ascending nodes of the TNO and P9 as the critical argument, $\Delta\Omega=\Omega-\Omega_9$, and governs the interactions between the orbital planes of Planet Nine and the TNOs it shepherds. The relative importance of this effect compared to the nodal regression forced by Jupiter, Saturn, Uranus and Neptune (equation \ref{omegadot}) dictates the equilibrium (Laplace) plane around which the orbits of the KBOs precess. That is, in the region of parameter space where forcing due to P9 dominates, the angular momentum vectors of the KBOs precesses around Planet Nine's orbit normal. Conversely, the angular momentum vectors of shorter period KBOs, whose evolution is primarily regulated by Neptune, will tend to precess around the total angular momentum of the canonical giant planets. This suggests that in order to perturb the planes of the distant KBOs, the orbit of P9 itself must be appreciably inclined with respect to the ecliptic\footnote{Notably, the harmonic term that follows is simply double the $\Delta\Omega$ term, and qualitatively achieves a similar effect. As suggested by its steeper dependence on the KBO's inclination, however, this $\propto\cos(2\,\Delta\Omega)$ correction only matters for highly inclined objects.}.

\paragraph{Eccentricity Coupling} The fifth term in the expansion contains the difference of longitudes of perihelion, $\Delta\varpi=\varpi-\varpi_9$, as its driving angle, and describes the coupling between the eccentricity (Runge-Lenz) vectors of P9 and the KBO orbits. Accordingly, oscillations in the KBO's eccentricity and orbital orientation relative to Planet Nine's apsidal line are regulated by this term. Unlike the preceding quadrupolar harmonics, this effect is octupolar in nature, and therefore explicitly depends on Planet Nine's eccentricity. This simple fact alone already implies that in order for P9 to facilitate any degree of apsidal confinement via bounded libration of $\Delta\varpi$, its orbit must have non-zero eccentricity.

\paragraph{High-Inclination Dynamics} Much like the $\Delta\varpi$ harmonic, the final term of equation (\ref{perfart}) is also octupolar in nature, but its dynamical consequences are considerably more subtle. Like the Kozai-Lidov resonance, this term mixes longitudes of perihelion and node, facilitating a complex dynamical evolution that simultaneously modulates the degrees of freedom related to the TNO's eccentricity and inclination (\citealt{batmorby,li2018}). As we will argue below, it is this harmonic that drives high-inclination dynamics and orbit-flipping behavior in the trans-Neptunian region of the solar system. \\
\\
With a rudimentary description of secular forcing in place, we are now in a position to construct a sequence of simplified analytical models for P9-induced evolution, and examine how they connect to the three primary features of the anomalous structure of the distant Kuiper belt. Particular emphasis will be placed on inspecting the dynamics qualitatively, with the aid of integrable Hamiltonians. As in the previous section, we begin by considering apsidal confinement of long-period KBOs.

\subsection{Apsidal Confinement}\label{subsection:apsconf}
Our examination of the functional form of $\bar{\mathcal{V}}_9$ implies that the apsidal confinement is dominantly driven by the secular angle containing the difference between the longitudes of perihelion. Let us now consider a perturbative model, characterized by this critical argument. In order to tease out the pertinent dynamics, it is advantageous to restrict the evolution to the plane, by assuming that $\sin(i_9),\sin(i)\rightarrow0$. From equation (\ref{perfart}), it is then evident that all but a single harmonic term -- namely $\cos(\varpi-\varpi_9)$ -- vanish from the disturbing function. Indeed, this simplification is sufficient for us to write down an integrable model for the secular motion of KBOs, perturbed by the canonical giant planets and Planet Nine.

\begin{figure}[tbp]
\centering
\includegraphics[width=1\textwidth]{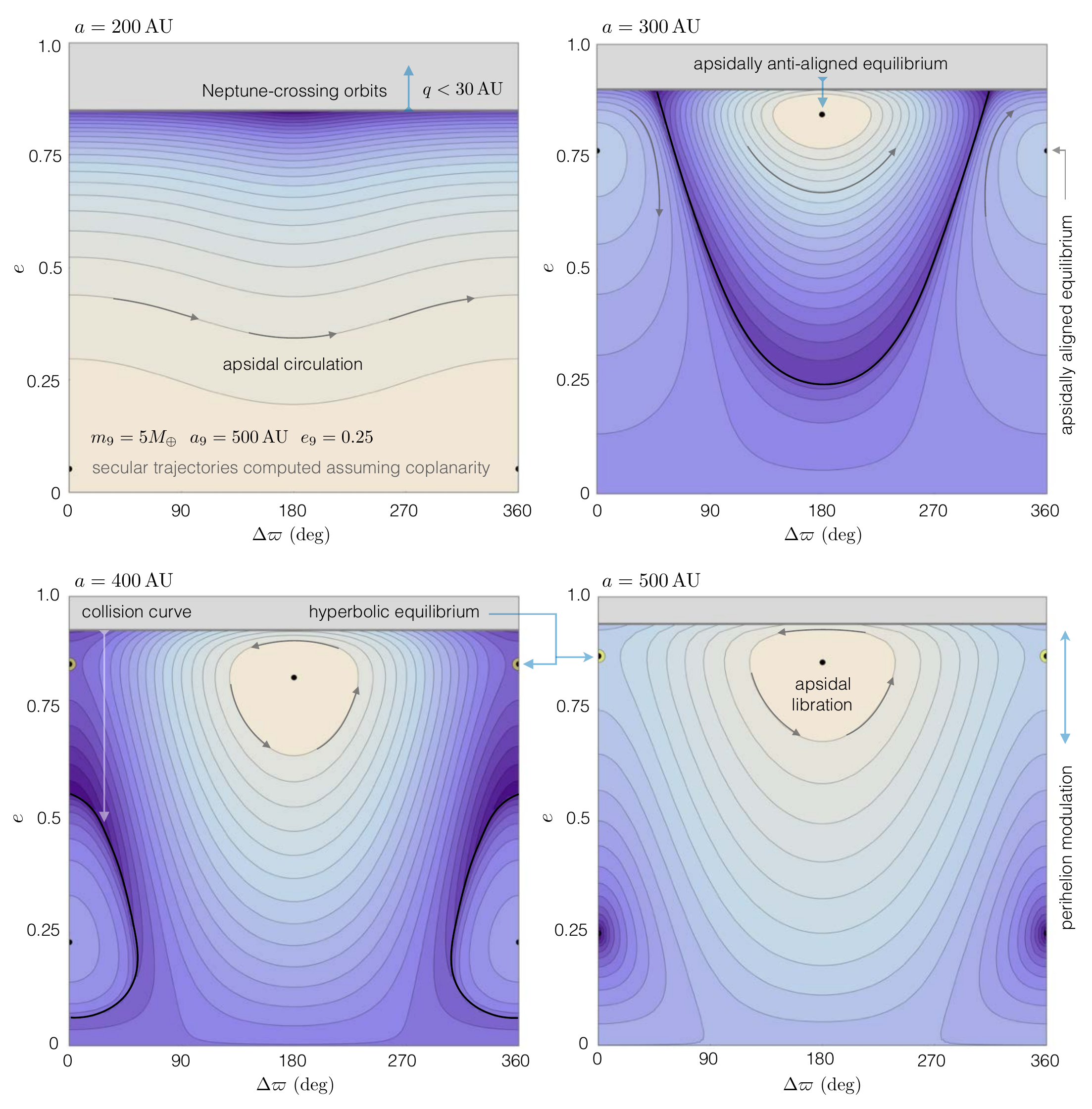}
\caption{Secular eccentricity dynamics induced by Planet Nine. Level curves of Hamiltonian (\ref{Hpuresec}) are projected onto the ($e,\Delta{\varpi}$) plane, at various choices of KBO semi-major axes. All cases use
canonical Planet Nine parameters of $m_9=5\,M_{\oplus}$, $a_9=500\,$AU, $e_9=0.25$, and $i_9=20\deg$. These equal-$\mathcal{H}$ contours represent analytic approximations to slow variations experienced by KBO orbits within the framework of the Planet Nine hypothesis. That is, over timescales much longer than an orbital period, the KBO eccentricity and longitude of perihelion gradually evolve along the depicted curves. Note that for a comparatively small KBO semi-major axis of $a=200\,$AU (upper left panel), the apsidal angle $\Delta\varpi$ circulates, implying a nearly uniform distribution of longitudes of perihelion. At $a=300\,$AU (upper right panel), however, a stable apsidally anti-aligned equilibrium point appears at high eccentricity. The prevalence of secular trajectories that encircle this $\Delta\varpi=180\deg$ fixed point signifies the emergence of perihelion confinement in the distant Kuiper belt. Note further that at even higher semi-major axes of $a=400\,$AU (lower left panel) and $a=500\,$AU (lower right panel), an additional hyperbolic equilibrium at $\Delta\varpi=0$ also emerges at high eccentricity.}
\label{fig:secular_e}
\end{figure}

The relevant Hamiltonian has the form (e.g., \citealt{beust,batmorby}):
\begin{align}
\Ham &= -\frac{1}{4}\frac{\G\,\Msun}{a}\frac{1}{\big(1-e^2\big)^{3/2}} \sum_{j=5}^{8}\frac{m_j\,a_j^2}{\Msun\,a^2} +\dot{\varpi}_9 \,\sqrt{\G\,\Msun a}\,\big(1-\sqrt{1-e^2} \big) \nonumber \\
&- \frac{1}{4\pi^2}\oint \oint \frac{\G\,m_9}{|\mathbf{r}-\mathbf{r}_9|}\bigg|_{i,i_9=0} \,d\lambda\,d\lambda_9.
\label{Hpuresec}
\end{align}
where $\mathbf{r}$ is the position vector, and $\lambda$ is the mean longitude (i.e., a fast angle that varies on the orbital timescale and informs the object's location on the orbit). 

The physical meaning of the three terms that comprise the above Hamiltonian can be understood as follows: the first term governs the slow precession of the KBO's longitude of perihelion due to the phase-averaged quadrupole fields of the canonical giant planets. The second term accounts for the fact that the reference frame as taken to be co-linear with the major axis of Planet Nine's orbit, and is therefore, also slowly precessing at the rate $\dot{\varpi}_9$, given by equation (\ref{varpidot}) (the generating function corresponding to the associated canonical contact transformation is spelled out in \ref{appB}). Most importantly, the third term governs the gravitational coupling between Planet Nine and the Kuiper belt object\footnote{Because in the assumed configuration the orbits can intersect, a series expansion of the form (\ref{perfart}) does not always provide a good representation of the dynamics. This complication is, however, easily circumvented by computing the averaged potential in closed form on the $e-\Delta\varpi$ plane \citep{Gronchi, beust}.}. 

Because the action conjugate to the fast angle $\lambda$ is a sole function of $a$, the process of averaging the Hamiltonian over the mean longitudes renders the semi-major axis of the KBO constant (e.g., \citealt{touma2009}). This means that within the framework of a purely secular model, $a$ acts as a \textit{parameter} of the problem, rather than a variable. Consequently, the contours of equation (\ref{Hpuresec}) projected onto the $e-\Delta\varpi$ plane at a given semi-major axis fully encompass the evolution entailed by the Hamiltonian.

Figure \ref{fig:secular_e} depicts a series of $e-\Delta\varpi$ phase-space portraits at KBO semi-major axes of $a=200,300,400$, and $500\,$AU, where we have adopted the Planet Nine parameters $a_9=500\,$AU, $e_9=0.25$ and $m_9=5\,M_{\oplus}$ (this particular choice of parameters is informed by the numerical simulations presented in section \ref{sec:numerical}). A key qualitative attribute of KBO dynamics that is immediately evident from Figure \ref{fig:secular_e} is that at large values of the semi-major axis (i.e., $a\gtrsim250\,$AU), Hamiltonian (\ref{Hpuresec}) is characterized by a pair of stable (elliptic\footnote{Note that at $a\gtrsim350\,$AU, an unstable (hyperbolic) $\Delta\varpi=0$ fixed point also emerges at large values of the eccentricity.}) equilibrium points: one at $\Delta\varpi=0$ and another at $\Delta\varpi=180\deg$. Meanwhile, only a single equilibrium point (at $\Delta\varpi=0$) exists for smaller semi-major axes. Note further that wherever present, the $\Delta\varpi=180\deg$ equilibrium resides at a higher eccentricity than its $\Delta\varpi=0$ counterpart. Importantly, the appearance of the stable $\Delta\varpi=180\deg$ fixed point beyond a critical value of $a$ is a robust feature of the model, and qualitatively explains the origin of the sharp changeover from randomized and apsidally confined sub-populations of the data.

The phase-plane portraits depicted in Figure \ref{fig:secular_e} demonstrate that if P9 exists, long-period orbits with perihelion distances close to Neptune will predominantly populate the apsidally anti-aligned island of libration. Moreover, once entrained in a stable mode of $\Delta\varpi\sim180\deg$ libration, orbits with initial $q\sim a_8$ will experience secular oscillations in eccentricity, periodically detaching from (and subsequently re-attaching to) Neptune. This effect provides a natural mechanism for generating the peculiar high-$q$ orbits of Sedna, 2012\,VP113, and 2015\,TG387 from typical scattered disk objects. Conversely, long-period orbits that are not locked into a stable $\Delta\varpi$ oscillation get driven to very high eccentricities, eventually crossing Neptune's orbit and leaving the system \citep{tali}. Figure \ref{fig:secular_e} further illustrates that comparatively short-period orbits occupy secular trajectories that simply circulate in longitude of perihelion. Thus, put simply, Figure \ref{fig:secular_e} shows that \textit{while the phase-averaged potential of Planet Nine is of little consequence for orbits with $a\lesssim250\,$AU, under its influence, scattered disk objects with $a\gtrsim250\,$AU will become dynamically organized into an apsidally confined configuration with a broad distribution of perihelion distances}.

\subsection{Clustering of the Orbital Planes}
Recall that in order to analytically describe apsidal confinement in the proceeding sub-section, we assumed exact coplanarity between the KBOs, Planet Nine, and the remainder of the solar system, which rendered inclination dynamics trivial. Let us now examine the consequences of abandoning this simplification, and introduce a small, but nevertheless finite tilt of P9's orbit with respect to the ecliptic. In particular, let us presume that the qualitative nature for $e-\Delta\varpi$ dynamics is not strongly affected by this development, and focus our attention on characterizing slow changes in $i$ and $\Omega$ in the distant Kuiper belt.

To approximately describe the secular evolution of the orbital planes of long-period KBOs, it suffices to consider the quadrupole component of the disturbing function (\ref{perfart}) under the assumption that inclinations remain small i.e., $\sin(i)\ll1$. Suitably, neglecting all terms with amplitudes that scale as $\propto\sin^2(i)$, we are left with an integrable Hamiltonian, containing only a single secular harmonic -- the difference between the longitudes of the KBO's and P9's ascending nodes (see \ref{appB}):
\begin{align}
\mathcal{H} &= -\frac{3}{8}\frac{\mathcal{G} M}{a} \frac{\cos(i)}{\left( 1-e^2 \right)^{3/2}} \sum_{i=1}^{4} \frac{m_i a_i^2}{M a^2} +\dot{\Omega}_9\, \sqrt{\mathcal{G} \, M \, a} \sqrt{1-e^2}  \left(1- \cos(i) \right) \nonumber \\
&- \frac{1}{4} \frac{\mathcal{G}\, m_9}{a_9} \left( \frac{a}{a_9} \right)^2 \big(1-(e_9)^2\big)^{-3/2} \bigg[ \bigg( \frac{1}{4} + \frac{3}{8} e^2\bigg) \left(3 \cos^2(i_9)-1\right)\left(3 \cos^2(i)-1\right) \nonumber \\
&+ \frac{3}{4}\, \sin(2i_9) \, \sin(2i) \, \cos(\Delta \Omega) \bigg].
\label{Hinc}
\end{align}

In direct parallel with equation (\ref{Hpuresec}), the three terms comprising the Hamiltonian have well-defined physical meanings. Specifically, the first and second terms describe the regression of the KBO's longitude of ascending node (equation \ref{omegadot}), and the slow rotation of the reference frame (such that $\Omega_9$ is always zero) respectively. Meanwhile, the third term governs the angular momentum exchange between P9 and the KBO. Moreover, because Hamiltonian (\ref{Hinc}) describes a system with only a single degree of freedom, we may examine the corresponding secular evolution simply by projecting level curves of $\Ham$ onto the $(i,\Delta\Omega)$ plane, as before.

\begin{figure}[tbp]
\centering
\includegraphics[width=\textwidth]{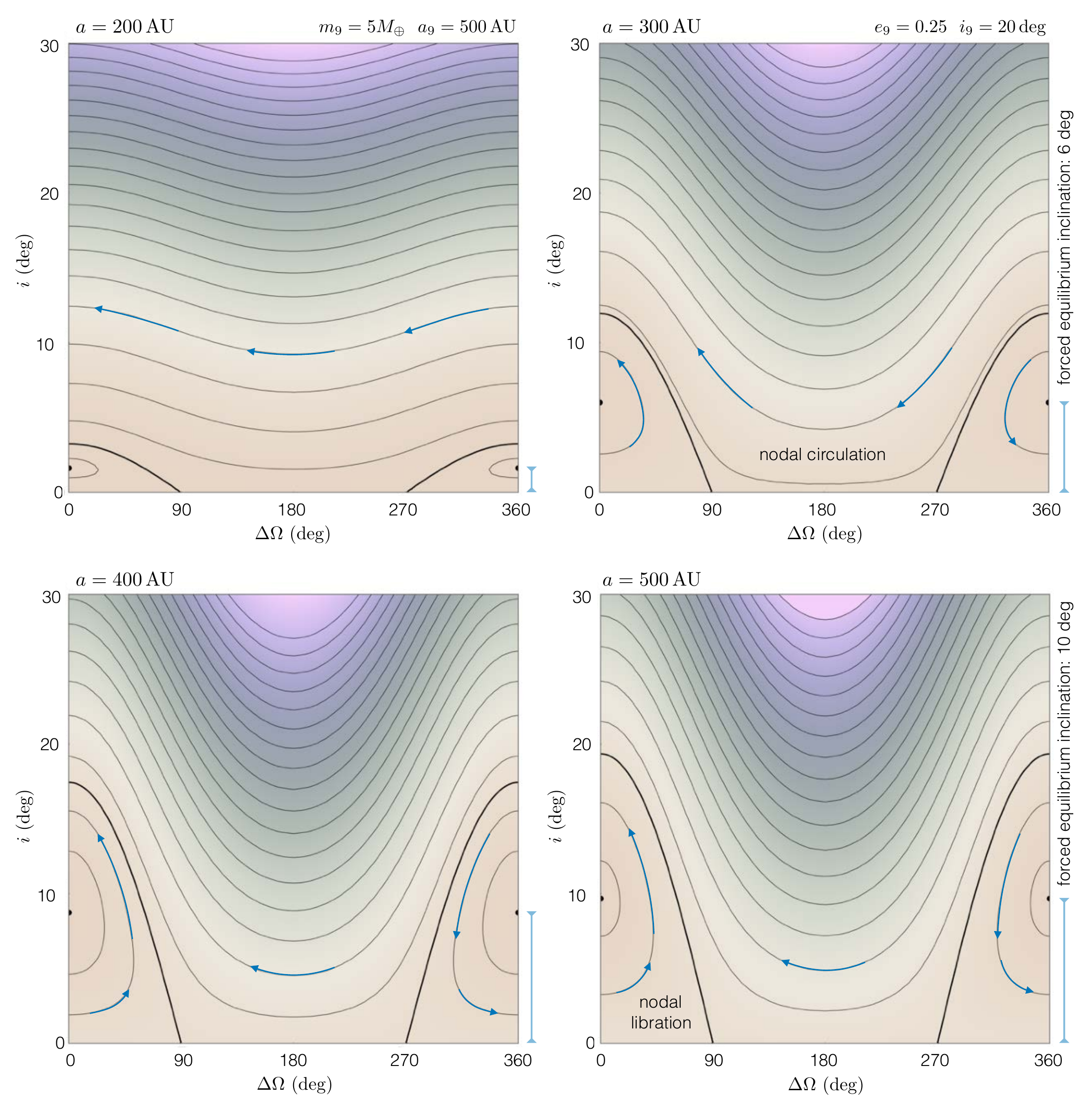}
\caption{Secular inclination dynamics induced by Planet Nine. Like Figure \ref{fig:secular_e}, these panels depict the contours of the secular Hamiltonian (\ref{Hinc}), which provides an analytic approximation to slow variations of KBO inclination and longitude of ascending node facilitated by Planet Nine. As can be discerned from the left-right, top-down progression of panels, prominent regions of nodal libration associated with clustering of the angular momentum vectors develop with increasing KBO semi-major axis. Unlike the case of apsidal confinement delineated in Figure \ref{fig:secular_e}, however, no new equilibrium points emerge with increasing $a$. Correspondingly, the clustering of the orbital poles ensues smoothly, and simply coincides with the bending of the distant solar system's equilibrium angular momentum plane by Planet Nine.}
\label{fig:secular_i}
\end{figure}

Figure \ref{fig:secular_i} shows the $i-\Delta\Omega$ analog of Figure \ref{fig:secular_e}, and exemplifies the dynamics of KBO orbital planes at different values of $a$. An important characteristic of P9's secular forcing that is clearly evident in the panels of Figure \ref{fig:secular_i} is the \textit{bending of the Laplace plane} i.e., the gradual change of the equilibrium inclination as a function of semi-major axis, and the appearance of a prominent island of $\Delta\Omega$ libration that surrounds this equilibrium. Outside of this island, relative longitude of ascending node circulates and KBOs experience considerable modulation of orbital inclination. Numerical simulations (section \ref{sec:numerical}) reveal that due to non-linear coupling between eccentricity and inclination dynamics, orbits residing far away from the $i-\Delta\Omega$ equilibrium either develop low perihelia and get ejected from the system, or experience large-scale $e-i$ excursions that remove them from the observable domain of the Kuiper belt (section \ref{sec:an_Theta}). As a result, apsidally-confined orbits can only maintain long-term stability if they lie close to the $i-\Delta\Omega$ fixed point in Figure \ref{fig:secular_i}, which approximately coincides with Planet Nine's orbital plane for large KBO semi-major axes. The emergence of the stable island of nodal libration at $a\gtrsim250\,$AU thus qualitatively explains the observed clustering of the orbital planes of long-period KBOs, as orbits residing on secular trajectories that encircle the $i-\Delta\Omega$ fixed point remain confined to this region of phase-space, causing the orbit poles to group together in physical space\footnote{Strictly speaking, Hamiltonian (\ref{Hinc}) is a leading order approximation to the full Hamiltonian, expanded as a power-series in $\alpha=a/a_9$, and is therefore not guaranteed to provide a good representation of the dynamics in the $\alpha\rightarrow1$ limit. Here, however, we are saved by the $\sin(i)\ll1$ approximation, which maintains the dominance of the leading-order term, rendering equation (\ref{Hinc}) an adequate model for understanding the relevant process (i.e., bending of the Laplace plane) in the context of the Planet Nine hypothesis.}.

While the transition from a randomized to clustered distribution of KBO orbital planes is keenly reminiscent of the onset of apsidal confinement discussed in the previous sub-section, it is worth noting that the two processes are theoretically distinct. In particular, the appearance of stable apsidally anti-aligned trajectories in Figure \ref{fig:secular_e} stems from the emergence of a new equilibrium point in phase space, and therefore occurs via a sharp transition. Moreover, the appearance of homoclinic curves in phase-space at $a\gtrsim350\,$AU renders libration of longitude of perihelion around $\Delta\varpi\sim180\deg$ a true \textit{secular resonance}. Conversely, the inclination-node fixed point shown in Figure \ref{fig:secular_i} is a \textit{forced equilibrium}, meaning that clustering of the orbital planes arises smoothly, as a function of $a$. Moreover, it is noteworthy that within the framework of the envisioned secular model, the existence of the apsidally confined population of KBOs requires Planet Nine to be eccentric but not necessarily inclined, whereas the clustering of the planes would ensue even if P9's orbit were circular. Therefore, to the extent that the perturbative Hamiltonians (\ref{Hpuresec}) and (\ref{Hinc}) approximate real motion, these two dynamical processes independently inform the necessary orbital eccentricity and inclination of Planet Nine, and only when considered simultaneously lead to an orbital distribution of long-period KBOs that exhibit large-scale clustering in physical space \citep{phattie}.

\subsection{Generation of Highly Inclined TNOs}
\label{sec:an_Theta}
Unlike the eccentricity-perihelion and inclination-node evolution described above, high-inclination dynamics induced by Planet Nine involves coupled modulation of two degrees of freedom (related to $e$ and $i$), with no clear symmetry inherent to the motion. Moreover, even in the simplified simulations of \citet{batmorby, li2018}, no discernible separation of timescales materializes, rendering the system intrinsically non-adiabatic. As a result of these complications, it is very difficult (if not impossible) to construct an integrable model for P9-driven high-inclination dynamics that will have any reasonable degree of quantitative accuracy. Thus, in the following discussion, we will limit ourselves to an essentially qualitative description of the secular evolution, with the hope of highlighting its most basic attributes. 

Despite being non-integrable in nature, high-inclination dynamics induced by P9 are primarily driven by a well-defined secular harmonic. This harmonic, $\theta$, is identified as being the octupole-level secular angle \citep{batmorby}
\begin{align}
&\theta= 2\omega - \Delta\varpi= \varpi+\varpi_9-2\Omega \nonumber \\
&\Theta = \frac{\sqrt{1-e^2}}{2}\big(1-\cos(i) \big),
\label{eqn:theta}
\end{align}
where $\Theta$ is the action conjugate to $\theta$, that emerges if we adopt $\Delta\varpi$ as the secular angle for the other degree of freedom (see \ref{appB}). We note that the physical meaning of $\theta$ is nothing other than the difference between the longitudes of perihelion of Planet Nine and a counter-revolving KBO, for which the retrograde longitude of perihelion is defined as $\varpi'=\Omega-\omega=2\Omega-\varpi$ \citep{Gayon2009}. In reality, the secular evolution in $(\theta,\Theta)$ is intimately coupled to motion in $(e,\Delta\varpi)$, meaning that during large-scale excursion of KBO inclination, the eccentricity changes in concert (while $\Delta\varpi$ executes complex librations around $180\deg$; see \citealt{LiCoplanar} for a related discussion). As a very crude illustration of the underlying dynamics, however, we can choose to ignore this reciprocity and freeze the evolution in ($e,\Delta\varpi$), which allows us to examine the level curves of the Hamiltonian as before.

\begin{figure}[t]
\centering
\includegraphics[width=0.65\textwidth]{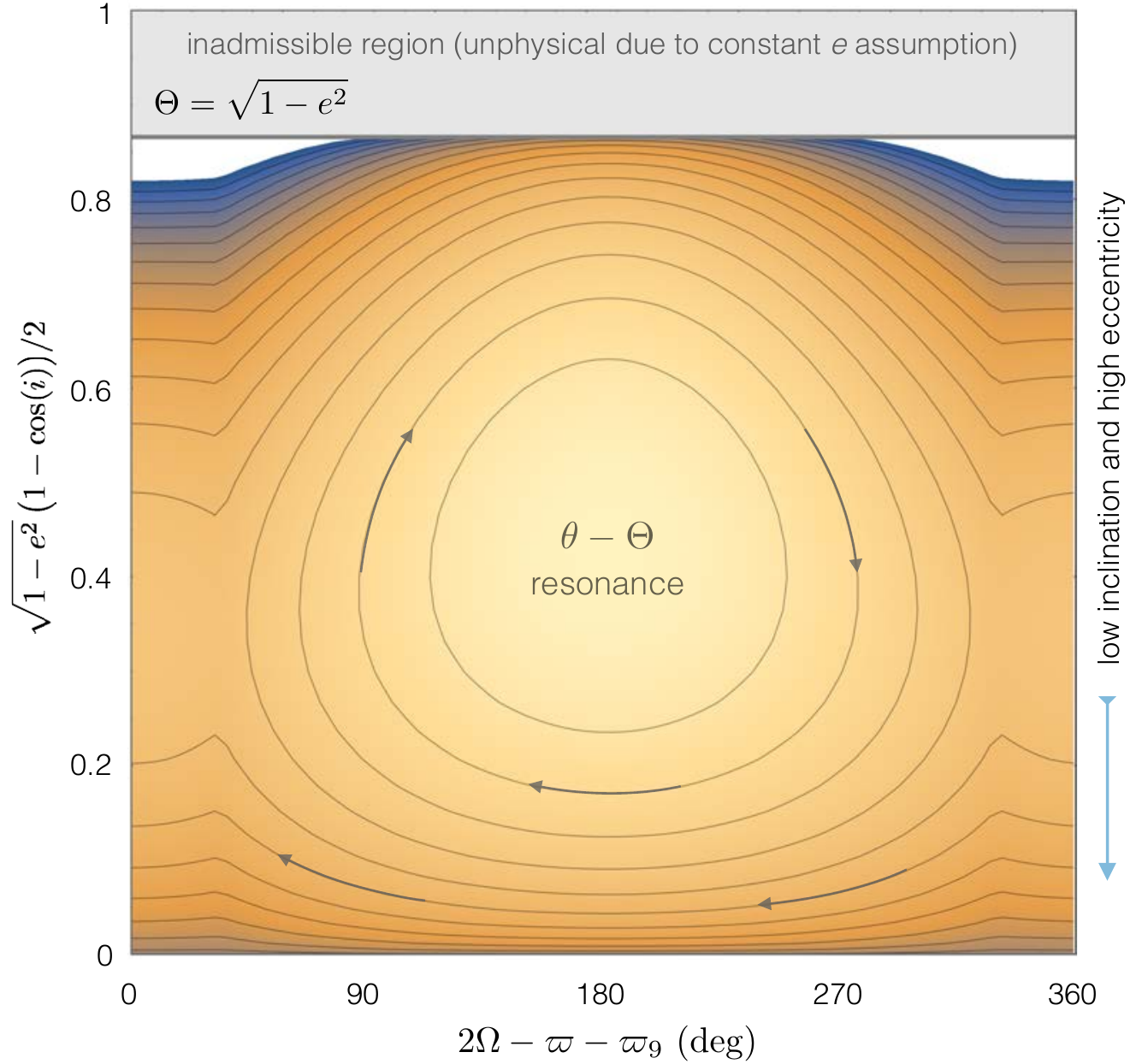}
\caption{Like Figures \ref{fig:secular_e} and \ref{fig:secular_i}, this diagram shows an analytic exemplification of large-inclination dynamics facilitated by Planet Nine. Unlike results depicted in Figures \ref{fig:secular_e} and \ref{fig:secular_i} however, this secular diagram only provides a very crude reproduction of the real high$-i$ behavior exhibited by TNOs within the framework of the Planet Nine hypothesis, and should be considered highly approximate (see text for details). Qualitatively, this figure illuminates the existence of two families of trajectories: at low values of the action $\Theta$ -- which corresponds to high eccentricity and low inclination -- the angle $\theta$ circulates, meaning that the $(\theta,\Theta)$ resonance plays little role in the dynamical evolution. If the action $\Theta$ is increased, however, the KBO may land on a secular trajectory characterized by libration of $\theta$, which in turn translates to large-scale variations of the orbital inclination.}
\label{fig:secular_theta}
\end{figure}

For definitiveness, let us take $\alpha=a/a_9=1$, $\Delta\varpi=180\deg$, and set $e=0.5$ (note from the right-bottom panel of Figure \ref{fig:secular_e} that secular eccentricity modulation in the plane can cover almost the entire $e\in(0,1)$ range). With these approximations in hand, we compute the function
\begin{align}
\Ham &= -\frac{1}{4}\frac{\G\,\Msun}{a}\frac{1}{\big(1-e^2\big)^{3/2}} \sum_{j=5}^{8}\frac{m_j\,a_j^2}{\Msun\,a^2} +\dot{\varpi}_9 \,\sqrt{\G\,\Msun a}\,\big(1-\sqrt{1-e^2} \, \cos(i) \big) \nonumber \\
&- \frac{1}{4\pi^2}\oint \oint \frac{\G\,m_9}{|\mathbf{r}-\mathbf{r}_9|}\bigg|_{e=0.5,\Delta\varpi=\pi} \,d\lambda\,d\lambda_9
\label{Hhighinc}
\end{align}
in closed form on the $(\theta,\Theta)$ plane, and project its contours on Figure \ref{fig:secular_theta}. 

A key feature that is immediately evident in Figure \ref{fig:secular_theta} is the existence of trivial circulating trajectories at low values of $\Theta$, and the emergence of a prominent island of $\theta$-libration at higher values of $\Theta$. Recalling the definition of $\Theta$ from equation (\ref{eqn:theta}), the picture outlined in Figure \ref{fig:secular_theta} implies that as long as the eccentricity is high and inclination is low -- which translates to low values of $\Theta$ -- the ($\theta,\Theta$) resonance plays no role in the dynamical evolution. In other words, the low-$\Theta$ regime of secular motion is the one where the integrable models outlined in the previous two sub-sections apply. 

Conversely, if $\Theta$ is allowed to reach a high enough value by some dynamical process, the system can transition into a regime where $\theta$ begins to undergo bounded oscillations, resulting in coupled variations in $\Theta$\footnote{The fact that $\Theta$ cannot exceed $\sqrt{1-e^2}$ in Figure \ref{fig:secular_theta} is an unphysical consequence of the constant eccentricity assumption inherent to our computation of equation (\ref{Hhighinc}).}. In turn, these excursions in $\Theta$ correspond to large changes in the orbital inclination. Although rudimentary and unsystematic, this interpretation of the dynamics is consistent with numerical results, where the orbit-flipping behavior of distant objects induced by Planet Nine is almost always triggered at the minimum of the KBO eccentricity cycle (which translates to a maximum in $\Theta$; \citealt{batmorby}).

\subsection{Mean-Motion Resonances}

The entirety of the theoretical discussion presented above is framed within the context of orbit-averaged perturbation theory. This approach to understanding dynamical evolution is sensible when the Keplerian motion of the interacting bodies is in essence, uncorrelated. Notably, however, this assumption is invalidated if the orbital periods of P9 and the KBO become commensurate, such that their ratio can be expressed as a fraction of two nearby integers. In this case, harmonics of the Hamiltonian involving the mean longitudes of P9 and the TNO can execute bounded oscillations, causing P9 perturbations to become resonant (as opposed to secular) in nature, and facilitate a coherent exchange of orbital energy and angular momentum.

Are mean-motion resonances (MMRs) with Planet Nine relevant to the dynamical evolution of the distant Kuiper belt? Strictly speaking, the answer is yes \citep{millholland, beckerp9, hadden, bailey2018}, although their observational consequences are thus far relatively insignificant. That is, even though resonant dynamics are implicated in ensuring long-term stability for KBOs that share Planet Nine's orbital plane, it is secular interactions that are dominantly responsible for sculpting the actual observed distribution of long-period KBOs \citep{batmorby}. Let us expand upon this point further.

Examination of particle evolution in the simulation suite of \citet{phattie} revealed that over the course of typical $4\,$Gyr integrations, KBO orbits can exhibit temporary capture into P9 MMRs, usually lasting $\sim10-100\,$Myr. This resonance-hopping behavior was more thoroughly investigated by \citet{beckerp9}, who further considered the long-term stability of observed KBOs as a constraint on P9's inferred orbit. Beyond purely theoretical considerations, the potential prevalence of P9 resonances in the distant belt inspired \citet{malhotra2016} to examine the numerological relationships between the orbital periods of observed KBOs, and note that 4 out of 6 objects known at the time lie close to N:1 and N:2 orbital period ratios with a putative perturber. Based upon these relationships, \citet{malhotra2016} suggested that P9 may reside on an orbit with $a\sim665\,$AU.

This line of reasoning was examined in a more quantitatively rigorous fashion by \citet{millholland}, who carried out a large-scale Monte-Carlo exploration of P9-forced dynamics of known KBOs, and derived a probability density function of P9 parameters that approximately maintains the KBOs in a clustered distribution, while temporarily preserving commensurabilities. However, \citet{millholland} simultaneously noted that resonant configurations generally have dynamical lifetimes that are considerably shorter than the age of the solar system, implying that even long-term stable KBOs do not remain bound to a single mean motion commensurability on $\sim$Gyr timescales (which is also seen in \citealt{beckerp9,hadden}). Practically, this means that even though some fraction of the observed KBOs can reasonably be expected to be locked resonance with Planet Nine, it is likely impossible to confidently identify P9 resonant vs non-resonant KBOs, or to determine the value of the specific resonant angle. Unfortunately, this tendency of long-period KBOs to stochastically skip between mean-motion commensurabilities essentially eliminates the promise of deriving P9's semi-major axis from the present-day orbital period distribution of observed KBOs \citep{bailey2018}.

If resonant effects can play a pronounced role in the dynamics of long-periods KBOs, then why is it sensible to adopt a purely secular Hamiltonian as a perturbative model for P9-induced evolution? Put simply, this is because the anomalous structure of the distant Kuiper belt is predominantly shaped by orbit-averaged interactions. More specifically, while MMRs may modulate the long-term stability of KBOs through the phase-locking mechanism \citep{Morbybook}, the clustering of orbits in physical space, as well as excitation of extreme inclinations of long-period TNOs is largely a secular effect. To demonstrate this separation of degrees of freedom, \citet{batmorby} considered a simplified 2D model of the solar system, where the gravitational fields of the canonical giant planets were replaced by an effective quadrupole ($J_2$) moment of the sun, leaving Planet Nine as the only direct perturber in the system. By restricting all orbital motion to a common plane, this physical setup renders all KBOs that are not phase-protected dynamically unstable, resulting in a fully resonant orbital distribution of distant KBOs.

\begin{figure}[tbp]
\centering
\includegraphics[width=1\textwidth]{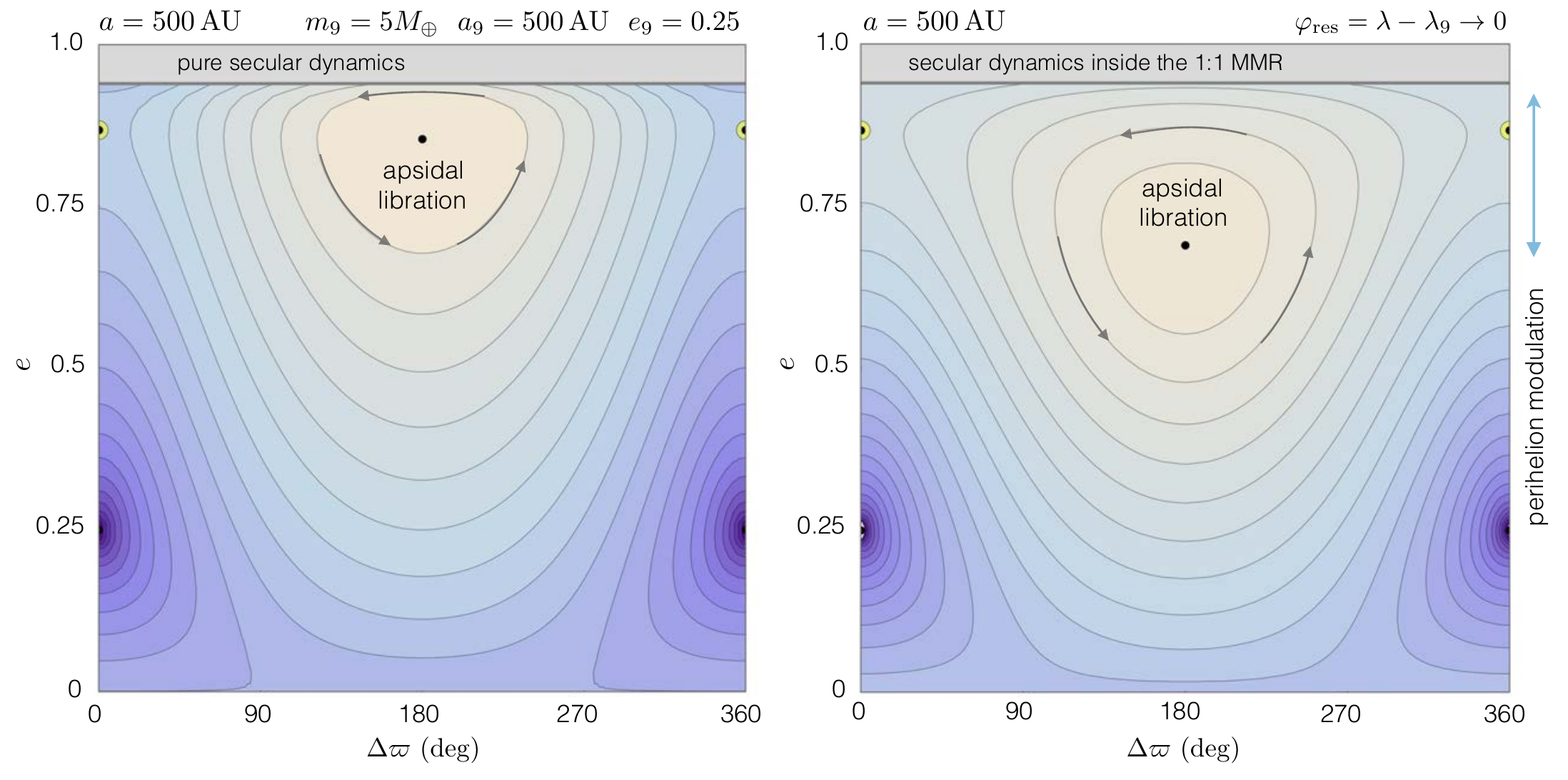}
\caption{A comparison between purely secular and resonant-secular dynamics induced by Planet Nine. The left panel shows the $a=500\,$AU (co-orbital) phase-space portrait, arising from equation ($\ref{Hpuresec}$) (also shown as the right bottom plot of Figure \ref{fig:secular_e}). The right panel depicts an equivalent $e-\Delta\varpi$ phase-space portrait corresponding to an object also with $a=500\,$AU, but residing in a 1:1 mean-motion resonance with Planet Nine, where the orbit-averaging procedure was carried out under the condition $\varphi_{\rm{res}}=\lambda-\lambda_9\rightarrow0$ (see \citealt{batmorby} for details). The similarity between the two panels highlights the inherent separation between the secular and resonant degrees of freedom.}
\label{fig:secular_e_mmr}
\end{figure}

Examining the dynamics of the simulated apsidally clustered objects, \citet{batmorby} pointed out that the resonant multiplets responsible for driving the KBO evolution typically have the form
\begin{align}
\varphi_{\textrm{res}}=p\lambda-q\lambda_9-(p-q)\varpi_9.
\label{resangle}
\end{align}
Resonances of this type (sometimes referred to as ``corotation" resonances) modulate the KBO's semi-major axis but not eccentricities (because $de/dt \propto \partial \cos{\varphi}/\partial \varpi$=0), leaving the $e-\Delta\varpi$ degree of freedom to be controlled by secular interactions. Accordingly, by carrying out the orbit-averaging procedure under the resonant condition, \citet{batmorby} demonstrated that secular dynamics embedded within P9 MMRs have a very similar phase-space topology to the purely secular phase-space portraits computed using Hamiltonian (\ref{Hpuresec}). As an example of this correspondence, Figure \ref{fig:secular_e_mmr} illustrates a comparison between purely secular co-orbital dynamics (left panel) and the $e-\Delta\varpi$ secular dynamics embedded within the 1:1 mean motion resonance with Planet Nine\footnote{Strictly speaking, the secular phase-space portrait shown on Figure \ref{fig:secular_e_mmr} should be computed by confining the value of $\varphi_{\rm{res}}$ to its equilibrium value, which is generally a function of both $e$ and $\Delta\varpi$. In the special case of the 1:1 MMR, however, secular modulation of $\phi_{\rm{res}}$ is very small, and simply assuming that $\phi_{\rm{res}}=0$ everywhere leads to an insignificant error in the computation of the resonant-secular Hamiltonian.}. The resemblance between purely secular and resonant-secular phase-space portraits in Figure \ref{fig:secular_e_mmr} is self-evident. This inherent separation of the degrees of freedom further qualitatively explains how KBOs can smoothly enter and exit MMRs facilitated by Planet Nine without noticeably affecting the orbital clustering of the long-period objects (see \citealt{bailey2018} for further discussion).

\section{The Planet Nine Hypothesis: Numerical Simulations}\label{sect2}
\label{sec:numerical}
A complete formulation of the Planet Nine hypothesis requires both a well-developed analytical understanding of the underlying physics and a detailed numerical description of the associated dynamics.  The previous section provides an analytic account of the effects of Planet Nine on the orbital structure and evolution of the distant Kuiper belt. However, a comprehensive comparison between the observations and P9-sculpted orbital structure requires a more detailed description of long-period KBO evolution. As a result, the Planet Nine hypothesis must be explored with the aid of realistic numerical simulations. The description, execution, and analysis of such numerical experimentation is the primary focus of this section.

Although the Planet Nine hypothesis (in its current incarnation) was put forth only a few years ago, a vast collection of N-body simulations has already been published. This body of work employs varying degrees of approximation in order to understand the dynamical mechanisms through which Planet Nine sculpts the distant Kuiper Belt. For example, one class of calculations \citep{beust, batmorby, hadden, bailey2018} employs a simplified model of the solar system, where the Keplerian motion of all four canonical giant planets is smoothed over. In this approximation, the solar system interior to 30 AU is replaced with a central body possessing a quadrupole field equivalent to the phase-averaged gravitational field of Jupiter, Saturn, Uranus, and Neptune. More realistic calculations \citep{phattie, brownbatygin2016, beckerp9, millholland, tali} include Neptune and Planet Nine directly as active bodies, while replacing the orbits of the three interior giants with massive wires; this scheme provides a compromise between capturing short-period effects in the Kuiper belt and retaining a long simulation time-step. Finally, numerical experiments that fully resolve the dynamics of the entire outer solar system have also been performed \citep{ batbrown2016b, sheptru2016, gomes2016, bailey2016, bp519, goblin}.

The successive approximations to the dynamics described above represent a trade-off between increasingly realistic orbital evolution and increased computational resources. The model that averages out the Keplerian motion of the canonical giant planets is clearly the fastest, but the least realistic. The intermediate simulations that directly account for the full dynamics of Neptune and Planet Nine, but smooth over the other planets, are expected to yield more precise results. This approach retains a diminished computational cost relative to the final approximation of keeping all major bodies active. To this end, note that the CPU run-time for a given integration is set by the resolution of the shortest dynamical timescale and the orbital period of Neptune is longer than that of Jupiter by a factor of $\sim14$. Although the approach of including only Neptune and Planet Nine as active perturbers is often used, the extent to which this approximation captures the relevant physics remains to be determined quantitatively. 

This section presents an overview of the numerical simulations of the Planet Nine hypothesis carried out to date.  In order to provide a consistent description of the numerical results, and to sort through the computational approximations used in previous work, we have carried out an additional ensemble of simulations of the solar system with the inclusion of Planet Nine. 
In the first set of calculations, we adopt the intermediate description of the dynamics, where only Planet Nine and Neptune are considered as active bodies (section \ref{SAS}). We then carry out an analogous set of simulations that self-consistently account for the N-body dynamics of all outer planets (section \ref{sec:fully_resolved}). With these additions, an analysis of all of the numerical simulations is presented in section \ref{sec:sim_enemble}, which assesses which properties of Planet Nine (mass and orbital elements) provide the best description of the observed solar system anomalies. The  particular case of TNOs on high inclination orbits is taken up in section \ref{sec:highi}. Finally, we discuss how interactions between Planet Nine and the known planets can affect the orientation of the orbital plane of the solar system (section \ref{sec:solarobl}). A complementary discussion that concerns the dynamical stability of the observed KBOs in the presence of Planet Nine is presented in \ref{appC}.

\subsection{Semi-Averaged Simulations}
\label{SAS}
As an initial numerical test of the P9 hypothesis, we simulate the effects of including Planet Nine and Neptune on a population of TNOs. The primary goal is to see which properties of Planet Nine control the onset of observed orbital anomalies of the TNOs. Following \citet{phattie}, in our preliminary suite of simulations we set the absorbing radius of the central body to $\mathcal{R}=29\,$AU, and endow the sun with a $J_2$ gravitational moment given by \citep{Burns1976}
\begin{align}
J_{2} = \frac{1}{2} \sum_{j=5}^{7} \frac{m_j\,a_j^2}{\Msun\,\mathcal{R}^2},
\label{Jeff}
\end{align}
where where the sum runs over contributions from Jupiter, Saturn, and Uranus. In the Figures and the following text, we refer to this simulation setup with the abbreviation \texttt{J2NP9}. 

Adopting a time-step of $\Delta t=$14 years ($\sim1/10)\,$th of Neptune's orbital period), we evolve an initially unstructured population of test-particles for $4\,$Gyr, allowing it to be sculpted by the combined gravitational influence of Planet Nine and Neptune as well as the mean-field effects of the solar system interior to $\mathcal{R}$. Following \citet{tali}, we initialize the Kuiper belt as a disk of $N=1000$ eccentric planetesimals, spanning the semi-major axis range $a \in (100, 800)\,$AU, perihelion range of $q\in (30,100)\,$AU, and adopt a half-gaussian inclination dispersion with standard deviation of $\sigma_i=15$ degrees, while drawing the other orbital angles $\varpi,\Omega,\mathcal{M}$ from a uniform distribution. With respect to the initial $q$-distribution in particular, we note that although secular dynamics driven by P9 alone can act to lift the perihelia of Neptune-attached KBOs (see section \ref{subsection:apsconf}), this process is also expected to naturally manifest within the solar system's birth cluster \citep{morbyhal2004}, justifying our adoption of an broad primordial spread of perihelion distances. In light of phase-space mixing associated with the chaotic nature of P9-facilitated evolution, however, we do not expect specific choices of initial conditions to affect the final results significantly. All numerical simulations reported herein were performed using the \texttt{mercury6} gravitational dynamics software package \citep{Chambers1999}, employing the hybrid Wisdom-Holman/Bulirsch-Stoer algorithm \citep{WisdomHolman, Press1992}.

In total, 1134 semi-averaged N-body integrations were performed, with the simulation grid corresponding to Planet Nine parameters $a_9 \in (300,1500)\,$AU, $e_9 \in (0.05,0.95)$, $i_9 \in (10,35)\,\deg$, with increments $\Delta a_9=100\,$AU, $\Delta e_9=0.1$, $\Delta i_9=5\deg$, and masses $m_9=5,10$, and $20 \,M_{\oplus}$. Drawing on previously published results (\citealt{brownbatygin2016, hadden, caceres2018} and the references therein), simulations in which Planet Nine's perihelion distance was smaller than $q_9\leqslant100\,$AU or greater than $q_9\geqslant500\,$AU were immediately discarded. Moreover, Planet Nine's starting orbital angles were chosen such that its argument of perihelion at the end of the simulation was always close to $\varpi_9-\Omega_9=\omega_9\approx140\,\deg$ i.e., approximately $180$ degrees away from the mean argument of perihelion of the dynamically stable KBOs with $a\geqslant250\,$AU.

\paragraph{Orbital Footprints of Stable KBOs}

Ideally, the number of particles in any given simulation would be so large that the synthetic Kuiper belt created after $4\,$Gyr of integration could be compared in a quantitatively sound manner with the observational dataset. Unfortunately, it is not computationally feasible to carry out such detailed simulations in sufficient numbers to meaningfully explore the P9 parameter space. As a result, rather than analyzing the final orbits of simulated long-term stable\footnote{For the purposes of this work, we define long-term stability as being equivalent to a dynamical lifetime in excess of $4\,$Gyr.} KBOs, we will instead examine the chaotic orbital footprints generated by these particles. In this context, the orbital footprints are generated simply by plotting the orbital state of the test particles every $1\,$Myr throughout the latter half of the integration. The resulting scatter of points delineates the overall orbital characteristics of surviving TNOs, including the range in angle $\Delta{\varpi}$ as a function of semi-major axis $a$. As outlined in section \ref{anomalous}, the observed TNOs show a distinct pattern for sufficiently large $a$ that must be consistent with successful Plant Nine models. We further note that owing to their fundamentally stochastic nature, these footprints trace out the confines of the available phase-space, and thus define a P9-sculpted \textit{distribution} of orbital parameters which can then be compared against the statistical properties of the observed census of distant KBOs. 

\begin{figure}[tbp]
\centering
\includegraphics[width=\textwidth]{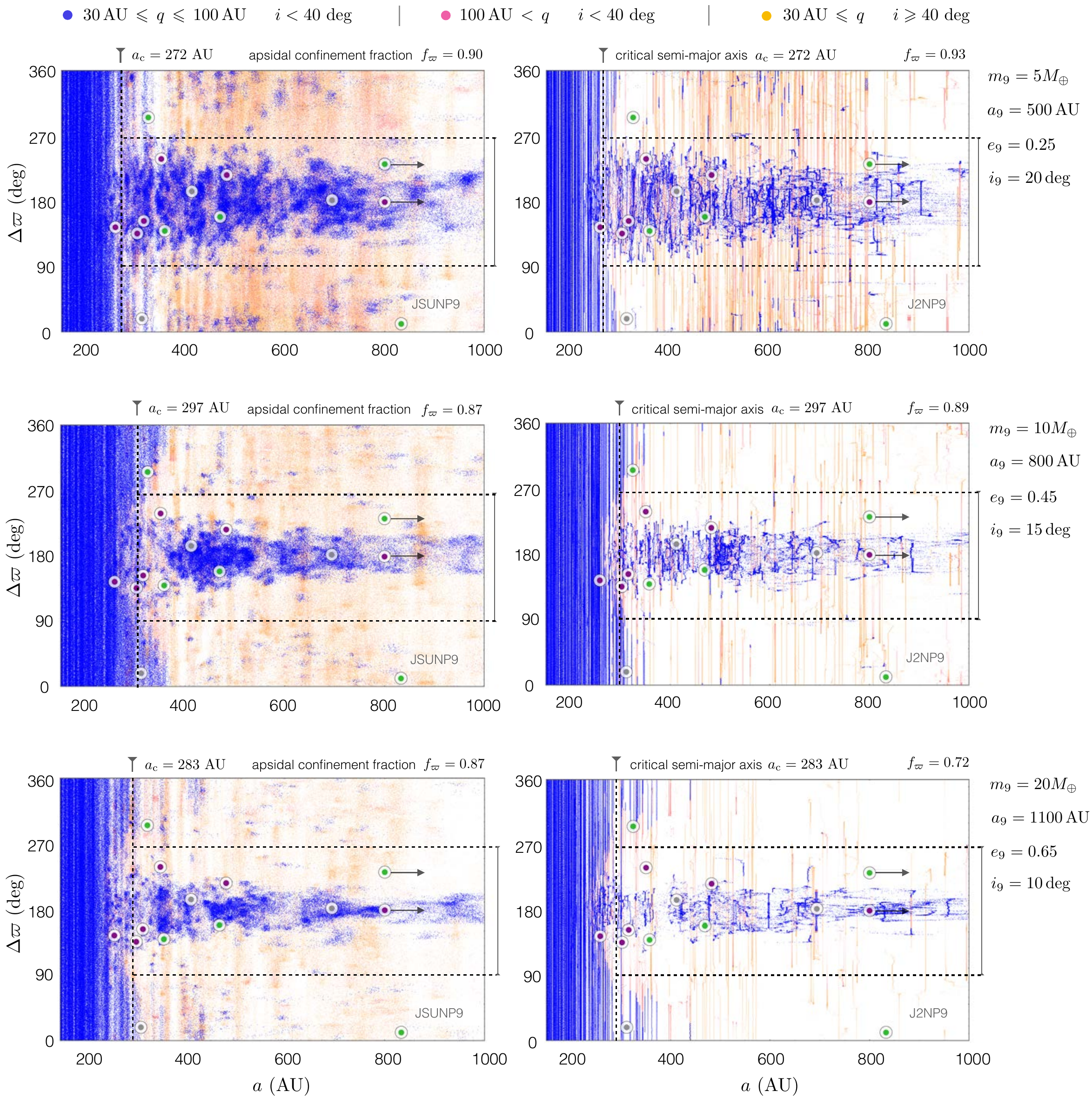}
\caption{Apsidal confinement of simulated KBOs. Each panel depicts the orbital footprints of longitude of perihelion of stable KBOs relative to Planet Nine, $\Delta\varpi$, as a function of semi-major axis, $a$. The top, middle and bottom pairs of plots depict synthetic Kuiper belts, generated with Planet Nine masses of $m_9=5\,M_{\oplus}$ (top panels), $10\,M_{\oplus}$ (middle panels), and $20\,M_{\oplus}$ (bottom panels). The corresponding values of P9 orbital elements are labeled above each row. The panels on the left-hand-side of the figure show the results of \texttt{JSUNP9} simulations, where the orbital motion of Jupiter, Saturn, and Uranus is resolved self-consistently. The right-hand-side panels show analogous results, obtained within \texttt{J2NP9} simulations that only treat Neptune and Planet Nine as active perturbers and replace the other planets with an effective quadrupole moment of the sun. The enhanced fuzziness of orbital footprints shown on the left panels stems primarily from our use of heliocentric rather than barycentric orbital elements. Blue points signify orbits that have $30\,$AU$\,\leqslant q\leqslant100\,$AU and $i<40\deg$, and thus satisfy our criteria for observability. Pink points mark unobservable orbits with $q>100\,$AU and $i<40\deg$. Orange dots denote high-inclination objects with $i\geqslant40\deg$. The large circular points depict the orbital elements of the observed KBOs, color-coded in the same way as in Figures \ref{fig_orbits} and \ref{fig_data}. In all cases, the value of $\Delta{\varpi}$ for the simulated TNOs is randomly distributed for small semi-major axes, and becomes confined to a much more limited range for larger $a$. The critical (transition) value of the semi-major axis is marked on each panel with a vertical dashed line. The top of each panel further reports the apsidal confinement fraction, $f_\varpi$, which is given by the ratio of blue points that fall within $\pm90\deg$ of exact perihelion anti-alignment with P9 (the domain between the two horizontal dashed lines) to the total number of blue points that reside outside of the critical semi-major axis.}
\label{fig:dw}
\end{figure}

The orbital footprints of $\Delta\varpi = \varpi-\varpi_9$, generated by stable particles over the $2-4\,$Gyr integration timespan are shown as a function of $a$ in Figure \ref{fig:dw}. Simulation results corresponding to three sets of P9 parameters are depicted, with \texttt{J2NP9} calculations illustrated on the right-hand-side. While the panels of this Figure contain the orbital footprints generated by \textit{all} long-lived particles over the last two billion years of their dynamical evolution, only a fraction of these synthetic data points would have been detectable by past and ongoing surveys. Thus, we require a proxy to distinguish between observationally detectable and undetectable objects. One possibility would be to develop a survey simulator to construct samples of observable data points from the simulations. Such a simulator, however, would require adequate characterization of the surveys which could have detected these distant objects. The number of objects detected with $a\gtrsim250$ AU is relatively small and they are found by several different surveys, each with differing detection efficiencies and biases. Unfortunately, a definitive assessment of the cumulative biases and other survey characteristics has not yet been carried out, although simulators tailored to specific surveys exist, and additional simulators are under construction (see, e.g., \citealt{shankmansurvsim}; Hamilton et al., \emph{in prep}). Here we instead make simple orbital element cuts to approximate observational biases.

Well known biases exist against the detection of both high perihelion and high inclination objects. Maintaining consistency with our discussion of the observations, we approximate these biases by classifying the simulation data into three categories, as follows. Objects deemed discoverable by conventional ecliptic surveys are characterized by $30\,\mathrm{AU}\leqslant q<100\,$AU and $i < 40\deg$, and are shown with blue points in Figure \ref{fig:dw}. The undiscoverable counterpart to this population has $q\geqslant100\,$AU and $i < 40\deg$, and is shown with pink points. Finally, the high-inclination component of the synthetic Kuiper belt is taken to be composed of objects with $q\geqslant30\,$AU and $i \geqslant 40\deg$, and is graphically depicted with orange points. Akin to Figure \ref{fig_data}, the aggregate of observed $a\geqslant250\,$AU KBOs is over-plotted on the panels of Figure \ref{fig:dw}, maintaining the same color-scheme as that employed in Figure \ref{fig_orbits}. 

\paragraph{The Critical Semi-Major Axis}
In each simulation depicted in Figure \ref{fig:dw}, the observationally discoverable subset of orbital footprints (blue points) exhibits a well-defined boundary, where low-$a$ objects experience apsidal circulation and large-$a$ particles cluster around $\Delta\varpi\sim180\deg$. Accordingly, we can characterize the suite of \texttt{J2NP9} calculations by delineating the critical semi-major axis beyond which orbital clustering emerges, $a_{\rm{c}}$, as a function of $a_9,e_9,i_9,$ and $m_9$. 

In order to map the dependence of $a_{\rm{c}}$ on P9 parameters, we examined the (observationally discoverable) distribution of $\Delta\varpi$ as a function of $a$ in each simulation.
We then computed the apsidal confinement fraction, $\zeta_\varpi$, defined as the fraction of particle footprints that lie within $\pm90\deg$ of exact anti-alignment with Planet Nine (shown with dashed horizontal lines in Figure \ref{fig:dw}) inside semi-major axis bins $\delta a=10\,$AU. As semi-major axis is increased from $150\,$AU outward, $\zeta_\varpi$ shifts from an initial value very close to $1/2$ (implying a fully randomized $\varpi-$distribution), to some higher value (typically close to unity), signifying a clustered distribution. To parameterize this transition in the apsidal confinement fraction, we fit an an error function to the resulting ($a,\zeta_\varpi$) sequence, and interpret the crossover point in the fit as $a_{\rm{c}}$. In each panel depicted in Figure \ref{fig:dw}, this transition point is marked with a vertical dashed line, and marked on top of each plot. 

\begin{figure}[tbp]
\centering
\includegraphics[width=\textwidth]{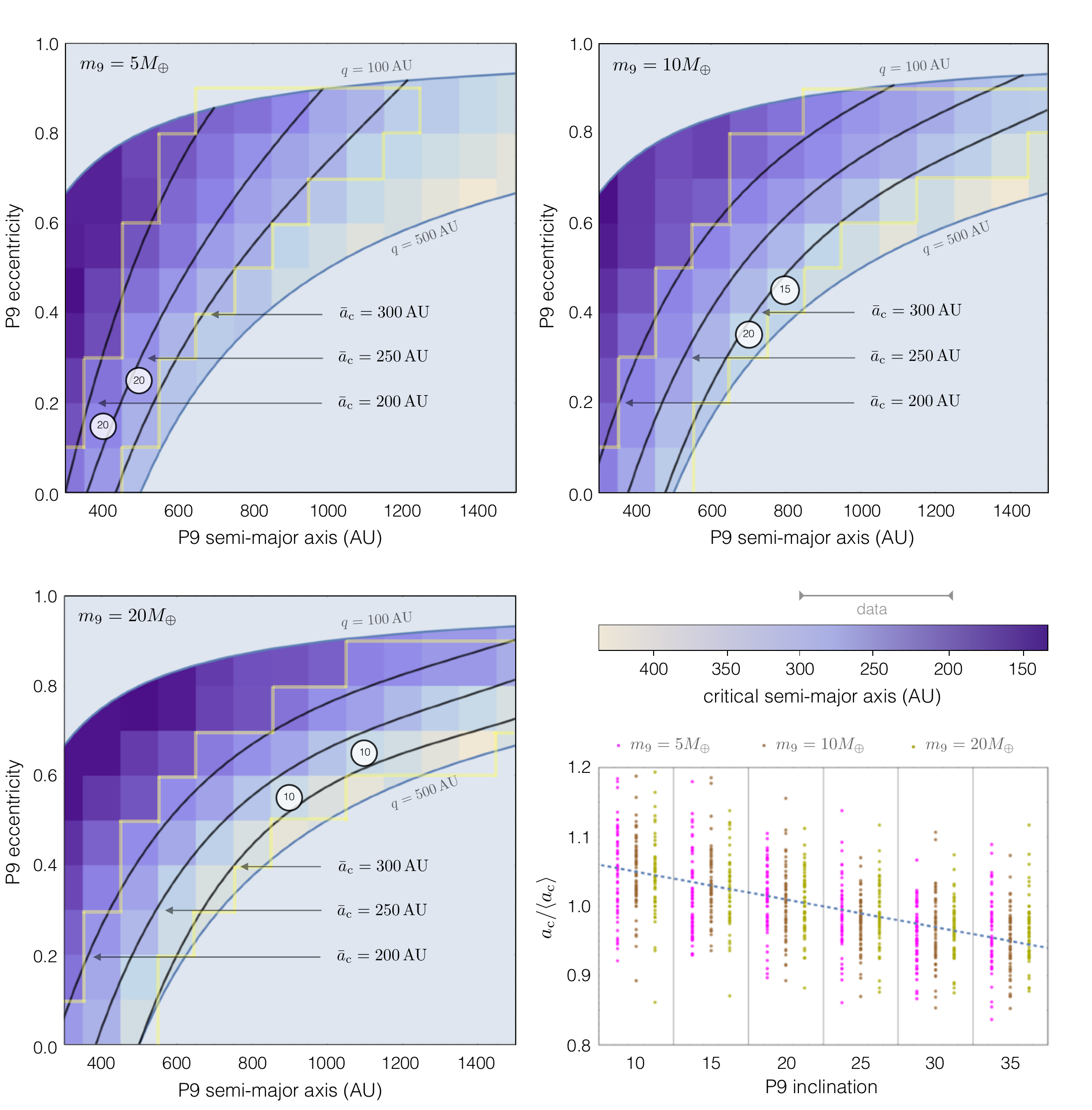}
\caption{Critical KBO semi-major axis, $a_{\rm{c}}$, corresponding to a transition from perihelion-randomized to apsidally confined orbital distribution, as a function of Planet Nine parameters. The lower right panel demonstrates the weak dependence of $a_{\rm{c}}$ on Planet Nine's inclination. Because this dependence is shallow, we model it out with a simple linear regression, and define a weighted average of the critical semi-major axis $\bar{a}_{\rm{c}}$. The remaining three panels portray this quantity on the $(a_9,e_9)$ plane, and explicitly mark the contours corresponding to $\bar{a}_{\rm{c}}=200\,$AU, $250\,$AU, and $300\,$AU. The yellow outlines on each graph delineate the domain over which full-fledged \texttt{JSUNP9} simulations were carried out. The two circles on each panel mark the ($a_9,e_9$) combination for the specific value of $m_9$ that yield optimal fits to the data, with the embedded numbers reporting the value of $i_9$ in degrees.}
\label{fig:wall}
\end{figure}

Remarkably, our semi-averaged simulation suite reveals that $a_{\rm{c}}$ exhibits only a weak dependence on Planet Nine's inclination. The bottom right panel of Figure \ref{fig:wall} depicts the fractional variation of the critical semi-major axis, $a_{\rm{c}}/\langle a_{\rm{c}} \rangle$ as a function of $i_9$ across the entire simulation suite, where $\langle a_{\rm{c}} \rangle$ represents a simple average of $a_{\rm{c}}$ over $i_9$ computed at constant values of $a_9$ and $e_9.$ Magenta, brown, and dark yellow points correspond to $m_9=5,10,$ and $20\,M_{\oplus}$ simulations respectively. Put simply, the results illustrated in this panel show that at a given combination of $m_9,$ $a_9,$ and $e_9$, changes in $a_{\rm{c}}$ due to variations in $i_9$ are only on the order of $\sim10\%$. While the dependence of $a_{\rm{c}}$ upon $i_9$ is so shallow that it can in principle be ignored, here we correct for it by fitting a linear regression to the numerical data shown in Figure \ref{fig:wall}, and computing a corresponding weighted average of the critical semi-major axis $\bar{a}_{\rm{c}}$.

Figure \ref{fig:wall} shows $\bar{a}_{\rm{c}}$ on a ($a_9,e_9$) simulation grid for our three choices of the mass of Planet Nine ($m_9$), with the solid lines on each plot corresponding to $\bar{a}_{\rm{c}}=200,250,$ and $300\,$AU as labeled. The simulations show a clear relationship, which bounds the P9 semi-major axis from below at $a_9\gtrsim300\,$AU, and requires progressively higher eccentricities to generate an observationally acceptable value of $\bar{a}_{\rm{c}}$ at larger semi-major axis. Moreover, in the $e_9\lesssim0.5$ range, the solid curves shown on each panel are essentially identical to one-another, meaning that for $q_9\gtrsim300\,$AU, the critical semi-major axis is approximately $m_9$-independent. This finding is quantitatively consistent with analytical estimates, wherein $\bar{a}_{\rm{c}}$ is interpreted as the minimum semi-major axis at which Hamiltonian (\ref{Hpuresec}) possess an elliptic equilibrium at $\Delta\varpi=180\deg$ \citep{batmorby}. 

The results summarized in Figure \ref{fig:wall} elucidate the acceptable range of $a_9$ and $e_9$ combinations for a given value of $m_9$ within the Planet Nine hypothesis. Simultaneously, however, the derived loci of $\bar{a}_{\rm{c}}$ point to a fundamental degeneracy between these two orbital elements i.e., the same value of critical semi-major axis can be reproduced by a broad array of $(a_9,e_9)$ pairs, meaning that additional information is needed to further constrain Planet Nine's orbit. The following section considers such additional constraints from a targeted set of fully resolved simulations.

\subsection{Fully Resolved Simulations}
\label{sec:fully_resolved}
With the P9 parameter range corresponding to $200\,\mathrm{AU}\lesssim \bar{a}_{\rm{c}} \lesssim 300\,$AU identified for the semi-averaged simulations, we have carried out a second suite of calculations that fully resolve the orbital evolution of Jupiter, Saturn, and Uranus, in addition to Planet Nine and Neptune (we will refer to calculations employing this physical setup as the \texttt{JSUNP9} suite of simulations). Given the higher computational cost associated with these numerical experiments, we limited our choice of P9 orbital elements to those generating values of $\bar{a}_{\rm{c}}$ that roughly fall into the desired range, as informed by \texttt{J2NP9} results. For each value of the mass $m_9$, this domain is shown with a yellow outline in Figure \ref{fig:wall} (corresponding to 660 additional simulations). With the exception of a smaller timestep necessary to properly model the orbit of Jupiter (which we set to $\Delta\,t=1\,$year, modifying the absorbing radius to $\mathcal{R}=4.5\,$AU and setting $J_2=0$), we adopted all other parameters of the \texttt{JSUNP9} calculations to be the same as those of the corresponding \texttt{J2NP9} runs.

\texttt{JSUNP9} analogues of the \texttt{J2NP9} calculations that are depicted on the RHS panels of Figure \ref{fig:dw} are shown on the LHS of the same Figure. Note that although \texttt{JSUNP9} and \texttt{J2NP9} results are qualitatively identical, panels corresponding to \texttt{JSUNP9} simulations appear more ``fuzzy" than their \texttt{J2NP9} counterparts. This difference dominantly stems from the fact that (to maintain consistency with the observations) we have chosen to plot the dynamical evolution in osculating heliocentric, rather than barycentric orbital elements. Because our semi-averaged calculations do not resolve the orbital motion of Jupiter, Saturn, and Uranus around the sun, much of the short-periodic jitter that is present in the \texttt{JSUNP9} results is filtered out in the \texttt{J2NP9} simulation suite. In other words, the underlying dynamical evolution detailed in the two sets of simulations is even more similar than it appears. 

As the next step in comparing the fully resolved and semi-averaged simulations, we re-computed the critical semi-major axis in the \texttt{JSUNP9} runs, and found that the resulting contours of $\bar{a}_{\rm{c}}$ on the $(a_9,e_e)$ plane are in close agreement with those obtained within the \texttt{J2NP9} set of numerical experiments. In fact, a hint of this agreement was already evident in Figure \ref{fig:dw}, where \texttt{JSUNP9} and \texttt{J2NP9} experiments employing the same parameters are illustrated side-by-side. Consequently, we conclude that the black curves delineated on Figure \ref{fig:wall} can be considered to be an accurate representation of the \texttt{J2NP9} and \texttt{JSUNP9} results alike.


Despite the considerable degeneracy that exists in P9 orbital elements that can generate a value of $\bar{a}_{\rm{c}}$ close to $250\,$AU, the actual characteristics of the distant belt produced in the simulations exhibit a sensitive dependence on Planet Nine's orbital properties. Therefore, a key objective of our analysis is to quantify the orbital architecture of the simulated $a>\bar{a}_{\rm{c}}$ Kuiper belt in each numerical experiment, and compare its statistical properties with the observations. For consistency, we will follow the procedure outlined in our discussion of the data (section \ref{anomalous}), and characterize the attributes of the synthetic Kuiper belt in terms of apsidal confinement, mean inclination, and the rms dispersion of the orbital poles. 

\begin{figure}[t]
\centering
\includegraphics[width=\textwidth]{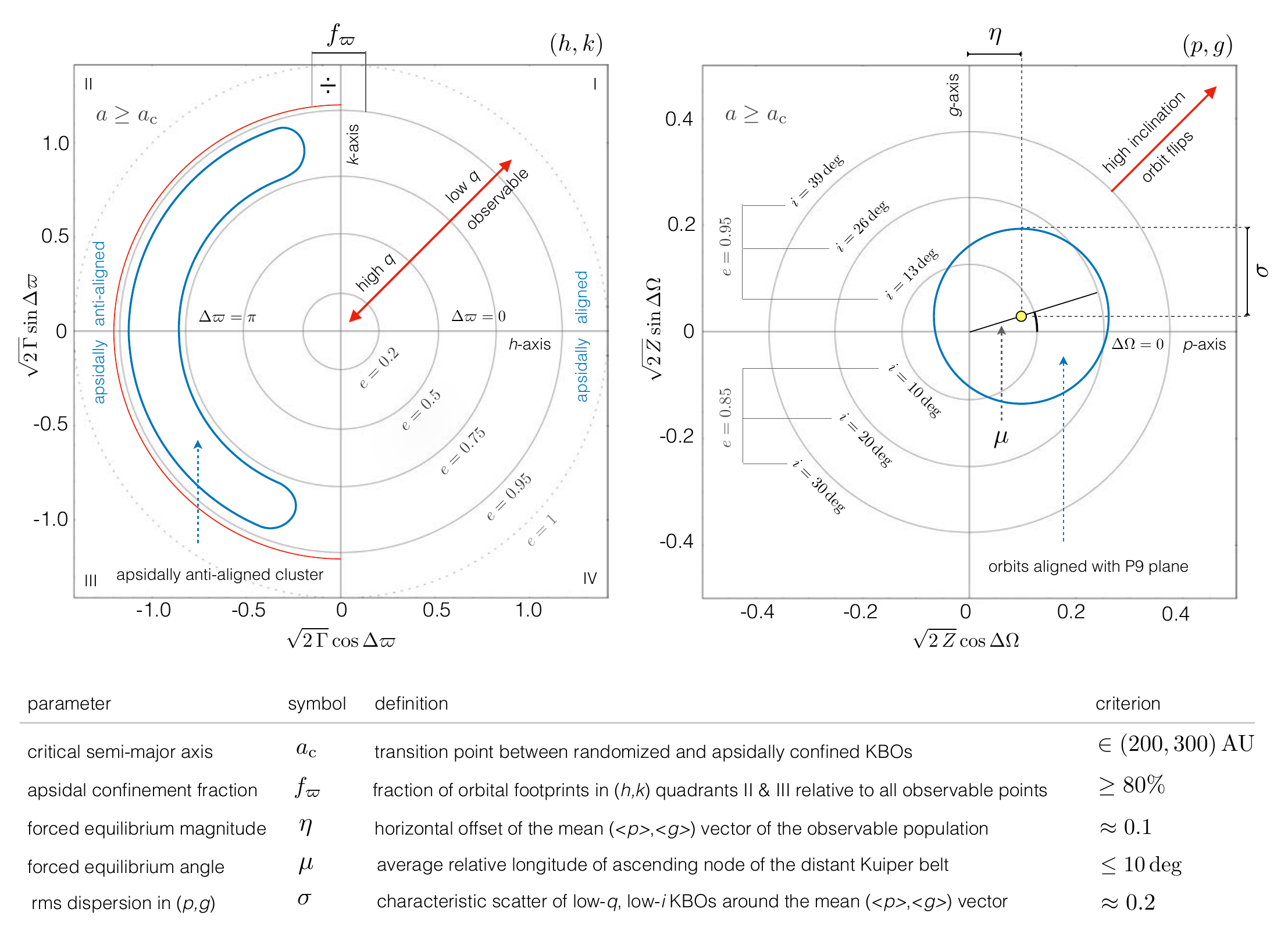}
\caption{Phase-space coordinates and statistical measures employed in our analysis of the distant ($a>a_{\rm{c}}$) Kuiper belt. The panel on the left-hand-side depicts the relationship between canonical cartesian $(h,k)$ variables and the orbital eccentricity. Analogous relationship between $(p,g)$ variables and inclination is shown on the right panel for $e=0.85$ and $e=0.95$ orbits (see equation \ref{Poincvar} for definition). Domains of phase-space occupied by observable low-$q$, low-$i$ KBOs that share the same orbital plane and are apsidally anti-aligned with respect to Planet Nine are marked on both panels by blue outlines. Additionally, the geometrical meanings of the statistical characteristics of the distant Kuiper belt $f_\varpi,\eta,\mu,\sigma$ are labeled on the panels, and their corresponding success criteria are summarized in the table below the plots.}
\label{fig:hkpg}
\end{figure}

\begin{figure}[tbp]
\centering
\includegraphics[width=\textwidth]{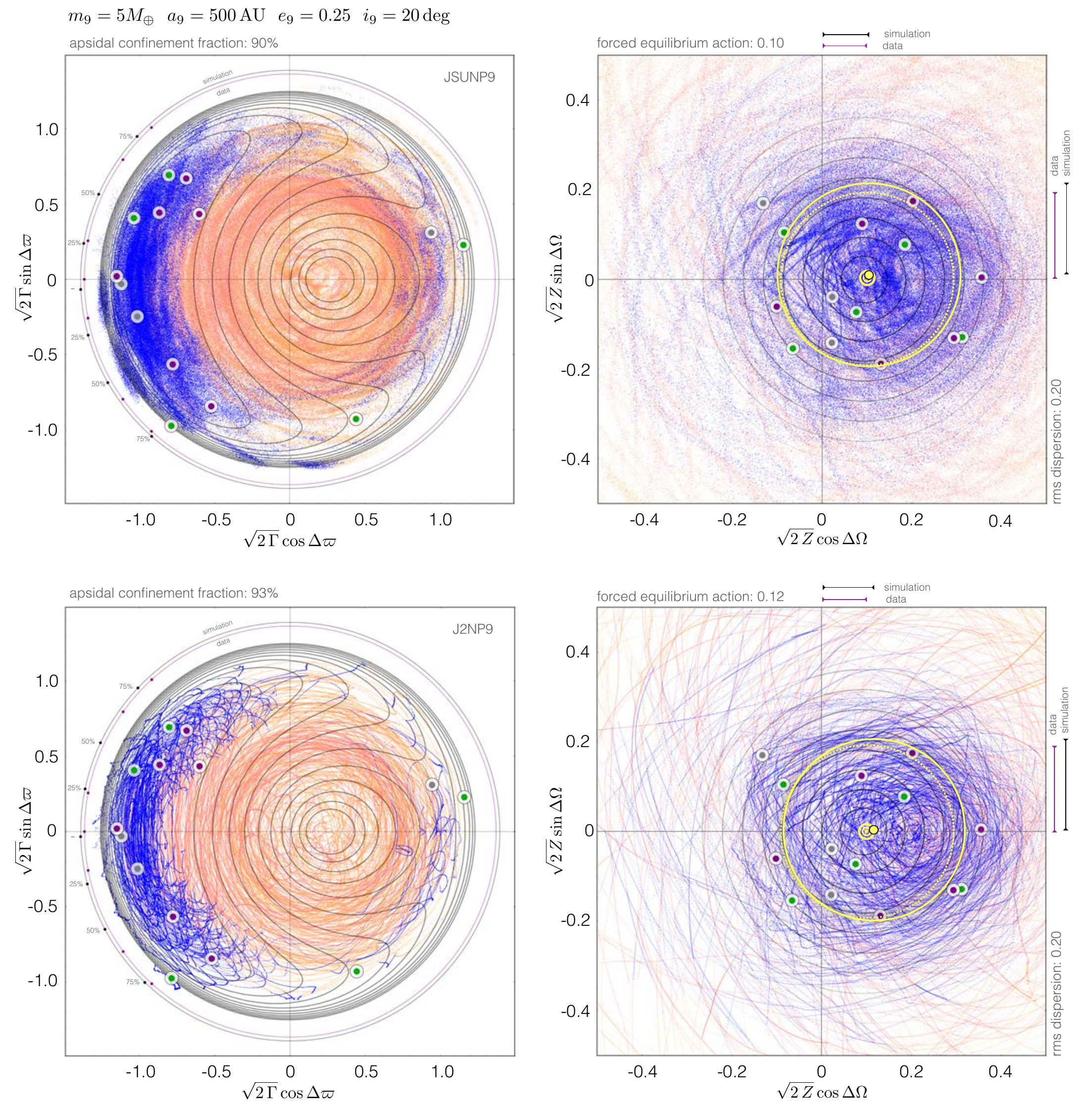}
\caption{Comparison between simulated and observed dynamical architecture of the distant Kuiper belt. Results are depicted in phase-space, defined by cartesian analogs of reduced \Poincare\ action-angle coordinates (equation \ref{Poincvar}; Figure \ref{fig:hkpg}). The panels on the left-hand-side show the degree of freedom related to the eccentricity, such that a circular orbit corresponds to the origin of the plot, and an $e=1$ orbit resides at a radial distance of $\sqrt{2}$. Meanwhile, the polar angle denotes the KBO longitude of perihelion, relative to that of Planet Nine. The panels show all stable $t\geqslant2\,$Gyr simulation data with $a\geqslant a_{\rm{c}}$, employing the same color scheme as that adopted in Figure \ref{fig:dw}. The strong concentration of observable particles (blue points) at $\Delta\varpi\sim180\deg$ is clearly evident on the left panels, and provides a good match to the observational data. The corresponding first, second, and third quartiles of the $\Delta\varpi$ distribution of simulated particles are marked on the outer black circle that encloses the Figure. For comparison, the first three quartiles corresponding to the observational data are shown on the inner purple circle (akin to that depicted in Figure \ref{fig_orbits}). The right-hand-side panels show the degree of freedom related to KBO inclination. Similarly to the panels on the LHS, $i=0$ orbits lie at the origin and the polar angle denotes the longitude of ascending node relative to that of P9. The quantitative attributes of the dynamics depicted in these panels are well summarized by the location of the forced equilibrium action (shown with a yellow dot) and the rms spread of the observable points about this equilibrium (shown with a solid yellow circle). For comparison, the correspondent characteristics of the observational dataset are shown with a target-like dot and a dashed circle. The amplitude of the forced equilibrium and the rms spread inherent to the simulation results as well as the observations are further shown with lines on top and on the right of each plot. In addition to numerical and observational data, the LHS and RHS panels also show the contours of Hamiltonians (\ref{Hpuresec}) and (\ref{Hinc}) respectively, evaluated at $\alpha=1$. The conformation of simulation results to analytical expectations inform the agreement between numerical experiments and secular perturbation theory.}
\label{fig:pspace5Me}
\end{figure}

\begin{figure}[t]
\centering
\includegraphics[width=\textwidth]{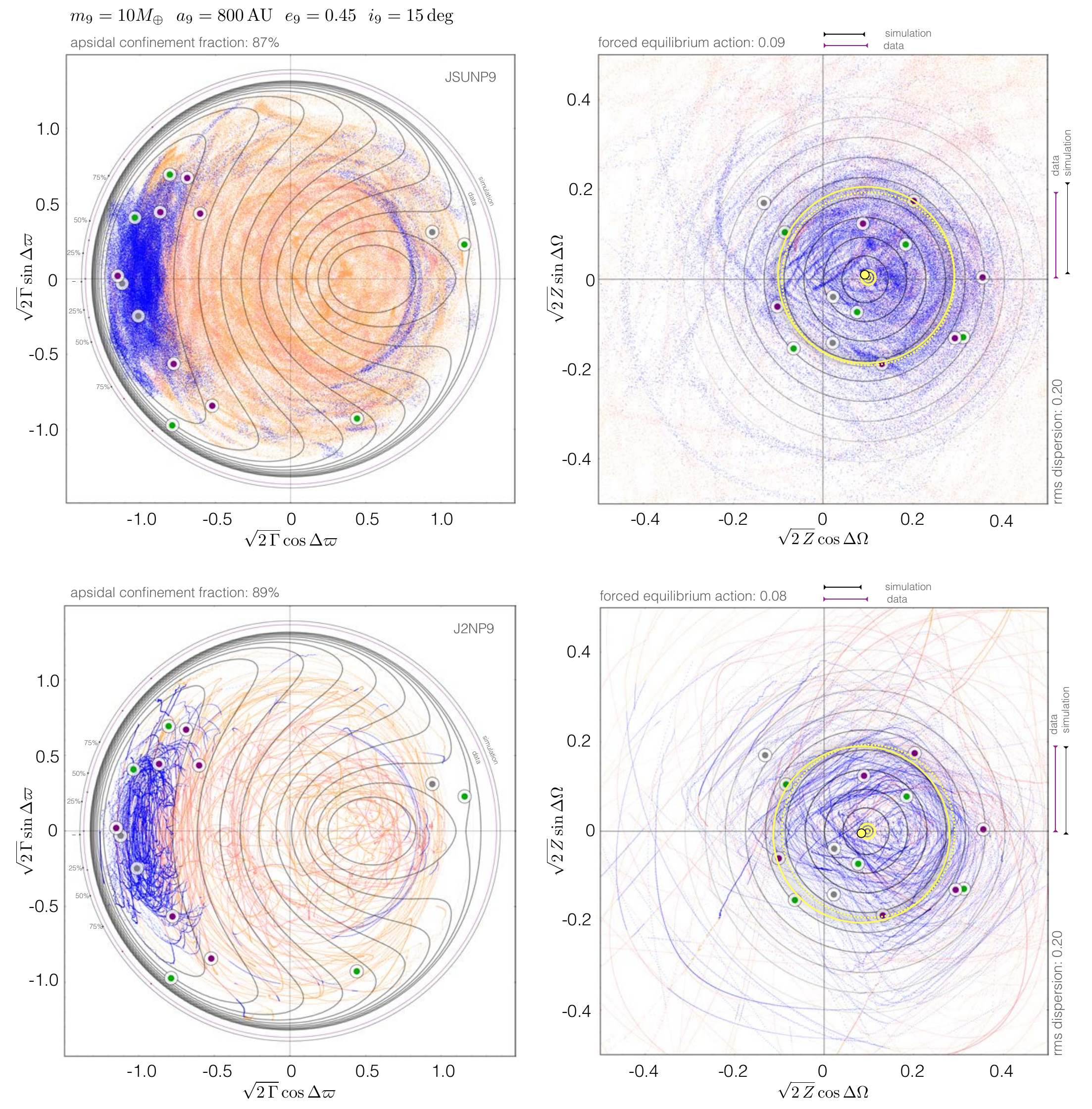}
\caption{Same as Figure \ref{fig:pspace5Me}, but for P9 parameters $m_9=10\,M_{\oplus}$, $a_9=800\,$AU, $e_9=0.45$, and $i_9=15\deg$.}
\label{fig:pspace10Me}
\end{figure}

\begin{figure}[t]
\centering
\includegraphics[width=\textwidth]{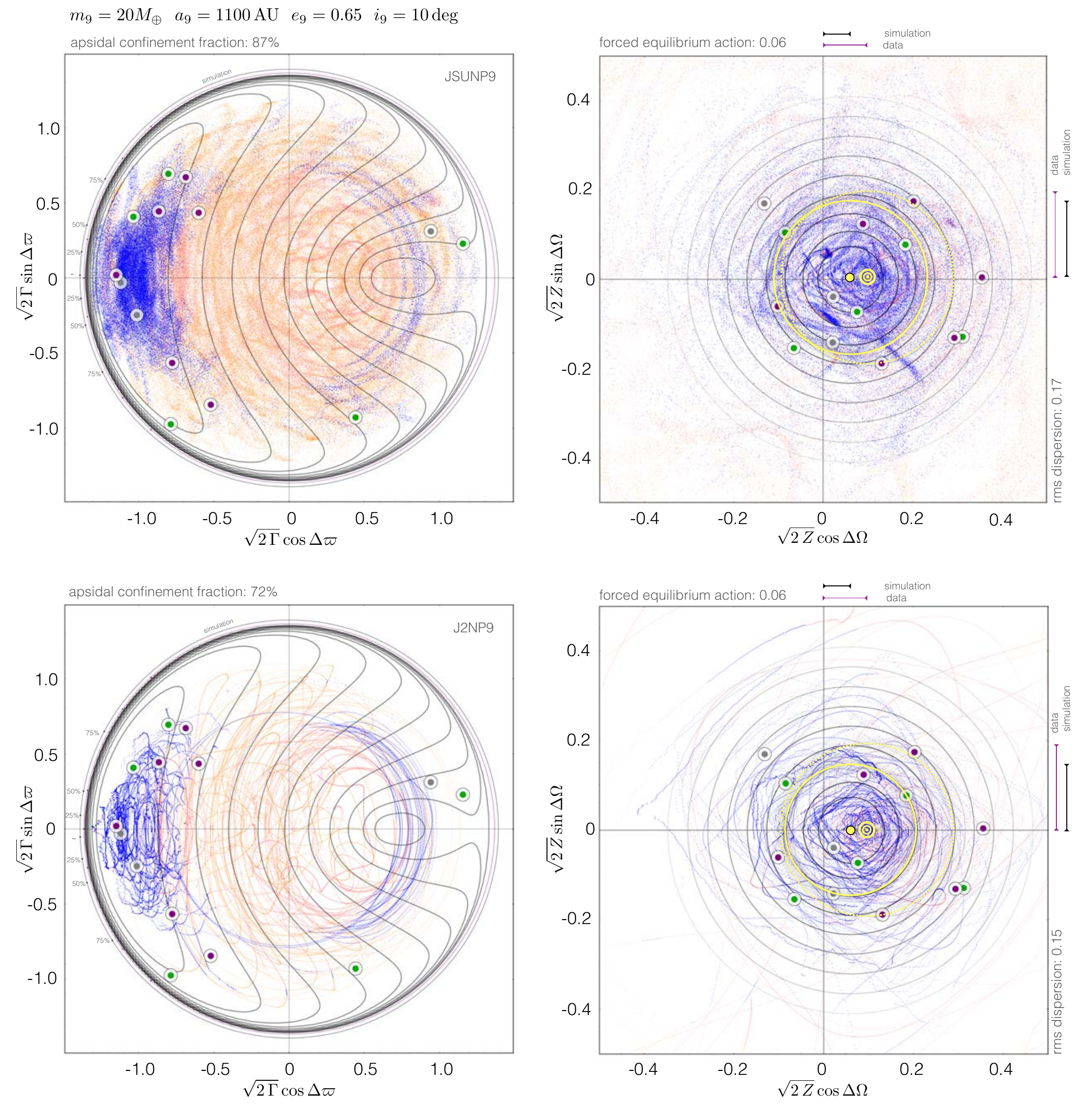}
\caption{Same as Figure \ref{fig:pspace5Me}, but for P9 parameters $m_9=20\,M_{\oplus}$, $a_9=1100\,$AU, $e_9=0.65$, and $i_9=10\deg$.}
\label{fig:pspace20Me}
\end{figure}

To ensure the orthonormality of our coordinate system, it is useful to introduce scaled \Poincare\ action-angle variables \citep{dm1999}:
\begin{align}
&\Gamma=\big(1-\sqrt{1-e^2} \big)   &\gamma=-\Delta\varpi \nonumber \\
&Z=\sqrt{1-e^2}\big(1-\cos(i) \big)   &z=-\Delta\Omega,
\label{Poincvar}
\end{align}
and carry out this analysis in phase-space, rather than in terms of Keplerian orbital elements. Note that in the above expressions, the conjugate angles are measured with respect to the corresponding parameters of Planet Nine's orbit (as mentioned already in section \ref{sec:analytical}, this choice leads to a trivial modification of the Hamiltonian, wherein additional terms are introduced to account for the slow apsidal precession and nodal regression of Planet Nine's orbit; see \ref{appB}). Although the actions $\Gamma$ and $Z$ have well-defined physical meanings and naturally connect to the definition of the angular momentum deficit \citep{Laskar1997}, one limitation of these variables is that at $i=0$, the angle $\Delta\Omega$ becomes ill-defined\footnote{The same issue technically arises for $\Delta\varpi$ in relation to circular orbits, but this poses no practical limitations for the representation of the distant Kuiper belt.}. A convenient way to sidestep this coordinate singularity is to transform to the cartesian canonical variables $h=\sqrt{2\Gamma}\cos(\Delta\varpi),k=\sqrt{2\Gamma}\sin(\Delta\varpi)$ and $p=\sqrt{2Z}\cos(\Delta\Omega),g=\sqrt{2Z}\sin(\Delta\Omega)$, such that radial distances from the origin on $(h,k)$ and $(p,g)$ plots become measures of eccentricity and inclination, while the polar angles translate to $\Delta\varpi$ and $\Delta\Omega$, respectively. The relationship between these variables and standard orbital elements is illustrated graphically in Figure \ref{fig:hkpg}. The dynamical footprints of long-term stable particles in the $a>\bar{a}_{\rm{c}}$ domain for the same three combinations of P9 parameters as in Figure \ref{fig:dw} are shown in Figures \ref{fig:pspace5Me}-\ref{fig:pspace20Me}, where \texttt{JSUNP9} and \texttt{J2NP9} simulations are plotted on the top and bottom pairs of panels respectively. For consistency, here we continue to employ the same color scheme as that used in Figure \ref{fig:dw}.

\paragraph{Apsidal Confinement} The panels depicted on the left-hand-side of Figures \ref{fig:pspace5Me}-\ref{fig:pspace20Me} illustrate apsidal confinement of synthetic long-period KBOs. The vast majority of simulated objects that satisfy our detectability criteria (blue points) cluster around $\Delta\varpi\sim180\deg$. Note that the degree of confinement is never perfect -- even at $a>\bar{a}_{\rm{c}},$ contamination from particles that circulate in $\Delta\varpi$ always exists. As with the observational data itself, we characterize the degree of $\varpi-$clustering by the \textit{total apsidal confinement fraction}, $f_\varpi$, (distinct from the previously mentioned running apsidal confinement fraction, $\zeta$) by splitting the observable output of the simulation into two bins, and computing the ratio (see Figure \ref{fig:hkpg}):
\begin{align}
f_\varpi= \frac{\mathcal{N}\big(30\mathrm{AU}\leqslant q\leqslant100\mathrm{AU},i<40\deg, \Delta\varpi\in[90,270)\deg\big)}{\mathcal{N}\big(30\mathrm{AU}\leqslant q\leqslant100\mathrm{AU},i<40\deg\big)}\Bigg|_{a> \bar{a}_{\rm{c}}}.
\label{fvarpi}
\end{align}
While the apsidal confinement fraction of all but one simulation shown in Figures \ref{fig:dw} and \ref{fig:pspace5Me}-\ref{fig:pspace20Me} is close to $\sim90\%$, in general $f_\varpi$ exhibits a rather sensitive dependence on P9 parameters, as will be discussed in greater detail below.

To complement $f_\varpi$, in Figures \ref{fig:pspace5Me}-\ref{fig:pspace20Me}, we also mark the first, second, and third quartiles of the simulated data on an exterior outer circle that encompasses the graph, and report the corresponding quartiles of the long-term stable real objects (also shown in Figure \ref{fig_orbits}) on the inner circle. Importantly, the characteristic spread of of the primary $\Delta\varpi\sim180\deg$ cluster exhibits considerable mass-dependence. That is, the $m_9=10$ and $20\,M_{\oplus}$ anti-aligned clusters (Figures \ref{fig:pspace10Me} and \ref{fig:pspace20Me}) have a noticeably smaller dispersion than the real data, implying that the $m_9=5\,M_{\oplus}$ simulation appears to be favored by the observations\footnote{The tendency towards a diminished range of $\Delta\varpi$ among the apsidally confined subset of points is also evident in Figure \ref{fig:dw}.}. Note further that at higher masses, a prominent collection of observable orbital footprints with $|\Delta\varpi | \lesssim90\deg$ also emerges, implying a comparatively diminished $f_\varpi$.

More generally, we acknowledge that in addition to $f_\varpi$ -- which is a relatively crude measure of the degree of apsidal confinement -- it is also possible to consider the angular width of the $\Delta\varpi$ cluster as a meaningful constraint on the simulation results. Paired with more precise modeling of the observational biases, such an analysis may yield added insight into the most optimal fits of Planet Nine's orbital parameters. For completeness, we overlay the simulation results on Figures \ref{fig:pspace5Me}-\ref{fig:pspace20Me} with observed data points, as well as the analytical contours of the Hamiltonian (\ref{Hpuresec}) (evaluated at $\alpha=a/a_9$ of unity, shown with black lines), which further elucidate the qualitative nature of secular trajectories executed by the particles in the simulations.

The left top panels of Figures \ref{fig:pspace5Me}-\ref{fig:pspace20Me} provide a natural point of comparison of apsidal confinement entailed by the observational data, analytical theory, and fully resolved numerical simulations. In section \ref{sec:analytical}, we interpreted the observed grouping in longitude of perihelion as a consequence of secular libration of particle trajectories around $\Delta\varpi=180\deg$, along contours of Hamiltonian (\ref{Hpuresec}). The prominence of this dynamical behavior is indeed clearly exemplified by the evolutionary tracks of the simulated KBOs. The majority of the stable observational data points, in turn, appear to be fully consistent with simulated objects that are randomly drawn from the vicinity of the $\Delta\varpi=180\deg$ island of libration. 

\paragraph{Clustering of the Orbital Planes}
In terms of canonical variables (\ref{Poincvar}), the dynamical state of the orbital planes of KBOs is described by the $(Z,z)$ degree of freedom, where physically, the action $Z$ represents a measure of an orbit's angular momentum deficit in the direction normal to the ecliptic. The dynamical footprints of stable particles, projected into cartesian analogs of the \Poincare\ action-angle variables are shown on the right-hand-side panels of Figures \ref{fig:pspace5Me}-\ref{fig:pspace20Me}. As with the panels depicting the dynamics in $(\Gamma,\gamma)$, we overlay the observational data points as well as analytical contours of Hamiltonian (\ref{Hinc}) evaluated at $\alpha=1$ and $e=0.85$ on the plots.

An advantage of portraying the simulation results, theory, and observations in the cartesian canonical coordinates ($p,g)$ is that the interpretation of the underlying dynamics becomes straightforward. In these variables, the analytic contours of Hamiltonian (\ref{Hinc}) simply appear as a succession of circles, centered on a point that resides on the $p-$axis, displaced to the right from the origin. Crucially, the distance between this point and the origin is the analytical approximation to the \textit{forced equilibrium}, $\eta=\sqrt{2Z_e}$ -- a quantity that characterizes the mean tilt of the orbital plane in P9-controlled domain of the solar system.

If the dynamics executed by stable KBOs in the numerical experiment conforms to the expectations from analytic theory, then orbital footprints of simulated particles should encircle this equilibrium point as well. With this notion in mind, we compute the location of the equilibrium point in each simulation as the average of the vector
\begin{align}
(\langle p\rangle,\langle g\rangle)=\frac{1}{\mathcal{N}}\sum_j^{\mathcal{N}}(p_j,g_j)\Bigg|_{30\mathrm{AU}\leqslant q\leqslant100\mathrm{AU},\,i<40\deg,\,a> \bar{a}_{\rm{c}}},
\label{pq}
\end{align}
where the sum runs over all observable (blue) points depicted in Figures \ref{fig:pspace5Me}-\ref{fig:pspace20Me}. Quantitatively, the magnitude of the forced equilibrium is simply the norm of the two quantities $\eta=\sqrt{\langle p\rangle^2+\langle g\rangle^2}$. In Figures \ref{fig:pspace5Me}-\ref{fig:pspace20Me}, the coordinates of the average vector $(\langle p\rangle,\langle g\rangle)$ are marked with a yellow dot, and the magnitude of $\eta$ is shown on top of each right-hand-side panel with a black line.


We caution that the norm of the average vector $(\langle p\rangle,\langle g\rangle)$ alone does not fully characterize the clustering of orbital planes, because it does not specify the direction (i.e., longitude) of the mean angular momentum vector of the distant KBOs. This complementary quantity is defined by the \textit{forced equilibrium angle} $\mu=\arctan(\langle g \rangle/\langle p \rangle)$. The analytic picture described in section \ref{sec:analytical} suggests that within the framework of the P9 hypothesis, $\mu$ should generally be close to zero. While this expectation is well-satisfied in simulations presented in Figures \ref{fig:pspace5Me}-\ref{fig:pspace20Me}, numerical experiments where $\mu$ is far away from zero are plentiful in our aggregate of simulations. Examining such large-$\mu$ simulations on an individual basis, we have determined that they routinely provide a poor match to the observational data. Typically, this is because they represent calculations where $\eta$ is so small that $\mu$ is essentially ill-defined, or because the low-inclination component of the distant belt is so depleted that the statistical measures $\eta$ and $\mu$ become unintelligible. As a result, simulations with $\mu\not\approx0$ are disfavored in our analysis. For this reason, in Figure \ref{fig:hkpg}, we simply denote $\eta$ as being the horizontal displacement of the average $(\langle p\rangle,\langle g\rangle)$ vector along the $p-$axis.

A second quantity that characterizes the extent to which simulated objects follow theoretically expected dynamics is the rms dispersion of the $(p,g)$ vector around the equilibrium point $(\langle p\rangle,\langle g\rangle)$:
\begin{align}
\sigma=\frac{1}{\mathcal{N}}\sum_j^{\mathcal{N}} \sqrt{\big(p_j-\langle p \rangle\big)^2+\big(g_j-\langle g \rangle\big)^2}\Bigg|_{30\mathrm{AU}\leqslant q\leqslant100\mathrm{AU},\,i<40\deg,\,a> \bar{a}_{\rm{c}}}.
\label{pqrms}
\end{align}
Graphically, the rms dispersion of numerical data is shown in Figures \ref{fig:pspace5Me}-\ref{fig:pspace20Me} with a yellow circle centered on the $(\langle p\rangle,\langle g\rangle)$ equilibrium, and its magnitude is depicted on the right of each panel with a black line. It is worth noting that a less precise way to characterize the orbital state of the distant Kuiper belt that has been employed in the literature draws attention to the clustering of the longitudes of ascending node (shown in Figure \ref{fig_data}). In terms of the above quantities, this clustering can be understood as a direct consequence of the equilibrium inclination forced by Planet Nine, and only arises when the rms dispersion of particle trajectories, $\sigma$, is comparable to, or smaller than $\eta$ itself.

To complement the simulation quantities $\eta$ and $\sigma$, on each right-hand-side panel of Figures \ref{fig:pspace5Me}-\ref{fig:pspace20Me}, we illustrate analogous quantities corresponding to the observed long-term stable objects as follows. The forced equilibrium point is marked with a large yellow target-shaped point that resides on the $p-$axis, and the extent of rms dispersion is shown with a dashed yellow circle. The two quantities are further depicted with purple lines on the top and the right of each panel, respectively. As is evident from examination of Figures \ref{fig:pspace5Me}-\ref{fig:pspace20Me}, among the depicted numerical experiments, the $m_9=5\,M_{\oplus}$ and $m_9=10\,M_{\oplus}$ simulations provide a notably better match to the current observational census of distant Kuiper belt objects than the $m_9=20\,M_{\oplus}$ simulation. As we will discuss further below, the incompatibility of $m_9=20\,M_{\oplus}$ simulations with the observations is a general result of our suite of numerical experiments.


\subsection{Analysis of Simulation Ensemble}
\label{sec:sim_enemble}
Having outlined a sequence of criteria by which a given simulation can be quantified, let us now evaluate our entire ensemble of calculations with an eye towards identifying trends between P9 orbital elements and the statistical properties of the synthetic KBOs. A key goal of this exercise is not only to delineate the dependence of $f_\varpi,\eta,\mu$ and $\sigma$ (summarized in Figure \ref{fig:hkpg}) on $a_9,e_9,i_9,$ and $m_9$, but also to identify P9 parameters that yield a distant population of small bodies that match the observations most closely. 

\begin{figure}[tbp]
\centering
\includegraphics[width=\textwidth]{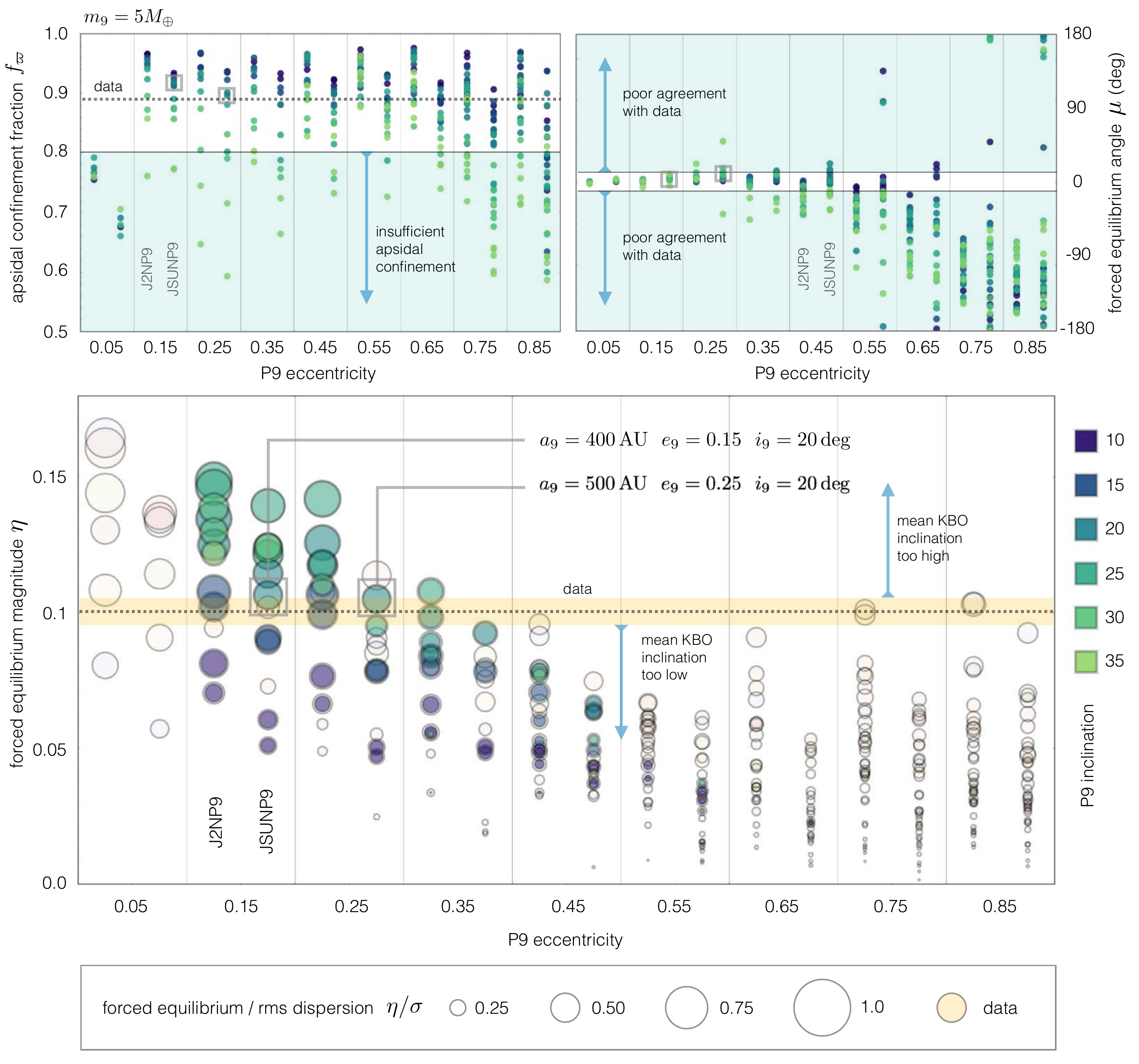}
\caption{A summary of the $m_9=5\,M_{\oplus}$ simulation ensemble. The top left plot shows the perihelion confinement fraction, $f_\varpi$ (equation \ref{fvarpi}), as a function of P9 eccentricity (recall from Figure \ref{fig:wall} that $e_9$ and $a_9$ are linked together by the requirement that the critical semi-major axis lies in the range $200\,\mathrm{AU}\lesssim\bar{a}_{\rm{c}}\lesssim300\,\mathrm{AU}$). While the current census of stable long-period KBOs is characterized by $(f_\varpi)_{\rm{data}}=8/9$, as a rudimentary cut on the results, we disregard any simulation that generates a distant Kuiper belt with $f_\varpi<80\%$. The top right panel depicts the forced equilibrium angle, $\mu$ (a measure of the vertical offset of the center of the yellow circle in the right panels of Figure \ref{fig:pspace5Me} away from the positive $p$-axis) of the simulated distant Kuiper belt, as a function of $e_9$. Reducing the aggregate of successful simulations further, we ignore any P9 parameter combination that produces a Kuiper belt with $|\mu|\geqslant10\deg$. The bottom panel shows a bubble chart where the y-axis corresponds to the magnitude of the forced equilibrium, $\eta$, and the size of the individual bubbles informs the ratio of the forced equilibrium amplitude to the rms dispersion (meaning that larger bubbles correspond to tighter clustering of the orbital poles). Only simulations that satisfy the aforementioned $f_\varpi$ and $|\mu|$ criteria are shown with colored circles (those that do not are shown with transparent bubbles), demonstrating that there exists only a limited eccentricity range that produces distant Kuiper belt architecture that is compatible with observations. The parameters of the two best-fit simulations for this choice of $m_9$ are labeled, although given observational uncertainties on the values of $f_\varpi$, $\eta$, and $\sigma$, it is clear that these parameter combinations are not unique. Note further that for each value of $e_9$, we plot the results from \texttt{JSUNP9} and \texttt{J2NP9} simulations side-to-side, and that semi-averaged simulations tend to systematically exhibit marginally better confinement in both $f_\varpi$, and $\eta$.}
\label{fig:5Me_adderall}
\end{figure}

\begin{figure}[t]
\centering
\includegraphics[width=\textwidth]{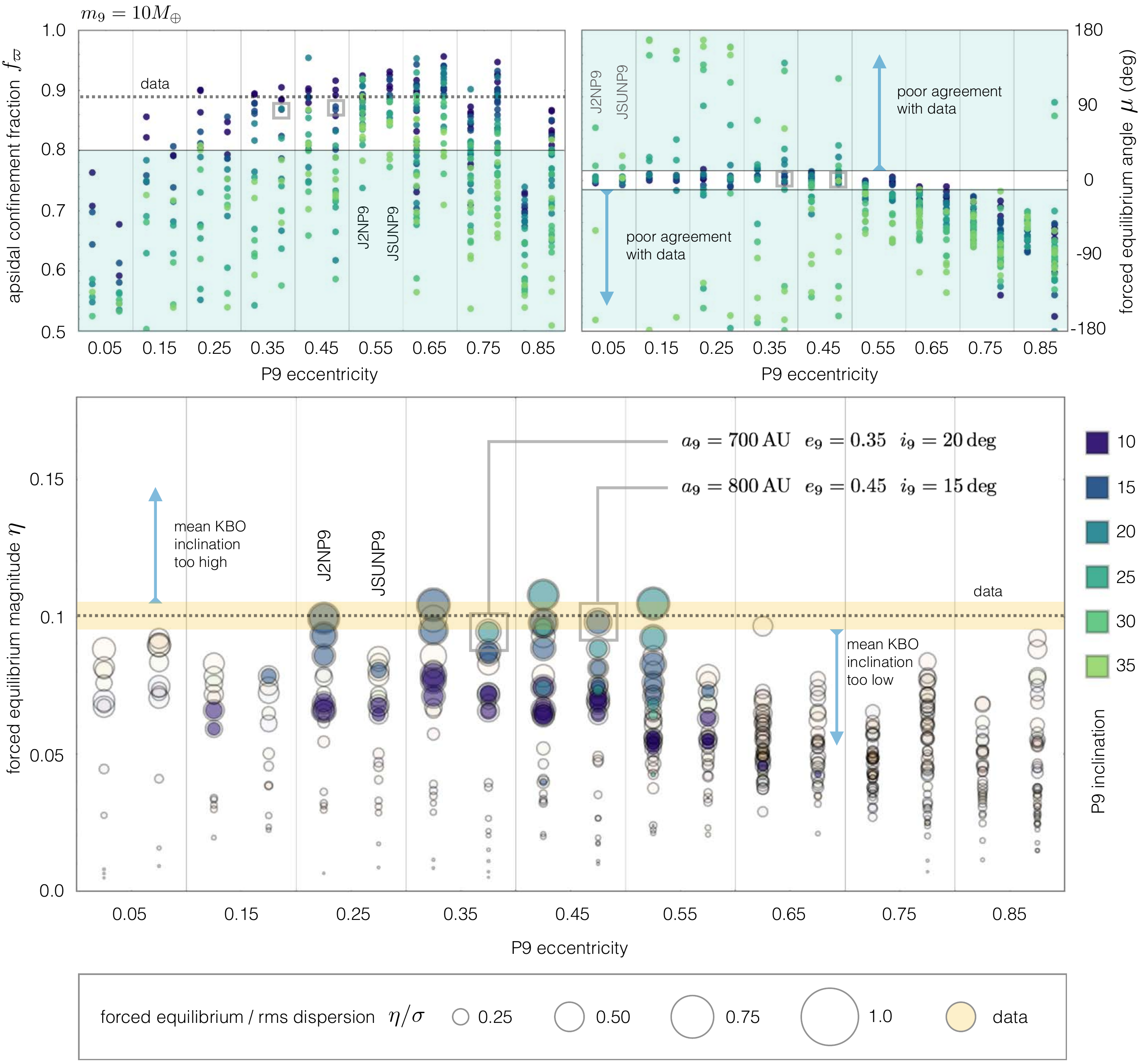}
\caption{Same as Figure \ref{fig:5Me_adderall}, but for $m_9=10\,M_{\oplus}$.}
\label{fig:10Me_adderall}
\end{figure}

\begin{figure}[t]
\centering
\includegraphics[width=\textwidth]{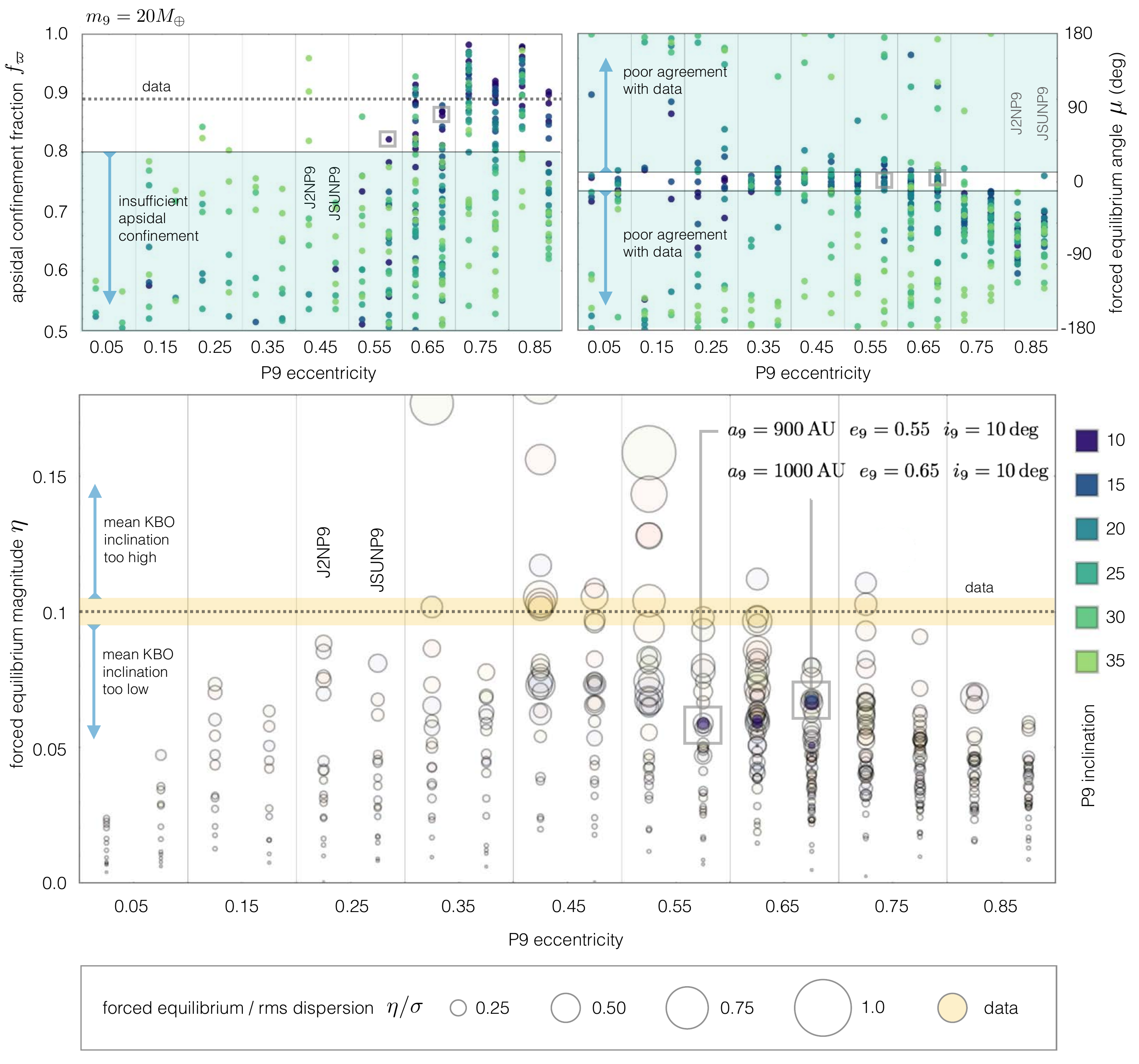}
\caption{Same as Figure \ref{fig:5Me_adderall}, but for $m_9=20\,M_{\oplus}$. Note that unlike the $m_9=5\,M_{\oplus}$ and $m_9=10\,M_{\oplus}$ cases, in this simulation suite, none of the successful simulations exhibit clustering of the orbital planes that is as good as the data, suggesting that the mass of Planet Nine is considerably smaller than $20\,M_{\oplus}$.}
\label{fig:20Me_adderall}
\end{figure}

We consider the total apsidal confinement fraction $f_\varpi$ as the first constraint. As a starting point, however, we draw on above results to reduce the number of independent variables by one. Recall from section \ref{SAS} that for a given P9 mass, a combined choice of $a_9$ and $e_9$ yields a value of $\bar{a}_{\rm{c}}$, which is delineated in Figure \ref{fig:wall}. Employing this relationship, we can restrict our analysis to systems characterized by $200\,\mathrm{AU}\lesssim \bar{a}_{\rm{c}} \lesssim 300\,$AU, and thus (approximately) eliminate $a_9$ as a free parameter in favor of $e_9$. The degree of apsidal confinement -- computed using the two-bin approach discussed above -- is shown as function of P9 eccentricity, and color-coded by P9 inclination in the top left panels of Figures \ref{fig:5Me_adderall}, \ref{fig:10Me_adderall}, and \ref{fig:20Me_adderall}, for the $m_9=5,10,$ and $20\,M_{\oplus}$ simulation suites, respectively. For each value of the eccentricity, \texttt{J2NP9} and \texttt{JSUNP9} results are portrayed side by side, with the former plotted on the left.

Three trends immediately emerge upon examination of these plots. First, it is evident that (perhaps counter-intuitively) simulations with $m_9=5\,M_{\oplus}$ generally produce better confinement of the longitude of perihelion than their higher-mass counterparts. In fact, values of $f_\varpi$ close to unity are readily achieved for $e_9\gtrsim0.15$ in $m_9=5\,M_{\oplus}$ experiments, while simulations with a more massive Planet Nine exhibit a more non-uniform dependence of $f_\varpi$ on P9 eccentricity, resulting in only select runs ($e_9\gtrsim0.35$ and $e_9\gtrsim0.55$ for $m_9=10$ and $20\,M_{\oplus}$ respectively) that attain satisfactory results. A second trend, already pointed out in \citet{brownbatygin2016}, is that the degree of apsidal confinement degrades with increasing P9 inclination. Particularly, on all three Figures, the color-gradient of the illustrated simulation points suggests that P9 orbital solutions with $i_9\gtrsim30\deg$ are simply not viable. Finally, although not universally true, provided the same P9 parameters, \texttt{J2NP9} simulations tend to exhibit marginally tighter apsidal confinement than those performed within the fully resolved \texttt{JSUNP9} calculations. Recalling from section \ref{anomalous} that among dynamically stable long-period KBOs $(f_\varpi)_{\rm{data}}=8/9\approx89\%$ (shown on the top left panels with a horizontal dashed line), here we adopt a slightly less stringent value of $f_\varpi\geqslant80\%$ as a criterion for success.

In the vast majority of our $f_\varpi\geqslant80\%$ simulations, apsidal confinement ensues sufficiently close to $\Delta\varpi\sim180\deg$ that the compatibility of the mean longitude of perihelion in numerical experiments with theory (section \ref{sec:analytical}) does not entail a practically useful constraint. On the contrary, the forced equilibrium angle $\mu$ -- which corresponds to the mean value of $\Delta\Omega$ -- varies significantly with $e_9$. As already discussed above, simulations with $\mu\not\approx0$ fail to generate an inclination dispersion of the distant belt that is in good agreement with the observations, and as a quantitative cut here we have chosen to disregard simulations with $|\mu|\geqslant10\deg$. We emphasize that this cut is not motivated by a specific observational constraint, and instead stems from empirical examination of simulation results.

The distribution of $\mu$ as a function of $e_9$ is shown on the right top panels of Figures \ref{fig:5Me_adderall}-\ref{fig:20Me_adderall}. Although the specifics of each figure differ considerably, a common thread emerges, wherein $\mu$ remains close to zero for nearly circular P9 orbits, but tends towards a strongly negative value at high Planet Nine eccentricities ($e_9\gtrsim0.65$ for $m_9=5,10\,M_{\oplus}$ and $e_9\gtrsim0.75$ for $m_9=20\,M_{\oplus}$). This limitation is particularly constraining for $m_9=20\,M_{\oplus}$ simulations because in this case the apsidal confinement criterion restricts P9 eccentricity from below at $e_9\gtrsim0.55$, leaving only a limited parameter range where such a massive planet can even approximately reproduce the real data. Moreover, even at modest values of $e_9$ and $m_9\leqslant10\,M_{\oplus}$, simulations with $i\gtrsim30\deg$ tend to produce $\mu$ significantly in excess of $10\deg$. We note however, that this behavior clashes with the apsidal confinement criterion, since high-inclination P9 simulations also generate particle disks with $f_\varpi$ considerably below $0.8$, and are therefore incompatible with observations anyway.

The large lower panels on Figures \ref{fig:5Me_adderall}-\ref{fig:20Me_adderall} encode the characteristics of the KBO inclination degree of freedom from the simulations, as a function of P9 eccentricity. The vertical coordinate of each simulation result in this plot denotes the value of the forced equilibrium, $\eta$. Meanwhile, the size of the individual bubbles represents the ratio of this magnitude to the rms dispersion of observable particles, $\eta/\sigma$. Consequently, smaller bubbles correspond to synthetic belts where the orbital poles of the KBOs are more randomly distributed, while larger bubbles correspond to simulations where clustering of orbital planes (and by extension, clustering of the longitude of the ascending node) is strong. Simulations that satisfy the criteria $f_\varpi\geqslant0.8$ and $|\mu|\leqslant10\deg$ are shown with colored bubbles (where as before, the color-scale informs P9's inclination), while those that do not conform to aforementioned benchmarks are depicted as nearly-transparent circles. As with the top plots in these Figures, \texttt{J2NP9} and \texttt{JSUNP9} results are illustrated next to one another, with the fully resolved calculations shown on the right.

The magnitude of the mean $(\langle p\rangle,\langle q\rangle)_{\rm{data}}$ vector corresponding to the long-term stable observed KBOs is shown with a dashed line, and the measure of rms dispersion inherent to the observational data $(\eta_{\rm{data}}/\sigma_{\rm{data}})$ is shown by the yellow band that encompasses this line. Accordingly, a numerical experiment that constitutes an ideal match to the data in the $(p,g)$ plane would be represented with a bubble that fits perfectly inside this yellow band. Examination of simulation data depicted in this sequence of plots reveals some of the same trends that we already highlighted for $f_\varpi$. Across the board, \texttt{J2NP9} simulations tend to yield slightly higher values of $\eta$ than their \texttt{JSUNP9} counterparts. More importantly, there exists a significant, and non-trivial dependence of the degree of orbital clustering on Planet Nine's mass and eccentricity. 

Among acceptable $m_9=5\,M_{\oplus}$ calculations (Figure \ref{fig:5Me_adderall}), $\eta$ exhibits a nearly monotonic inverse dependence on $e_9$, such that simulations in the $e_9\sim0.15-0.25$, $a_9\sim400-500\,$AU and $i_9\sim15-25\deg$ range result in the closest agreement between numerical experiments and data. The characteristic range of best-fit P9 parameters shifts to higher eccentricities and somewhat lower inclinations for $m_9=10\,M_{\oplus}$ runs, with $e_9\sim0.35-0.45$, $a_9\sim600-800\,$AU and $i_9\sim15-20\deg$ appearing most favorable. We note, however, that in this set of calculations, synthetic disks that exhibit the tightest clustering of the orbital planes are only as clustered as the real data. This suggests that at higher masses, the degree of angular momentum vector clustering displayed by the real objects simply cannot be reproduced, implying that ten Earth masses should be viewed as a working upper limit on Planet Nine's mass. Accordingly, the $m_9=20\,M_{\oplus}$ simulation suite indeed exhibits rather poor agreement with the observations. Even when the degree of apsidal confinement is satisfactory -- which already constrains the eccentricity and semi-major axis to the $e_9\sim0.55-0.65$ and $a_9\sim900-1200\,$AU range -- inclination dynamics of observed long-period KBOs are not well reproduced by the numerical experiments since the $\mu\sim0$ requirement restricts P9 inclination to $i_9\sim10\deg$, preventing adequate excitation of $\eta$.

For each P9 mass, we emphasize two simulations in Figures \ref{fig:5Me_adderall}-\ref{fig:20Me_adderall} that generate synthetic KBOs that are in closest agreement with the real distant Kuiper belt. These orbital parameters are highlighted on $a_9-e_9$ diagrams in Figure \ref{fig:wall} with circles, and are marked by their corresponding values of $i_9$. We further note that because the $(\Gamma,\gamma)$ and $(Z,z)$ phase space structure of each pair of simulations (corresponding to a given value of $m_9$) is rather similar, we only depict the larger semi-major axis calculations in Figures \ref{fig:pspace5Me}-\ref{fig:pspace20Me} to avoid unnecessary redundancy. Overall, the analysis carried out above points to an orbital solution where a $m_9\sim5-10\,M_{\oplus}$ Planet Nine resides on a mildly eccentric ($e_9\sim0.1-0.5$) and moderately inclined ($i_9\sim15-25\deg$) trajectory. This set of properties results in good agreement between theory, simulation, and data, in contrast to solutions with higher values of $m_9$. Let us now examine one additional aspect of P9-induced evolution in these simulations -- namely, the generation of highly inclined TNOs.

\subsection{High-Inclination Dynamics}
\label{sec:highi}
\begin{figure}[tbp]
\centering 
\includegraphics[width=1.0\textwidth]{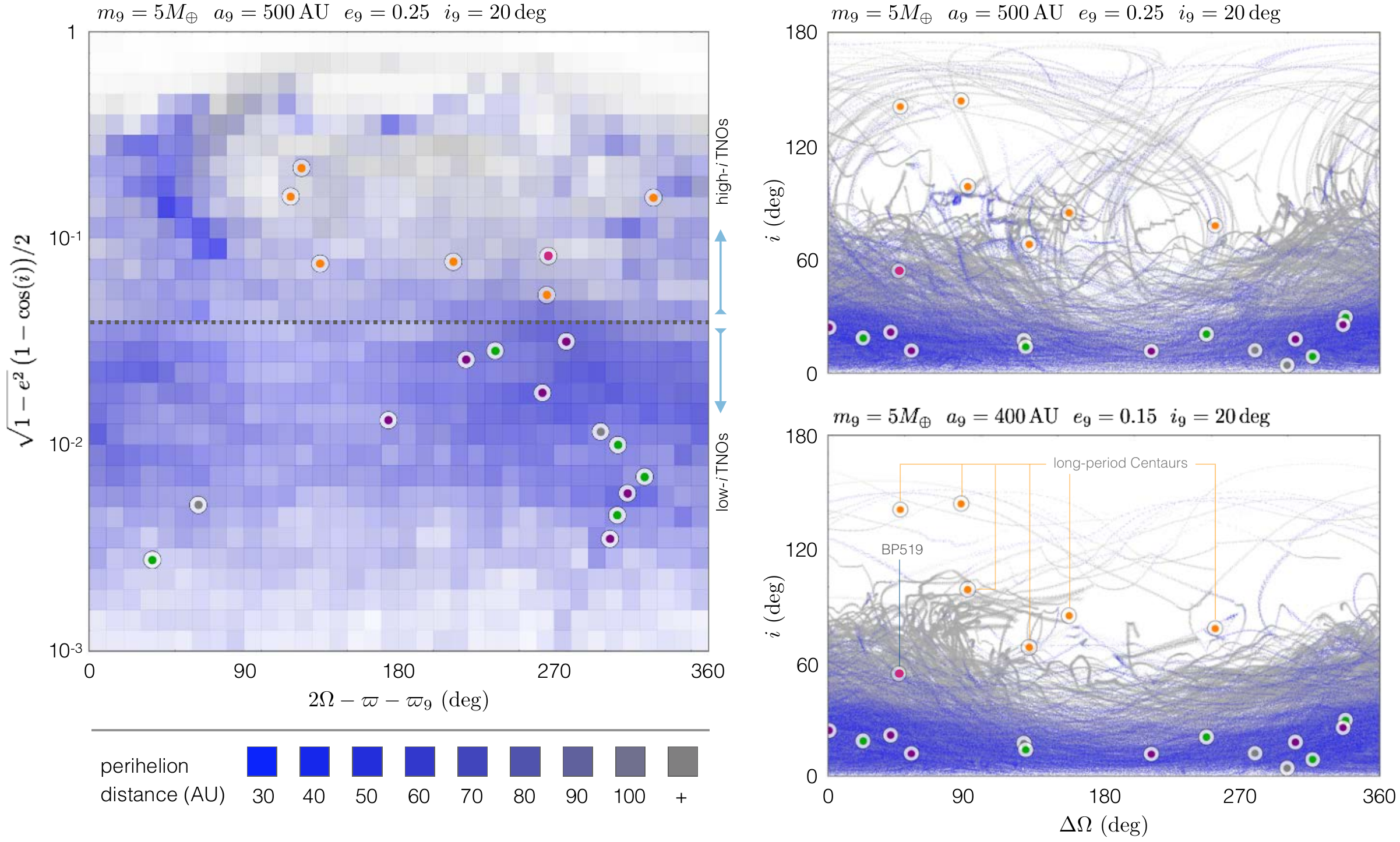}
\caption{High-inclination dynamics induced by a $m_9=5\,M_{\oplus}$ Planet Nine. Akin to plots illustrated in Figure \ref{fig:secular_i}, the right panels show the orbital footprints of simulated KBOs on the ($i,\Delta\Omega$) plane for the two optimal P9 orbital solutions emphasized in Figure \ref{fig:5Me_adderall}. Note that the $a_9=500\,$AU, $e_9=0.25$ solution produces high-inclination objects more readily than its lower eccentricity counterpart. The panel on the left shows a density histogram of particles footprints, projected into ($\theta,\Theta$) phase-space (see equation \ref{eqn:theta} and Figure \ref{fig:secular_theta}). Unlike the color-scheme employed in Figures \ref{fig:dw} and \ref{fig:pspace5Me}, here we adopt a blue-gray gradient to exclusively represent perihelion distance.} 
\label{fig:high_i_num} 
\end{figure}

As a first step in examining the high-inclination component of the distant Kuiper belt generated within the simulations, we inspect the orbital footprints of the long-term stable particles on the $i-\Delta\Omega$ plane. This projection is shown on the right-hand-side of Figure \ref{fig:high_i_num} for both of the aforementioned $m_9=5\,M_{\oplus}, a_9=500\,$AU, $e=0.25$, $i=20\deg$ and $m_9=5\,M_{\oplus}, a_9=400\,$AU, $e=0.15$, $i=20\deg$ calculations, with the more eccentric simulation plotted on the top panel\footnote{Equivalent $i-\Delta\Omega$ plots for the favorable $m_9=10\,M_{\oplus}$ solutions look quite similar to the top right panel of Figure \ref{fig:high_i_num}, so we omit them to avoid redundancy.}. Unlike Figures \ref{fig:pspace5Me}-\ref{fig:pspace20Me}, here we employ a color-scale that is solely a measure of the perihelion distance, with blue points corresponding to $q=30\,$AU, and gray points signifying $q\geqslant100\,$AU. The full census of observed $a\geqslant250\,$AU TNOs is also shown on the panels, with high-inclination ($i>40\deg$) Centaurs plotted as large orange points, and the KBO 2015\,BP$_{519}$ shown with a pink dot, as in Figure \ref{fig:highincTNOs}. 

Because we are showing only long-term stable particles in Figure \ref{fig:high_i_num}, simulated orbits that achieve $q<30\,$AU are essentially absent from the plot. This renders the comparison between the numerical experiments and the present-day orbits of high-inclination Centaurs inexact, leaving 2015\,BP$_{519}$ as the only high-inclination TNO which conforms strictly to the depicted numerical results (recall that severe observational biases exist against detecting high-inclination, high-perihelion KBOs). Nevertheless, it is very likely that the observed Centaurs originate as strongly inclined $q>30\,$AU Kuiper belt objects that are scattered inwards by Neptune, meaning that the numerical data depicted in Figure \ref{fig:high_i_num} represents the simulated source population of high-$i$ KBOs from which the observed Centaurs are derived. 

As can be deduced from examination of Figure \ref{fig:high_i_num}, the simulation with $e=0.25$ produces retrograde TNOs in much greater proportions than its lower eccentricity counterpart. Qualitatively, this can be attributed to the fact that the harmonic term responsible for orbit-flipping behavior is octupolar in nature (section \ref{sec:secularforcing}) and thus necessitates a significant eccentricity to drive high-inclination dynamics. In light of the strong observational biases that act against the detection of high-inclination TNOs, the true occurrence ratio of distant high-$i$ objects to low-$i$ objects is not known. Unfortunately, this means that at present, we cannot observationally favor either of the two shown simulations, since they both produce strongly inclined TNOs in some proportion. To complement the discussion presented in section \ref{sec:an_Theta}, the panel on the left-hand-side of Figure \ref{fig:high_i_num} presents the simulation results as a density histogram in phase-space, where the transparency of the cells corresponds to a logarithmic measure of the number of orbital footprints contained within each cell and the color corresponds to typical values $q$ of the constituent points. The regions of phase-space occupied by real TNOs is also heavily populated by simulated particles, suggesting that the agreement between the observed and synthetic populations of distant TNOs is satisfactory.

Although highly inclined long-period Centaurs embody an unexpected consequence of P9-induced dynamics, TNOs residing on nearly orthogonal and retrograde orbits are not confined to the distant edges of the Kuiper belt, and are observed on shorter-period orbits as well (Figure \ref{fig:highincTNOs}). As with the large semi-major axis Centaurs themselves, these objects are envisioned to have acquired large inclinations through interactions with Planet Nine at large semi-major axes, and to have subsequently been scattered inwards by Neptune. N-body simulations that elucidate this process for a $m_9=10\,M_{\oplus}$, $a_9=600\,$AU, $e_9=0.5$ Planet Nine were reported in \citet{batbrown2016b}. Accordingly, let us now examine the generation of high-inclination TNOs with $a\leqslant100\,$AU by a lower mass, lower eccentricity Planet Nine derived from the preceding analysis.

\begin{figure}[tbp]
\centering 
\includegraphics[width=1.0\textwidth]{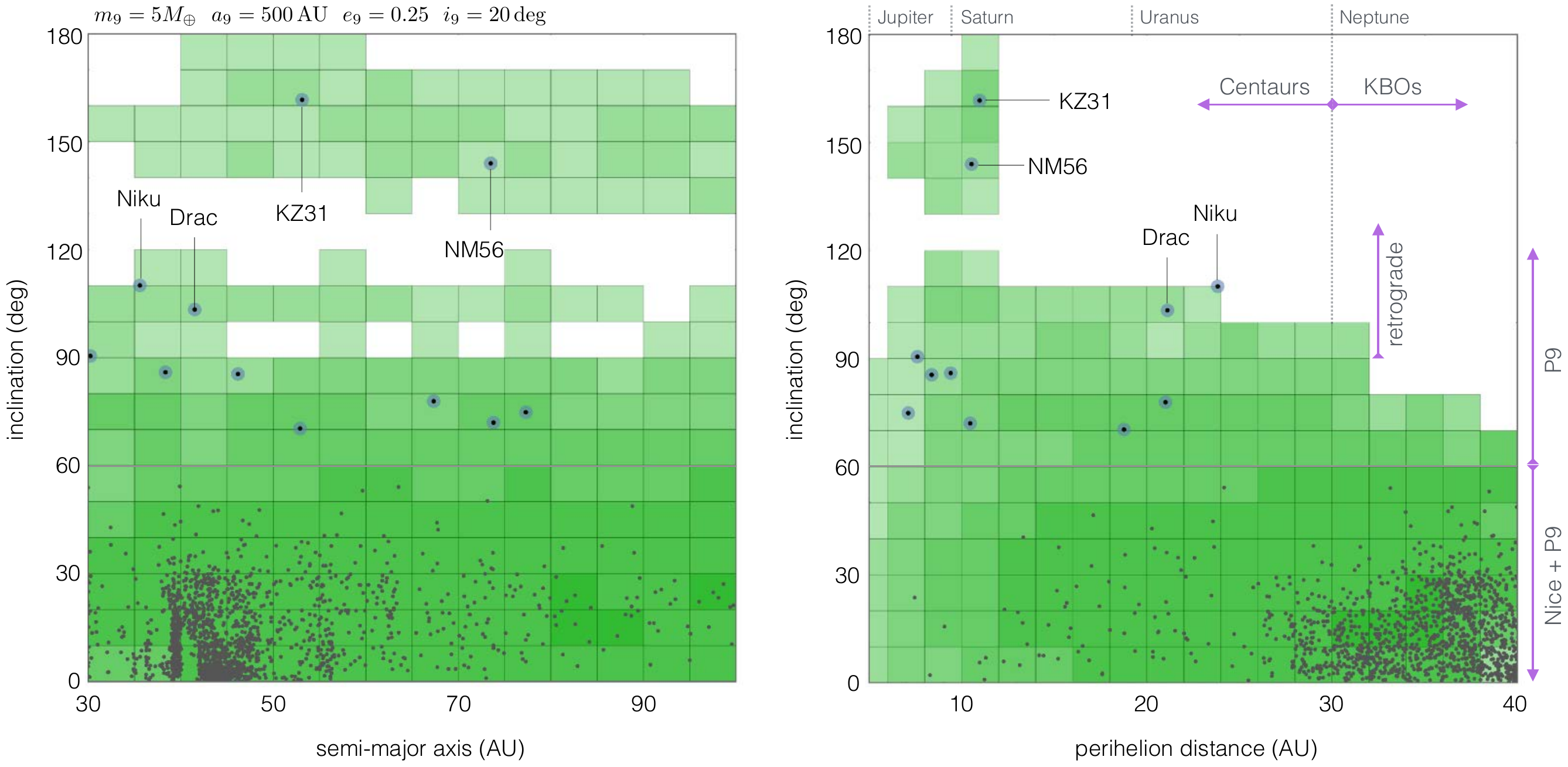}
\caption{Generation of $a\leqslant100\,$AU high-inclination Centaurs by inward scattering of P9-perturbed distant objects. The current census of TNOs is shown with black points, and the $i\geqslant60\deg$ objects that cannot be explained within the standard model of solar system evolution are emphasized. The underlying density histogram delineates the regions of parameter space onto which long-period, P9-influenced particles are emplaced when they are scattered inwards by Neptune. The specific domains of $a-q-i$ space most heavily populated by Planet Nine signal a strong consistency with the observational dataset.} 
\label{fig:high_i_centaurs} 
\end{figure}

In order to populate the inner trans-Neptunian region with a sufficient number of synthetic KBOs, we re-ran the $m_9=5\,M_{\oplus}$, $a_9=500\,$AU, $e_9=0.25$, $i_9=20\deg$ numerical experiment, increasing the particle count to $N=20,000$. The simulated orbital distribution of particles that attain $a\leqslant100\,$AU throughout any point in the simulations is shown in Figure \ref{fig:high_i_centaurs} as a density histogram, where once again transparency represents a logarithmic measure of the density of points. Recalling that all initial conditions of our numerical experiments (including this one) are drawn from the $a\in(100\,$AU$,800\,$AU) range, every particle that comprises the density histogram shown in Figure \ref{fig:high_i_centaurs} has been emplaced into the $a\leqslant100\,$AU region from a more distant orbital domain. The current $a\leqslant100\,$AU observational sample of TNOs is over-plotted on the Figure, and objects with $i\geqslant60\deg$ are emphasized.

Upon examination of Figure \ref{fig:high_i_centaurs}, two qualitative features are immediately evident. First, P9-induced dynamics significantly boosts the inclination dispersion of comparatively short-period TNOs, providing a natural explanation for the existence of nearly orthogonal orbits, such as those of Drac \citep{gladman2009} and Niku \citep{Chen2016}. Second, a distinct population of strongly retrograde particles with perihelion distances in the $6\,$AU\ $\lesssim q\lesssim12\,$AU range is also produced. These synthetic objects seamlessly explain the almost planar, but backward orbits of the TNOs 2016\,NM$_{56}$ and 2017\,KZ$_{31}$. Comprehensively, the union of the two panels shown on Figure \ref{fig:high_i_centaurs} demonstrates that the observed data is in excellent agreement with synthetic population of TNOs produced within the simulation.

More generally, the inability of standard Nice model simulations to produce KBOs with inclinations in excess of $i\gtrsim40\deg$ has long been recognized as a shortcoming of the instability-driven scenario of outer solar system evolution \citep{Levisonetal2008, Nesvorny2015b}. Although the conflict between the observed aggregate of KBOs and the Nice model can in principle be resolved either by P9-facilitated pollution of the Kuiper belt with high-inclination objects that originate further out, or by inward delivery of Oort cloud objects, the two hypotheses make distinct predictions regarding the orbital structure of the resulting high-inclination Kuiper belt. That is, while dynamical emplacement driven by Planet Nine generates a rather specific orbital architecture (as detailed in Figures \ref{fig:high_i_num} and \ref{fig:high_i_centaurs}), high-inclinations objects sourced from the Oort cloud should stem from a more uniform distribution of orbital elements. Continued mapping of the high-$i$ component of the Kuiper belt thus offers a direct avenue towards observational differentiation between these two hypotheses.

\subsection{Solar Obliquity}
\label{sec:solarobl}
A final piece of information that is readily informed by our aggregate of \texttt{JSUNP9} simulations concerns the interactions between Planet Nine and the mean plane of the known eight-planet solar system (often referred to as the ``invariable plane"; \citealt{souami2012}). Shortly after the initial formulation of the Planet Nine hypothesis, it was pointed out by \citet{bailey2016, gomes2016, lai2016} that the secular gravitational torque exerted by P9 upon the canonical giant planets would slowly perturb the orbital plane of the planets away from its initial state, thereby exciting a spin-orbit misalignment between the total angular momentum vector of the canonical giant planets and the spin-axis of the sun. Moreover, these authors found that given plausible P9 parameters (e.g., $m_9=15\,M_{\oplus},a_9=500\,$AU, $e_9=0.5,i=20\deg$; \citealt{bailey2016}), the entire $6$-degree obliquity of the sun could be accounted for by P9 perturbations alone.

This effect is self-consistently captured in our simulations, and it is worthwhile to examine if the revised orbital properties of Planet Nine derived above remain consistent with a scenario where Planet Nine plays a dominant role in the excitation of solar obliquity. For the two best-fit P9 parameters combinations shown in Figures \ref{fig:pspace5Me} and \ref{fig:pspace10Me}, the answer is a resounding 'no.' Specifically, for $m_9=5\,M_{\oplus}$, $a_9=500\,$AU, $e_9=0.25$, $i_9=20\deg$, P9's secular torque only leads to a $\psi=1.1\deg$ change in the inclination of the solar system's invariable plane over $4\,$Gyr. For a $m_9=10\,M_{\oplus}$ perturber on a $a_9=800\,$AU, $e_9=0.45$, $i_9=15\deg$ orbit, the induced solar obliquity is similarly small, evaluating to only $\psi=0.7\deg$. This implies that some other process, unrelated to the existence of Planet Nine, must be responsible the present-day obliquity of the sun.

To elaborate on this result further, it is useful to contextualize the sun's spin-orbit misalignment within its broader, galactic context. Over the last decade, observations of the Rossiter-McLaughlin effect (see for example \citealt{winn2010,triaudreview}) and doppler tomography (e.g., \citealt{marsh2001,johnson2017}) have revealed that exoplanetary systems around sun-like stars generically exhibit a very broad range of spin-orbit misalignments, with projected stellar obliquities ranging from $0$ to $\sim180\deg$. Importantly, this finding applies both to singly-transiting planets as well as to multi-transiting systems wherein the planets themselves are almost exactly coplanar, but are cumulatively inclined with respect to the spin-axis of their host star (e.g., the inner planets of Kepler-56, \citealt{likep56,liwinn}). Although the exact mechanism through which extrasolar spin-orbit misalignments are excited remains an active area of research, viable theories for the generation of planet-star misalignments include turbulence within the protostellar core \citep{2010MNRAS.401.1505B,2014ApJ...797L..29S,2015MNRAS.450.3306F}, gravitational torques arising from primordial stellar companions \citep{batygin2012, batyginadms13, lai2014, 2014ApJ...790...42S, 2015ApJ...811...82S}, as well as magnetohydrodynamic interactions between the stellar magnetospheres and the inner edges of their circumstellar disks \citep{Lai2011}. In light of these results, we can comfortably attribute the $6$-degree obliquity of the sun to the same primordial process that shapes the broad distribution of spin-orbit misalignments of generic sun-like stars throughout the Galaxy, decoupling it from the Planet Nine hypothesis.

\bigskip 
\section{Prospects for Detection} 
\label{sec:detection}

\subsection{Optical Surveys}
An important aspect of the Planet Nine hypothesis is that all of its theoretical attributes are directly testable, through the astronomical detection, and characterization of P9 itself. Arguably, the most straightforward approach towards direct detection of Planet Nine is via conventional observations in reflected visible light. The reference Planet Nine envisioned by \citet{phattie,batbrown2016b} and \citep{brownbatygin2016} had a moderately high semi-major axis, eccentricity, and mass, making Planet Nine potentially as faint as $\sim$25th magnitude. The more detailed model comparisons shown here suggest a Planet Nine that is lower in all of these parameters. Let us examine the effect of this refinement on the expected brightness. 

The brightness of Planet Nine depends on its size, albedo, and distance. For a mass range between $5$ and $10\,M_{\oplus}$, exoplanets follow an approximate power law where $R\approx R_\oplus (M/M_\oplus)^{0.55}$, suggesting radii of $2.4$ and $3.5R_{\oplus}$, respectively \citep{weiss2014}. Such estimates are further consistent with the interior modeling efforts of \citet{linder2016} who compute physical radii of $\sim 1.9-3.7 R_{\oplus}$ for an isolated $10\,M_{\oplus}$ object (see also \citealt{ginzburg16}). The albedo of Planet Nine is unknown, but modeling by \citet{fortney2016} suggests that at the inferred heliocentric distance, the planetary H/He envelope will be free of all potential condensibles, rendering the atmosphere an essentially pure Rayleigh scatterer with a V-band albedo of almost unity. Neptune, in contrast, has an albedo of approximately $40\%$, which we take as a plausible lower limit. 

\begin{figure}
\centering
\includegraphics[width=0.8\textwidth]{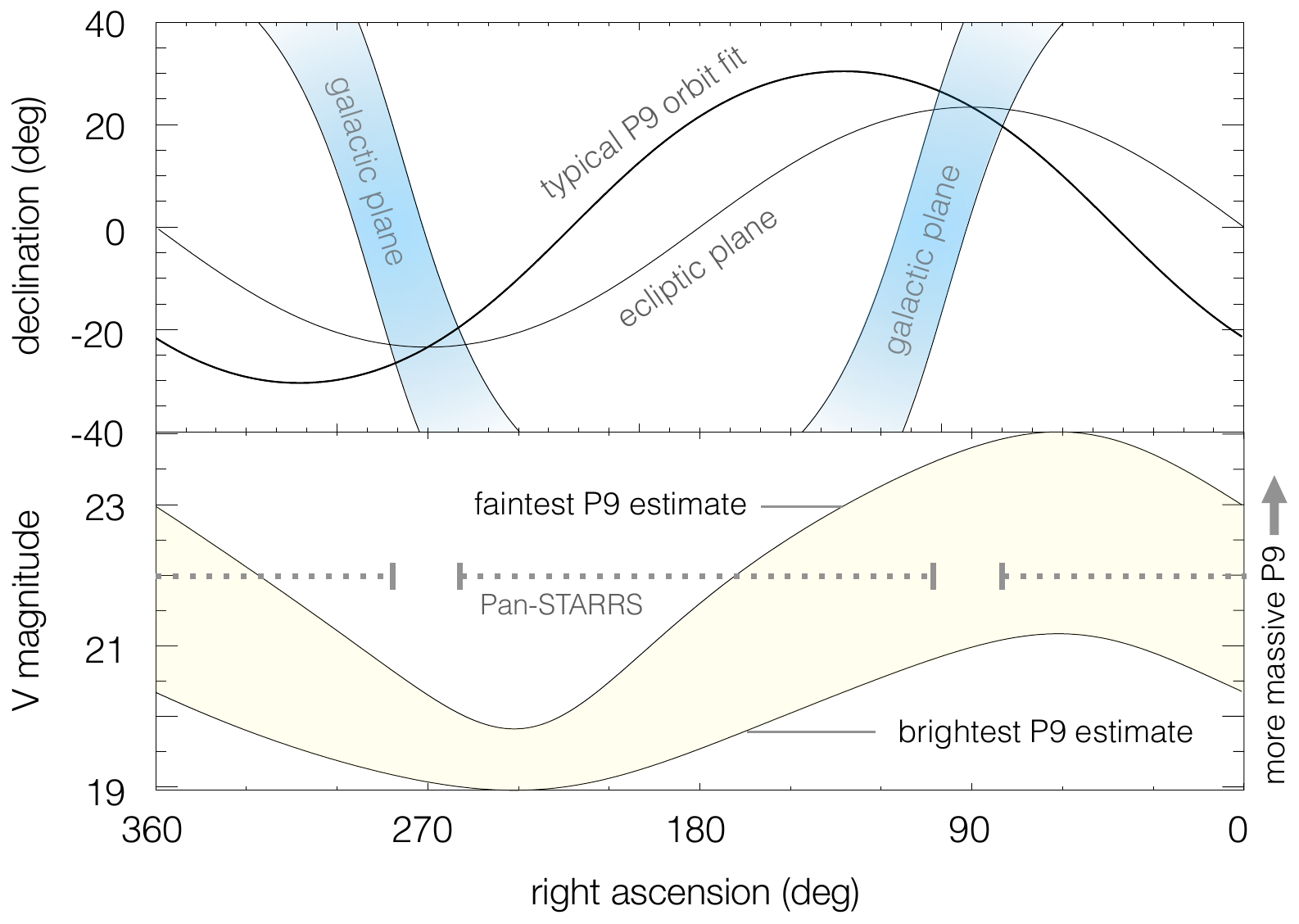}
\caption{On-sky properties of a typical P9 orbital fit. The top panel depicts an example path of Planet Nine, along with the RA-DEC projection of the ecliptic and galactic planes. The bottom panel reports the expected range of the visual magnitude of Planet Nine as a function of the right ascension. While ongoing observational surveys such as Pan-STARRS yield important limits on the location of P9 along its orbit, such constraints are absent from the galactic plane, which remains scarcely explored by solar system surveys.}
\label{fig:radec}
\end{figure}

As discussed above, for $5\,M_{\oplus}$, the best-fit orbital solution is characterized by $a_9=500$ AU and $e_9=0.25$ which corresponds to an aphelion distance of $625\,$AU, where it would be between magnitude $V\approx21.2$ and 22.2 (it would be between magnitude 19.0 and 20.0 at perihelion), depending on the albedo. For 10 M$_\oplus$, which we consider an upper limit to the mass of Planet Nine, the most distant acceptable orbit has $a_9=800$ AU, $e_9=0.45$, and an aphelion of $1160\,$AU. Such an object would have an aphelion magnitude between 23.0 and 24.0 (perihelion magnitude between 19.9 and 20.8).

Counterintuitively, a lower mass Planet Nine is brighter owing to its requirement for a smaller heliocentric distance to have the same dynamical effect. To this end, a $5 \,M_{\oplus}$ Planet Nine, for example, is bright enough to be detectable by wide-field surveys such as Pan-STARRS throughout most of its orbit. It is not yet known how complete the moving object search for Pan-STARRS is, but with the survey still ongoing a discovery could in principle happen at any time. An important complication, however, lies in that the inferred aphelion of Planet Nine's orbit is close to the intersection between P9's orbital path and the galactic plane, where higher source density can impede detection (see Figure \ref{fig:radec}).

A higher mass and thus more distant Planet Nine will require a dedicated survey along the predicted orbital path \citep{brownbatygin2016}, but with a lower limit to the brightness of 24th magnitude, such an object is readily observable by the current generation of telescopes with wide field cameras such as the Dark Energy Camera on the Blanco 4m telescope in Chile and the Hyper-Suprime Camera on the Subaru telescope in Hawaii. Finally, all but the very faintest possible Planet Nine will be observable with the Large Scale Synoptic Telescope (LSST), currently under construction in Chile and scheduled for operations in 2022. Therefore, Planet Nine -- if it exists as described here -- is likely to be discovered within a decade.

\subsection{Infrared and Microwave Surveys}

While detecting sunlight reflected at optical wavelengths might seem like the natural method for searching for Planet Nine, the $1/r^{4}$ dependency of the flux of reflected sunlight makes the brightness of any object drop precipitously with distance. At long wavelengths, thermal emission, on the other hand, drops as only $T/r^{2}$, where $T$ could be approximately constant with distance (or drop as $1/r^{2}$ if the planet is in thermal equilibrium with the sun). For sufficiently distant planets, thermal emission constitutes a potentially preferable avenue towards direct detection. While current cosmological experiments at millimeter wavelengths have sufficient sensitivity to detect Planet Nine \citep{cowan2016}, a systematic search would require millimeter telescopes with both high sensitivity and high angular resolution to robustly detect moving sources. The proposed CMB S-4, a next-generation cosmic microwave background experiment \citep{2016arXiv161002743A} could fulfill these requirements. Such an observatory would be sensitive not only to Planet Nine, but to putative even more distant bodies that might be present in the solar system.

It is worth noting that Planet Nine could emit more strongly than a blackbody at some wavelengths. \citet{fortney2016} found that in extreme cases the atmosphere of Planet Nine could be depleted of methane and have order of magnitude more emission in the 3-4 $\mu$m range. \citet{meisner2017b} exploited this possibility to search for Planet Nine in data from the Wide-field Infrared Survey Explorer (WISE) dataset. In particular, they examined 3$\pi$ steradians and placed a strongly atmospheric-model-dependent constraint on the presence of a high-mass Planet Nine at high galactic latitudes.

\subsection{Gravitational Detection}

A separate approach towards indirect detection of Planet Nine was explored by \citet{fienga2016}. Employing ranging data from the Cassini spacecraft, these authors sought to detect Planet Nine's direct gravitational signature in the solar system ephemeris (somewhat akin to the technique \citealt{leverriera, leverrierb} used to discover Neptune; section \ref{sec:findneptune}). Adopting the P9 parameters from \citep{phattie}, they were able to immediately constrain P9 to the outer $\sim50\%$ of the orbit (in agreement with observational constraints; \citealt{brownbatygin2016}). Moreover, the calculations of \citet{fienga2016} point to a small reduction in the residuals of the ranging data if the true anomaly of P9 is taken to be $\nu_9\approx118\deg$. This line of reasoning was further explored by \citet{holmanpayne16a, holmanpayne16b}, who additionally considered the long baseline ephemerides of Pluto to place additional constraints on the sky-location of Planet Nine. 

The reanalysis of the ephemeris carried out by \citet{folknerdps}, however, highlights the sensitivity of Fienga et al.'s results to the specifics of the underlying dynamical model, suggesting that the gravitational determinations of Planet Nine's on-sky location are in reality considerably less precise than advocated by \citet{fienga2016, holmanpayne16a, holmanpayne16b}. An additional complication pertinent to this approach was recently pointed out by \citet{pitjevapitjev}, who caution that failure to properly account for the mass contained within the resonant and classical Kuiper belt (which they determine to be on the order of $2 \times 10^{-2}\,M_{\oplus}$) can further obscure P9's gravitational signal in the solar system's ephemeris. Particularly, \citet{pitjevapitjev} find that the anomalous acceleration due to a $0.02\,M_{\oplus}$ Kuiper belt is essentially equivalent to that arising from a $m_9=10\,M_{\oplus}$ planet at a heliocentric distance of $r=540\,$AU (for orbits around Saturn), further discouraging the promise of teasing out Planet Nine's gravitational signal from spacecraft data.

\section{Formation Scenarios} 
\label{sec:formation} 

In terms of both physical and orbital characteristics, the inferred properties of Planet Nine are certainly unlike those of any other planet of the solar system. Recent photometric and spectroscopic surveys of planets around other stars \citep{kepler2010,kepler2013}, however, have conclusively demonstrated that $m\sim5-10\,M_{\oplus}$ planets are exceedingly common around solar-type stars, and likely represent one of the dominant outcomes of the planet conglomeration process\footnote{Jovian-class objects like Jupiter and Saturn, on the other hand, are comparatively rare and are believed to reside within 20 AU in only $\sim 20\%$ of sun-like stars \citep{cumming2008}.}. A moderately excited orbital state (and in particular, a high eccentricity) is also not uncommon among long-period extrasolar planets, and is a relatively well-established byproduct of post-nebular dynamical relaxation of planetary systems \citep{juric2008}. Nevertheless, the formation of Planet Nine represents a formidable problem, primarily due to its large distance from the sun. 

\begin{figure}[tbp]
\centering 
\includegraphics[width=1.0\textwidth]{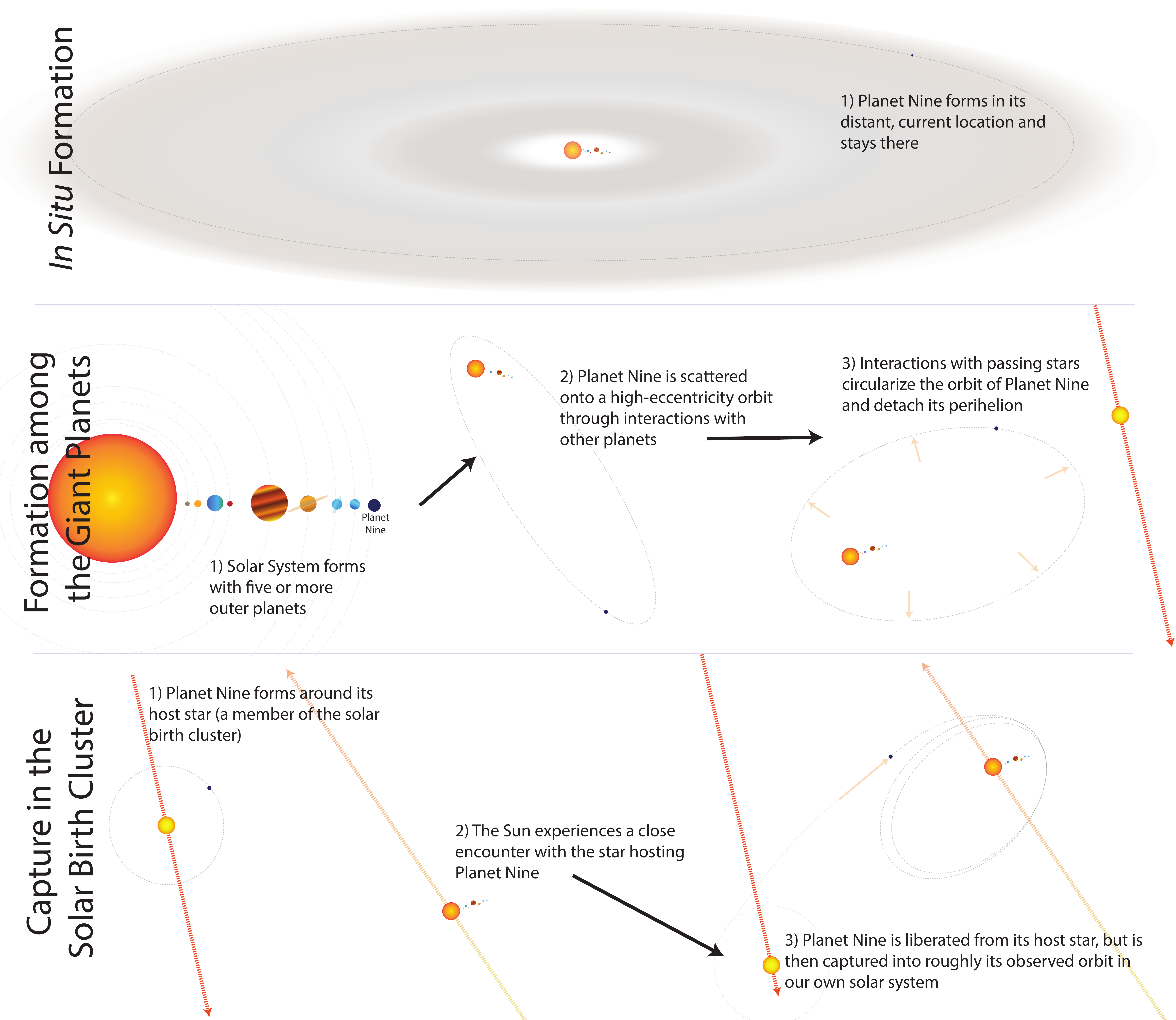}
\caption{Above, we show not-to-scale schematics for the three possible mechanisms by which Planet Nine could have been formed and placed in its current orbit in the solar system. (top panel) In \emph{in situ} formation, Planet Nine forms in its current distant orbit while the protoplanetary disk is still present, and resides there throughout the history of the solar system. (middle panel) If Planet Nine forms among the outer planets in the solar system, it could subsequently be scattered outwards onto a high-eccentricity orbit through interactions with the other solar system planets. Then, its orbit could be circularized through interactions with passing stars. (bottom panel) If Planet Nine originally formed around a host star other than the sun, a subsequent close encounter between this other star and the sun could result in Planet Nine being captured into its current-day long-period orbit around the sun.} 
\label{fig:formation} 
\end{figure}

To attack this speculative issue, several different origin scenarios have been proposed and are discussed in this section. The first option is for the planet to form {\sl in situ}, via analogous formation mechanism(s) responsible for the known giant planets (section \ref{sec:insitu}). Another option is for Planet Nine to form in the same annular region as the other giant planets, and then be scattered outward into its present orbit (section \ref{sec:forminside}). Yet another possibility is for the planet to originate from another planetary system within the solar birth cluster, and then be captured during the early evolution of the solar system (section \ref{sec:capture}). While all of these scenarios remain in play (Figure \ref{fig:formation}), each is characterized by non-trivial shortcomings, as discussed below. 


\subsection{In Situ Formation} 
\label{sec:insitu} 

Perhaps the most straightforward model for the origin of Planet Nine is for it to form {\sl in situ}, at its present orbital location. An attractive feature of this scenario is the fact that it does not require any physical processes beyond conglomeration itself. The advantages, however, stop there. Generally speaking, the timescale over which planetary building blocks (pebbles, planetesimals) amass into multi-Earth mass objects\footnote{While gravitational instability of the early solar nebula provides an alternate means of forming planets, it is irrelevant to the problem at hand, since the mass scale for objects generated through this channel is of order $10\,M_{J}$ (or higher, e.g., \citealt{rafikov2005})} is set by the orbital period at the location of the forming planet \citep{pebblereview}. With a $\sim10,000\,$year orbital period (corresponding to $a\sim500\,$AU), a forming Planet Nine would only complete $\sim300$ revolutions around the sun within the typical lifetime of a protoplanetary disk \citep{jesus}. The corresponding impedance of growth by the slowness of the orbital clock is illuminated by the calculations of \citet{kenyonbromley2016}, who find that even under exceptionally favorable conditions, formation of super-Earths at hundreds of AU requires billions of years.

Another shortcoming of {\sl in situ} formation concerns the availability of planet-forming material at large heliocentric distances. Various lines of evidence indicate the solar system did not originate in isolation, and instead formed within a cluster of $10^3-10^4$ stars (\citealt{al2001,zwart2009,adams2010,pfalzner2015}). Such a cluster environment can be highly disruptive to the outer regions of circumstellar disks and hence to planet formation. At minimum, a number of authors have shown that over a timescale of $\sim10\,$Myr, passing stars are expected to truncate the disk down to a radius of $\sim300$ AU, about one third of the minimum impact parameter \citep{heller,ostriker}. More importantly, these clusters also produce intense FUV radiation fields that evaporate circumstellar disks, removing all of the material beyond $\sim30-40$ AU over a time scale of 10 Myr \citep{adams2004}. Observational evidence supports this picture and indicates that disks in cluster environments experience some radiation-driven truncation (e.g., \citealt{kanderson}). Moreover, even in regions of distributed star formation, where external photoevaporation is unlikely to play a defining role, observations find that typical disk radii are only of order 100 AU \citep{haisch2001,andrews2009,andrews2010}. As a result, both observational and theoretical considerations suggest that the early Solar Nebula was unlikely to have extended much farther than the current orbit of Neptune at $a\sim30\,$AU (see also \citealt{kretke2012}). Forming Planet Nine at a radius of $\sim500\,$AU is thus strongly disfavored.

\subsection{Formation Among the Giant Planets} 
\label{sec:forminside} 

A somewhat more natural origin scenario is for Planet Nine to form within the region that produces the known giant planets (i.e., the annulus defined roughly by 5\,AU\,$\simless a \simless$\,30\,AU), and to be scattered out later. The physics of the planet formation process is notoriously stochastic, and the number of planets produced in a given planet-forming region cannot be calculated in a deterministic manner \citep{morbyraymond}, meaning that additional ice giants other than Uranus and Neptune may have occupied the outer solar system during its infancy. To this end, analytic arguments put forth by \citet{Peter2004} suggest that the outer solar system could have started out with as many as five ice giants. Along similar lines, numerical models for the agglomeration of Neptune and Uranus through collisions among large planetary embryos find that $\sim5\,M_{\oplus}$ objects routinely scatter away from the primary planet forming region \citep{izidoro}. Calculations of ancillary ice giant ejection during the outer solar system's transient epoch of dynamical instability are also presented in \citet{Nesvorny2011,batygin2012,nm212}.

It is important to recognize that this model of Planet Nine formation must necessarily involve a two-step process. This is because outward scattering of Planet Nine facilitated by the giant planets places it onto a temporary, high-eccentricity ($q\sim5\,$AU) orbit, which must subsequently be circularized (thus lifting its perihelion out of the planet-forming region) by additional gravitational perturbations arising from the cluster. The difficulty with this scenario, however, is that the likelihood of producing the required orbit for Planet Nine is low. \citet{liadams2016} estimated the scattering probability for this process by initializing Planet Nine on orbits with zero eccentricity and semi-major axis $a\approx100-200\,$AU, and found that stellar fly-by encounters produce final states with orbital elements $a=400-1500$ AU, $e=0.4-0.9$, and $i<60\deg$ only a few percent of the time. We note however, that the calculations of \citet{brasser2006,brasser2012} obtain considerably more favorable odds of decoupling a scattered planetary embryo from the canonical giant planets and trapping it in the outer solar system, with reported probability of success as high as $\sim15\%$, depending on the specifics of the adopted cluster model.

As an alternative to invoking cluster dynamics, \citet{ericksson2018} considered the circularization of a freshly scattered Planet Nine through dynamical friction arising from a massive disk of planetesimals, extending far beyond the orbit of Neptune. Unlike the aforementioned cluster calculations, in this scenario the chances of producing a favorable Planet Nine orbit can be as high as $\sim30\%$. An important drawback of this model, however, is that it suffers from the same issues of disk truncation outlined in the previous sub-section. Moreover, simulations of the Nice model \citep{tsiganis2005} require the massive component of the solar system's primordial planetesimal disk to end at $a\sim30\,$AU, to prevent Neptune from radially migrating beyond its current orbit. Therefore, within the framework of \citet{ericksson2018}, some additional physical process would be required to create an immense gap ranging from $30\,$AU to $100\,$AU in the solar system's primordial planetesimal disk.

\subsection{Ejection and Capture Within the Solar Birth Cluster} 
\label{sec:capture} 

Although gravitational perturbations arising from passing stars can act to alter the orbital properties of Planet Nine, as discussed above, the possibilities do not end there -- the birth environment of the solar system can also lead to the disruption of Planet Nine from its wide orbit. Several groups have worked to quantify the effects of scattering encounters in young stellar clusters using N-body methods over the last decade (e.g., \citealt{zwart2009,malmberg2011,pfalzner2013,pfalzner2015,pfalzner2018}). Another way to study this class of disruption events is to separately calculate the cross sections for fly-by encounters to ionize the solar system, and combine these results with the rate of stellar encounters to determine an optical depth, $\tau_{\rm s}$, for scattering \citep{al2001,liadams2015,adams2006,proszkow}. This quantity can be written in the form 
\be
\tau_{\rm s} = \int n_\ast\, \langle{v\,\sigma_{\rm{int}}}\rangle \,dt\,,
\label{tauscatter} 
\ee
where $n_\ast$ is the number density of stars in the system, $v$ is the relative speed between the Sun and other systems, and $\sigma_{\rm{int}}$ is the interaction cross section. The angular brackets denote averages over both the velocity distribution and the stellar initial mass function for the cluster members. 

For interactions in a cluster with velocity dispersion of $v_c = 1$\,km/s, and a planet with an aphelion distance of $640\,$AU (comparable to our best-fit P9 orbital solution obtained in section \ref{sec:numerical}), \citet{liadams2016} derive an ejection cross-section of $\sigma_{\rm int}\approx2.5\times10^6$\,AU$^2$. Adopting a stellar number density of $n_{\ast}=100$\,pc$^{-3}$ and a cluster lifetime of $\Delta t= 100\,$Myr, we obtain $\tau_{\rm s}\approx0.4$\footnote{Following \citet{liadams2015}, we have accounted for the fraction of single (vs binary) stellar encounters by reducing $\tau_{\rm s}$ by a factor of $2/3$.}, which translates to a P9 survival probability of $\exp(-\tau_{\rm{s}})\approx2/3$. In other words, integrated over the lifetime of the cluster, the probability of ejecting Planet Nine is of order $\sim30\%$.

Once the solar system emerges from its birth cluster, the probability that Planet Nine can be stripped from the solar system by a passing star diminishes dramatically. In particular, for scattering interactions in the field \citep{bintrem}, the stellar density is much lower ($\langle n_\ast \rangle \approx 0.1$ pc$^{-3}$) and the velocity dispersion is higher ($v_c\approx40$ km/s). The higher interaction speeds lead to much lower ejection cross sections (see \citealt{liadams2015}), which more than compensates for the longer residence time (about 4.5 Gyr). As a result, the corresponding scattering optical depth for the ejection of Planet Nine in the solar neighborhood is essentially negligible -- $\tau_{\rm s}$ $\approx$ 0.002 -- 0.02. In other words, once the solar system emerges from its birth cluster, and Planet Nine attains its required orbit, it has an excellent chance of continued survival.

In light of the fact that ejection of planets constitutes a distinct possibility within the early evolution of the solar system, the same reasoning applies to other members of the solar system's birth cluster. This opens up the possibility that rather than originating within the solar system, Planet Nine was ejected from some other planetary system, and was subsequently captured by the sun's gravitational field. This P9 capture scenario has been recently considered by a number of authors \citep{liadams2016,mustill2016,parker2017}. However, a multitude of requirements must be met in order for successful capture to take place. 

One limiting factor is that the scattering encounter must be sufficiently distant so that the cold classical population of the Kuiper Belt (which is likely primordial; \citealt{batygin2012}) is not destroyed. This demand implies that the impact parameter exceeds $b\simgreat150-200$ AU. More constraining is the requirement that the final orbital elements of the captured planet be consistent with those inferred for Planet Nine. For the conditions expected in the solar birth cluster, the probability for successful capture is of order a few percent, both for freely floating planets \citep{liadams2016}, and for planets captured from other planetary systems \citep{mustill2016}. More recent work \citep{parker2017} finds even lower probabilities, with successful captures of freely floating planets estimated to be only 5 -- 10 out of $10^4$. In addition, planet capture is enhanced for expanding, unbound clusters. In contrast, solar system enrichment of short-lived radiogenic isotopes is enhanced for subvirial (partially collapsing) clusters, so some tension exists between the requirement of capturing Planet Nine and explaining meteoritic enrichment.

\section{Conclusion}
\label{sec:conclude}

Over the past decade and a half, continued detection of minor bodies in the distant solar system has brought the intricate dynamical architecture of the distant Kuiper belt into much sharper focus. Staggeringly, the collective orbital structure of the current census of long-period trans-Neptunian objects offers a number of tantalizing hints at the possibility that an additional massive object -- Planet Nine -- may be lurking beyond the current observational horizon of the distant solar system. In this work, we have presented a comprehensive review of the observational evidence, as well as the analytical and numerical formulation of the Planet Nine hypothesis. The emergent theoretical picture is summarized in section \ref{sec:summary}, followed by a discussion of alternate explanations for the observed orbital anomalies (section \ref{sec:altideas}) and prospects for future work (section \ref{sec:future}). 

\subsection{Summary of Results} 
\label{sec:summary} 
The case for the existence of Planet Nine can be organized into the following four primary lines of evidence.

\medskip \noindent $\bullet$ {\sl Orbital Alignment of Long-Period KBOs.} 
The observed collection of dynamically stable Kuiper belt objects with semi-major axes in excess of $a\simgreat250$ AU (orbital periods greater than $P\gtrsim 4000\,$years), reside on orbits that are clustered together in physical space. This clustering is statistically significant at the $99.8 \%$ significance level, and ensues from the simultaneous alignment of the apsidal lines (equivalently, longitudes of perihelion) as well as the orbital planes (which are dictated by the inclinations and the longitudes of ascending node). Because the orbits precess differentially under perturbations from the known giant planets, a sustained restoring gravitational torque is required to preserve the orbital grouping. Dynamical evolution induced by Planet Nine can fully account for the generation and maintenance of the observed alignment of long-period KBOs, while simultaneously allowing for their orbital stability. 

\medskip \noindent $\bullet$ {\sl Detachment of Perihelia.} Long-period KBOs exhibit a broad range of perihelion distances, with a substantial fraction possessing $q > 40\,$AU. Such objects are dynamically decoupled from Neptune, and cannot be created during Kuiper belt formation via interactions with the known planets alone. As a result, this population of KBOs requires external gravitational perturbations to lift their perihelia. Within the framework of the Planet Nine hypothesis, the same dynamics that are responsible for the aforementioned apsidal alignment also provide a means of generating Neptune-detached Kuiper belt objects, thereby explaining their origins.

\medskip \noindent $\bullet$ {\sl Excitation of Extreme TNO Inclinations.} Trans-Neptunian objects with inclinations in excess of $i\gtrsim50\deg$ are not a natural outcome of the solar system formation process. Nevertheless, several objects with inclinations well above $50\deg$ have been detected on wide ($a>250\,$AU) orbits, including the recently discovered KBO 2015\,BP$_{519}$ which is the only known member of this class characterized by $q>30\,$AU. Despite their puzzling nature, such highly inclined objects are routinely produced within the framework of the P9 hypothesis, via a high order (octupolar) secular resonance with Planet Nine.

\medskip \noindent $\bullet$ {\sl Production of Retrograde Centaurs.} In addition to long-period TNOs with $i>50\deg$, the solar system also hosts a multitude of highly inclined, and even strongly retrograde shorter-period ($a<100\,$AU) objects. As with their distant counterparts, the dynamical origins of such objects is inexplicable through perturbations from the known giant planets alone. Although these objects are decoupled from Planet Nine's gravitational influence today, numerical simulations demonstrate that an intricate interplay between P9-induced dynamical evolution and Neptune scattering can deliver highly inclined long-period TNOs onto shorter period orbits, polluting the classical region of the Kuiper belt with highly inclined Centaurs. More generally, this process allows for the injection of objects that trace through orbits in the retrograde direction, into the trans-Jovian domain of the solar system. 

\medskip \noindent 
Each of the above dynamical effects can be understood from purely analytic grounds within the framework of the Planet Nine hypothesis. Simplified models of this sort, based on secular perturbation theory, are presented in section \ref{sec:analytical}. Detailed comparison with the data, however, requires the fabrication of a synthetic population of long-period KBOs using large-scale N-body simulations. The results of thousands of such simulations are described in section \ref{sec:numerical}, and collectively point to a revised set of physical and orbital parameters for Planet Nine. Specifically, compared to the original results \citep{phattie}, where P9 was reported to have $m_9\approx10M_{\oplus}$ and occupy an $a_9=700\,$AU orbit with $e_9=0.6$, the current simulations (reviewed in section \ref{sec:numerical}), point towards a marginally lower-mass planet that resides on a somewhat more proximate and less dynamically excited orbit, with $m_9 \approx 5-10 M_{\oplus}$, $a_9\sim400-800\,$AU, $e_9\sim0.2-0.5$, and $i_9\sim15-25\deg$. Perhaps counterintuitively, the increase in brightness due to a smaller heliocentric distance more than makes up for the decrease in brightness due to a slightly diminished physical radius, suggesting that Planet Nine is more readily discoverable by conventional optical surveys than previously thought. 

\subsection{Alternative Explanations} 
\label{sec:altideas} 
As a formulated dynamical model, the Planet Nine hypothesis provides a satisfactory account for the orbital anomalies of the distant solar system. Nevertheless, until the existence of Planet Nine is confirmed observationally, the possibility that the envisioned theory is incorrect will continue to linger. The history of proposed planets based on dynamical anomalies suggests that this option should be taken seriously. In this vein, it is useful to recall the cautionary tale surrounding the prediction and subsequent abandonment of planet Vulcan (see section \ref{sec:vulcan}), as it illuminates the vulnerability of even the most well-formulated theoretical models. While it is unlikely that the asymmetries of the orbital structure of the distant solar system point to fundamentally new physics (as in the case of Vulcan), we must acknowledge alternative explanations for the observations that do not invoke the existence of Planet Nine. These ideas generally fall into two distinct categories, and we briefly discuss their attributes below.

\paragraph{Observational Bias}
If the Planet Nine hypothesis proves to be unnecessary, perhaps the most likely explanation is that no explanation is required. 
One can envision a scenario where the observational strategy and serendipity conspire to produce the observed patterns in the data. In this case, the apparent clustering in longitudes of perihelion and orbital poles would be spurious, and the underlying distribution of orbital angles of the distant solar system objects would be uniform \citep[e.g.,][]{lawler2017, shankmansurvsim}.

The upshot here is that this interpretation can be quantified by a well-defined statistical likelihood. As discussed in section \ref{anomalous}, the associated false-alarm probability is estimated to be less than one percent \citep{brownbatygin2018}. Beyond angular clustering alone, the statistical significance of the Planet Nine hypothesis is further reinforced by the fact that multiple lines of evidence all point to the {\it same} Planet Nine model. Specifically, not only can the orbital alignment of the distant TNOs be understood in terms of a new planet, the existence of perihelion-detached objects, as well as the generation of the observed high inclination objects also naturally emerge within the framework of the same theoretical model. While observational bias can never be completely ruled out as an explanation for the data, continued detection of long-period Kuiper belt objects (e.g., \citealt{goblin,triplet}, etc) will lead to further refinement of the false-alarm probability.


\paragraph{Self-Gravity of the Distant Kuiper Belt} 
A separate class of models posit that the observed structure of the distant solar system is real but is not sculpted by a planet, and instead arises from the collective gravity of the distant Kuiper belt. The first of these theories is due to \citet{madigan}, who propose that the distant solar system contains $1-10M_{\oplus}$ of unseen material. Over $\sim$Gyr timescales, this population of objects experiences coupled inclination-eccentricity evolution -- a process the authors refer to as the ``inclination instability" (see \citealt{madigan18} for further discussion). The development of this instability results in a clustered distribution of the arguments of perihelion of the constituent bodies, in agreement with the anomalous pattern first pointed out by \citet{trushep2014}. However, this model provides no explanation for the observed confinement in the \textit{longitudes} of perihelion as well as the angular momentum vectors, and thus fails to reproduce the actual observed structure of the distant belt. Moreover, the inclination instability does not naturally manifest in numerical simulations of Kuiper belt emplacement that self-consistently account for self-gravity of the planetesimal disk, calling into question the specificity of initial conditions required for this model to operate \citep{fan2017}. 

A different self-gravity model has been recently proposed by \citet{2018arXiv180406859S}. In this theory, instead of the observed Kuiper belt contributing to the collective potential of the trans-Neptunian region, there exists a moderately eccentric ($e\sim0.2$) shepherding disk that extends from $a_{\rm{in}}=40\,$AU to $a_{\rm{out}}=750\,$AU, comprising $\sim10M_{\oplus}$ of material in total. The gravitational potential of this disk creates a stable, apsidally anti-aligned secular equilibrium at high eccentricity that facilitates the confinement of the longitudes of perihelion of the observed long-period KBOs (which act as test particles, enslaved by the disk's potential). In light of the fact that the primary mode of dynamical evolution induced by Planet Nine itself is secular in nature (and in an orbit-averaged sense, the gravitational potential of Planet Nine is not different from a massive wire that traces its orbit), there are clear mathematical similarities that ensue between this model and the analytical formulation of the P9 hypothesis (section \ref{sec:analytical}). Nevertheless, from the point of view of planet formation, the prolonged existence of the envisioned shepherding disk suffers from many of the same drawbacks outlined in section \ref{sec:insitu} (e.g., radiative stripping of $r\gtrsim50\,$AU protoplanetary disk material, disruptive perturbations due to passing stars in the birth cluster, etc.) as well as the apparent incompatibility with the striking decrease in observed number of KBOs beyond $a\gtrsim48\,$AU.

\subsection{Future Directions} 
\label{sec:future} 

One result of this review is that numerical simulations containing Planet Nine can produce orbits that are in good agreement with the observed structure of the distant Kuiper belt. Nonetheless, a number of theoretical questions remain unanswered, and their resolution may help illuminate the path toward the direct detection of the most distant planetary member of the solar system. As a result, we conclude this article with a brief (and incomplete) discussion of some problems that remain open within the broader framework of the Planet Nine hypothesis.

\paragraph{The Inclination Dispersion of Jupiter-Family Comets}

The inward scattering of TNOs that have been strongly influenced by Planet Nine constitutes a dynamical pathway through which highly inclined icy bodies can be delivered to much smaller heliocentric distances. In this way, Planet Nine can affect not only the orbital properties of the Kuiper belt itself, but the inclination dispersion of Jupiter-Family comets, which are sourced from the trans-Neptunian region \citep{1988ApJ...328L..69D}. In a recent study, \citet{spcomets} carried out detailed simulations of the generation and evolution of short-period comets, and found that inclusion of a $m_9=15M_{\oplus}$ Planet Nine that resides on a $a=700\,$AU, $e=0.6$ orbit leads to a simulated inclination dispersion of Jupiter-Family comets that is broader than its observed counterpart. 

Importantly, however, the simulation suite presented in section \ref{sec:numerical} of this work favors a lower mass, more circular Planet Nine over the parameters adopted by \citet{spcomets}. Because the inclination excitation of long-period TNOs is driven by octupole-level secular interactions with P9, we can reasonably expect that our revised P9 parameters would significantly diminish both the rate, and the efficiency of high-$i$ comet generation, potentially eliminating any discrepancy between the observed and modeled orbital distributions. A quantitative evaluation of the possibility is of substantial interest.

\paragraph{Contamination of the Distant Kuiper Belt}

Within the current observational sample of distant KBOs, orbital clustering exhibited by long-term (meta)stable KBOs is considerably tighter than that of the unstable objects. Qualitatively, this disparity is sensible since unstable objects actively interact with Neptune, and can be rapidly perturbed away from P9-sculpted dynamical states. Moreover, given their rapid excursions in semi-major axes, such objects could represent relatively recent additions to the distant Kuiper belt drawn from the more proximate (and therefore P9-unperturbed) region of trans-Neptunian space. At the same time, observational bias strongly favors the detection of low-perihelion objects, meaning that this dynamically unstable sub-sample of KBOs is significantly over-represented within the current dataset.

Because the typical lifetimes of unstable objects are short compared to the lifetime of the solar system, they are almost entirely absent from the simulation suite presented in section \ref{sec:numerical}, limiting the scope of our comparison between simulation and data only to (meta)stable objects. Overcoming this limitation, and better quantifying the P9-facilitated evolution of objects that experience rapid orbital diffusion due to strong coupling with Neptune with an eye towards better understanding the orbital patterns exhibited by unstable long-period KBOs would nicely complement the calculations presented in this work.

\paragraph{Interactions With the Galaxy}

A subtle limitation of almost all Planet Nine calculations that have been carried out to date, lies in that they treat the solar system as an isolated object, thus ignoring the effects of passing stars as well as the gravitational tide of the Galaxy (exceptions to this rule include the recent works of \citealt{spcomets, goblin}). This approximation is perfectly reasonable for simulating the evolution of objects with semi-major axes less than a few thousand AU, and for now, the vast majority of known long-period TNOs lies in this domain. This picture is however slowly changing, and recent detections of very long-period trans-Neptunian objects \citep{gomes2015, sheptru2016, goblin} increasingly point to an over-abundance of long-period minor bodies with semi-major axes larger than $\sim1000\,$AU. 

Because there exists a strong bias towards detection of shorter-period objects, it is likely that the prevalence of these extremely distant TNOs cannot be fully attributed to the standard model of scattered KBO generation, wherein long-period orbits are created by outward scattering facilitated by Neptune. If these objects did not originate from more proximate orbits, then where did they come from? A riveting hypothesis is that they are injected into the distant solar system from the inner Oort cloud, via a complex interchange between Galactic effects and Planet Nine's gravity. Moreover, because these long-period KBOs generally conform to the pattern of orbital alignment dictated by Planet Nine, a careful characterization of their dynamical evolution may yield an excellent handle on Planet Nine's gravitational sculpting at exceptionally large heliocentric distances.

\paragraph{Alternate Orbital Solutions} 

All simulations of P9-induced dynamical evolution carried out in this work (section \ref{sec:numerical}) were founded on a series of analytical models delineated in section \ref{sec:analytical}. While this aggregate of calculations cumulatively points to a specific range of P9 orbital parameters that provide a satisfactory match to the data, it is important to keep in mind that we have not strictly ruled out the possibility that there could exist other, more exotic orbital configurations that might match the data equally well. For instance, we have not thoroughly examined the possibility that Planet Nine's orbit itself could be very highly inclined (or even retrograde), and that a seemingly strange orbital architecture of the distant Kuiper belt generated by such a planet could be rendered compatible with the current dataset by observational biases. Continued numerical exploration of P9-sculpted orbital structure outside of the parameter range considered in this review will help quantify this possibility.

Even if the qualitative aspects of our theoretical models are correct, we must acknowledge that the dataset which informs our understanding of the distant Kuiper belt remains sparse, and new detections of large-$a$ KBOs may significantly alter the population's inferred statistical properties. As an example, we note that even basic attributes of the distant belt, such as the critical semi-major axis beyond which orbital clustering ensues, $a_{\rm{c}}$, are characterized by considerable uncertainties as well as a non-trivial dependence on $q$. While in this work we have tentatively adopted $200\rm{AU}\,$$\lesssim a_{\rm{c}}\lesssim300\rm{AU}$ as a criterion for simulation success, if additional data reveals that the true value of $a_{\rm{c}}$ lies interior or exterior to this range, P9 orbital fits with respectively lower or higher semi-major axes, $a_9$ would be rendered acceptable. Thus, as the census of well-characterized long-period TNOs continues to grow, alteration of this (and other) statistical characteristics of the distant Kuiper belt will inevitably lead to further refinement of Planet Nine's orbital elements.

\bigskip
Further theoretical curiosities aside, arguably the most practically attractive aspect of the P9 hypothesis is the prospect of near-term observational confirmation (or falsification) of the results discussed in this review. Not only would the astronomical detection of Planet Nine instigate a dramatic expansion of the Sun's planetary album, it would shed light on the physical properties of a Super-Earth class planet, while evoking extraordinary new constraints on the dramatic early evolution of the solar system. The search for Planet Nine is already in full swing, and it is likely that if Planet Nine -- as envisioned here -- exists, it will be discovered within the coming decade.

\bigskip
\noindent 
{\bf Acknowledgments:} This review benefited from discussions and additional input from many people, and we would especially like to thank Elizabeth Bailey, Tony Bloch, David Gerdes, Stephanie Hamilton, Jake Ketchum, Tali Khain, and Chris Spalding. We are indebted to Greg Laughlin, Erik Petigura, Alessandro Morbidelli, Gongjie Li and an anonymous referee for critical readings of the text and for providing insightful comments that led to a considerable improvement of the manuscript. We also thank Caltech's Division of Geological \& Planetary Sciences for hosting F.C.A. during his sabbatical visit, January -- April 2018, when work on this manuscript was initiated. K.B. is grateful to the David and Lucile Packard Foundation and the Alfred P. Sloan Foundation for their generous support.

\newpage
\appendix
\section{A Historical Remark} \label{appA}

Besides the false alarm of Vulcan described in section \ref{sec:vulcan}, the discovery of Pluto itself provides another cautionary tale that illustrates the potential power of
dynamical arguments -- this time, a missed opportunity. During the search for Planet X, Lowell Observatory was also being used to measure the recession velocities of spiral nebulae. These entities are now known to be external galaxies, although the spatial extent of the galaxy was not fully specified at the time. Vesto Slipher measured recession
velocities for 25 such spiral nebulae in the range $V=300-1120\,$km/s during the decade of 1910 -- 1920 \citep{slipher1917a,slipher1917b}. 

These measurements were published before the famous debate between Harlow Shapley and Heber Curtis \citep{curtis1921,shapley1921}, which considered the extent of the galaxy, and well before the publication of the Hubble relation \citep{hubble1929}. In order for a nebula with recession speed $v$ to be bound to the Milky Way, and {\it not} be an external entity, the mass of The Galaxy must be bounded from below by 
\be
M_{\rm mw} \simgreat {V^2 R_{\rm mw} \over G} > 
{V^2 R_{\rm min} \over G} \, .
\label{milkymass} 
\ee
During the debate of 1921, many issues were in dispute, but both sides agreed that the minimum size of the Galaxy was $\sim30,000$ light years \citep{curtis1921,shapley1921}; the point of contention was whether or not it was much larger. With this minimum size, the mass limit of equation (\ref{milkymass}) becomes
\be
M_{\rm mw} \simgreat 3 \times 10^{12} M_\odot \,, 
\ee 
where we have used the larger measured recession velocities (1100\,km/s). The same debate held that the Milky Way contained ``about one billion suns'', which falls short of the above limit by a factor of $\sim3000$. The debaters thus missed the opportunity to make a dynamical argument for the existence of external galaxies, and thereby resolve the
controversy i.e., the enormous recession speeds pointed strongly to the conclusion that spiral nebulae were not bound to our Galaxy. In order to avoid this result, the mass of the Galaxy would have to be thousands of times larger than expected (which would also have been an interesting possibility).

\section{Variable Transformations} \label{appB}

In section \ref{sec:analytical}, we employed a series of integrable Hamiltonians in order to elucidate the dynamical mechanisms through which Planet Nine sculpts the distant Kuiper belt. In addition to terms that represent orbit-averaged gravitational coupling between the KBO and the outer planets of the solar system, each of these Hamiltonians (\ref{Hpuresec},\ref{Hinc},\ref{Hhighinc}) also contains terms that arise from variable transformations that remove explicit time-dependence from the potential. Let us outline these variable transformations, starting with the one relevant to equation (\ref{Hpuresec}).

To begin with, let us switch from Keplerian orbital elements to a set of canonically conjugated variables. Correspondingly, we define the \Poincare\ action-angle coordinates \citep{dm1999}:
\begin{align}
&\Lambda=\sqrt{\G\,\Msun a}&\lambda=\nu+\varpi \nonumber \\
&\Gamma=\sqrt{\G\,\Msun a}\,\big(1-\sqrt{1-e^2} \big)   &\gamma=-\varpi \nonumber \\
&Z=\sqrt{\G\,\Msun a}\,\sqrt{1-e^2}\,\big(1-\cos(i) \big)   &z=-\Omega,
\label{Poincvarfull}
\end{align}
where $\nu$ is the mean anomaly. Additionally, let us explicitly define the precession and regression rates of Planet Nine's longitude of perihelion and longitude of ascending node:
\begin{align}
&\dot{\varpi}_9 \approx - \dot{\Omega}_9 \approx \frac{3}{4}\sqrt{\frac{\G\,\Msun}{a_9^3}}\frac{1}{\big(1-e_9^2\big)^{2}} \sum_{j=5}^{8}\frac{m_j\,a_j^2}{\Msun\,a_9^2}
\label{P9varsdt}
\end{align}
With these expressions in hand, we take $\varpi_9=\dot{\varpi}_9\,t$ and $\Omega_9=\dot{\Omega}_9\,t$. 

Within the framework of secular perturbation theory, the full Hamiltonian, $\mathcal{K}$ is canonically smoothed over the mean longitudes $\lambda$, to yield an averaged Hamiltonian $\bar{\mathcal{K}}$. Because equations (\ref{P9varsdt}) entail a temporal dependence of P9 angles (e.g., via $\Delta\varpi=\dot{\varpi}_9\,t-\varpi$, etc), the Hamiltonian $\bar{\mathcal{K}}$ is non-autonomous. To formally circumvent this time-dependence, we extend the phase-space \citep{Morbybook} and introduce a dummy action $\mathcal{T}$, conjugate to time, such that
\begin{align}
\mathcal{H}= \bar{\mathcal{K}} + \mathcal{T}.
\label{Hautonom}
\end{align}
In each case, to carry out the variable transformation, we define a generating function of the second type, $\mathcal{F}_2$, and derive the action transformation equations from the following relations:
\begin{align}
&&\Gamma=\frac{\partial\,\mathcal{F}_2}{\partial\,\gamma}   &&Z=\frac{\partial\,\mathcal{F}_2}{\partial\,z}   &&\mathcal{T}=\frac{\partial\,\mathcal{F}_2}{\partial\,t}.
\label{type2}
\end{align}

The first change of variables we consider is a transformation to a frame where the reference apsidal line co-precesses with the orbit of Planet Nine. The relevant generating function has the form
\begin{align}
\mathcal{F}_2=(\underbrace{\dot{\varpi}_9\,t+\gamma}_{\phi})\,\Phi+(\underbrace{t}_{\xi})\,\Xi.
\label{F2Phi}
\end{align}
The resulting transformation is then 
\begin{align}
&\Phi=\sqrt{\G\,\Msun\,a}\,\big(1-\sqrt{1-e^2}\big)    &\phi=\varpi_9-\varpi \nonumber \\
&\mathcal{T}=\dot{\varpi}_9\sqrt{\G\,\Msun\,a}\,\big( 1-\sqrt{1-e^2}\big)+\Xi   &\xi=t.
\label{finalvars}
\end{align}
This transformation elucidates the origin of the second term in equation (\ref{Hpuresec}). Moreover, because the Hamiltonian (\ref{Hpuresec}) is now dependent only on $\phi$ and not on $\xi$, $\Xi$ is a constant of motion and can be dropped all together from $\mathcal{H}$. 

The second transformation is in essence identical to the first, with the exception of the fact that we now seek to transform to a frame where the longitude of ascending node is measured from the lines of nodes of Planet Nine. In parallel with equation (\ref{F2Phi}), we define the generation function as follows:
\begin{align}
\mathcal{F}_2=(\underbrace{\dot{\Omega}_9\,t+z}_{\psi})\,\Psi+(\underbrace{t}_{\xi})\,\Xi.
\label{F2Psi}
\end{align}
Correspondingly, the transformation equations take the form:
\begin{align}
&\Psi=\sqrt{\G\,\Msun\,a}\,\sqrt{1-e^2}\big(1-\cos(i)\big)    &\psi=\Omega_9-\Omega \nonumber \\
&\mathcal{T}=\dot{\Omega}_9\sqrt{\G\,\Msun\,a}\,\sqrt{1-e^2}\,\big( 1-\cos(i)\big)+\Xi   &\xi=t.
\label{finalvars2}
\end{align}
As with the case of eccentricity coupling described above, the second term in Hamiltonian (\ref{Hinc}) arises from the transformation of the dummy action $\mathcal{T}$.

A final transformation we need to outline is one to variables (\ref{eqn:theta}). To do this, consider the generating function
\begin{align}
\mathcal{F}_2=(\underbrace{\dot{\varpi}_9\,t+\gamma}_{w})\,\mathcal{W}+(\underbrace{2z-\gamma+\dot{\varpi}_9\,t}_{\theta})\,\Theta+(\underbrace{t}_{\xi})\,\Xi.
\label{F2Theta}
\end{align}
Upon direct substitution of $\mathcal{F}_2$ into equations (\ref{type2}), we obtain
\begin{align}
&\mathcal{W}=\sqrt{\G\,\Msun\,a}\,\big(2-\sqrt{1-e^2}\,(1+\cos(i)) \big)    &w=\varpi_9-\varpi \nonumber \\
&\Theta=\sqrt{\G\,\Msun\,a}\,\sqrt{1-e^2}\,\big(1-\cos(i) \big)/2   &\theta=\varpi+\varpi_9-2\Omega \nonumber \\
&\mathcal{T}=\dot{\varpi}_9\sqrt{\G\,\Msun\,a}\,\big( 1-\sqrt{1-e^2}\,\cos(i) \big)+\Xi   &\xi=t.
\label{finalvars3}
\end{align}
In section \ref{sec:analytical}, we evaluated $\mathcal{H}$, keeping $e$ (and $\Delta\varpi$) constant (which allowed us to project contours of $\mathcal{H}$ onto a $\theta-\Theta$ plane). We note that an equally crude assumption would be to keep the action $\mathcal{W}$ constant instead. Either way, it is important to keep in mind that within the framework of a more complete model of P9-induced dynamics, the actions $\mathcal{W}$ and $\Theta$ evolve on a comparable timescale, although the ($\Theta,\theta$) degree of freedom plays a more dominant role in facilitating the excitation of high-inclination dynamics in the distant Kuiper belt.

\section{Dynamics of Observed KBOs Subject to P9 Perturbations} \label{appC}

The results discussed in the main text point toward the existence of a new solar system member, Planet Nine, with particular properties. More specifically, the observed orbits of the extreme KBOs are best explained for a new planet with mass $m_9\approx5M_\oplus$, semi-major axis $a_9\approx$ 500 AU, orbital eccentricity $e_9\approx0.25$, and
inclination $i_9\approx$ 20 deg. Compared with most previous Planet Nine scenarios in the literature, this updated version has mass near the lower end of the usually quoted range, with a somewhat closer aphelion distance.  As a consistency check on this emerging solution, this Appendix presents the results of numerical simulations of the observed
KBOs orbiting under the influence of this particular Planet Nine (along with the rest of the known solar system). As outlined below, the resulting dynamics of the extreme KBOs are consistent with expectations.

It is important to note that the analysis presented in the text (see section \ref{sec:numerical}) is essentially a forward model. The initial states of the simulations consist of an unstructured population of KBOs, and a candidate Planet Nine is introduced to sculpt the synthetic disk of icy bodies. The `best' versions of Planet Nine are then taken to be those that produce collections of KBO orbits that resemble those that are observed. In addition, one would like the proposed new planet to be demonstrably consistent with the orbital properties of the actual long-term stable KBOs that are observed. To this end, let us examine an ancillary numerical simulation that starts with the orbits of observed long-period KBOs and the version of Planet Nine that is preferred from the previous set of analyses.


The goal of this consistency check is to constrain two requirements. First, we demand that the introduction of Planet Nine does not rapidly destabilize the observed objects. Second, Planet Nine should cause the observed objects to behave in a matter similar to the test-particles of the simulations, so that they execute the same type of phase-space evolution as depicted in Figure \ref{fig:pspace5Me}. Rather than carrying out a full parameter search employing the long-term stability of the observed objects, here we choose the same Planet Nine properties as those described in the main text ($m_9=5M_\oplus$, $a_9=500$ AU, $e_9=0.25$, $i_9=20$ deg). Moreover, the numerical treatment is the same as that described in section \ref{anomalous}. That is, for each of the observed $a>250\,$AU TNOs, we cloned the object 10 times and allowed their orbits to evolve over a time scale of 4 Gyr, under perturbations from the giant planets and Planet Nine.

\begin{figure}[tbp]
\centering 
\includegraphics[width=1.0\textwidth]{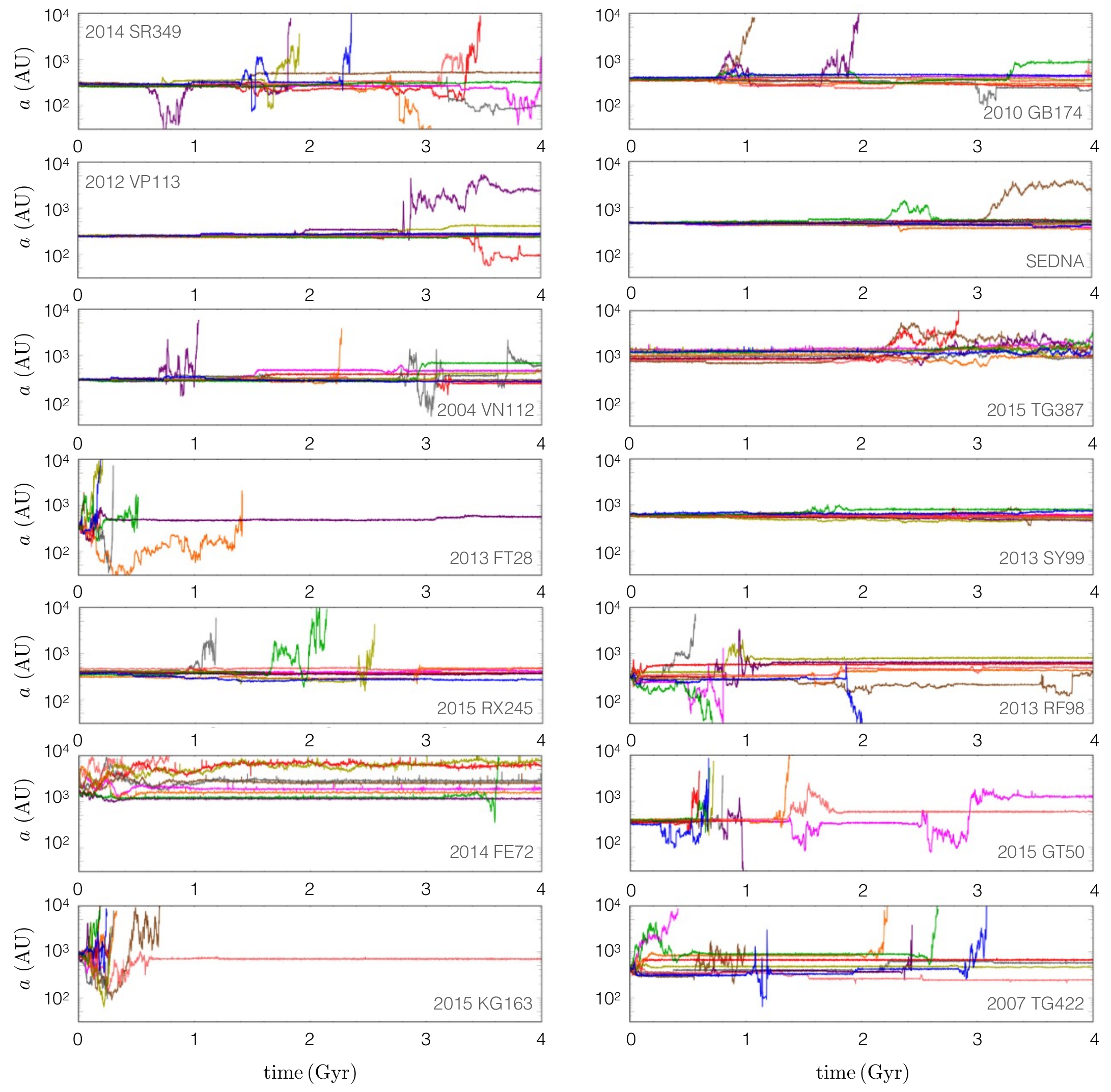}
\caption{The semi-major axis time-series of long-period KBOs. Each panel shows the time evolution for ten clones of a given KBO as labeled. These simulations were run assuming the updated baseline parameters for Planet Nine: $m_9=5M_\oplus$, $a_9=500$ AU, $e_9=0.25$, $i_9=20$ deg. Most of the clones of objects deemed dynamically stable in section \ref{anomalous} remain stable over $\sim\,$Gyr timescales in presence of Planet Nine. This result signals a consistency between a the P9 orbital fit that allows the observed extreme solar system objects to remain dynamically stable and the P9 parameters required to optimize the physical alignment of their orbits. } 
\label{fig:tnotime} 
\end{figure}   

One set of results from these integrations is depicted in Figure \ref{fig:tnotime}, which shows the semi-major axis of each long-period KBO as a function of time. The vast majority of the objects that are dynamically stable in absence of Planet Nine also exhibit only mild semi-major axis evolution over $\sim$Gyr timescales, and a large fraction of the clones survive the full integration. This finding indicates that the presence of Planet Nine, with the stated properties, does not destabilize most of the objects. Moreover, the particular objects that have motivated the majority of the analysis in the literature (namely, 2014\,SR$_{349}$, 2010\,GB$_{174}$, 2012\,VP$_{113}$, Sedna, 2004\,VN$_{112}$, 2015\,TG$_{387}$, 2013\,SY$_{99}$ and 2015\,RX$_{245}$) tend to exhibit long-term stability. Only a small number of clones of the aforementioned objects get ejected from the system during the simulations. 

One outlier in this group, 2013\,FT$_{28}$, deserves further discussion. This object is the only KBO from our analysis that does not show orbital alignment. In presence of Planet Nine, most realizations of this object survive for only $\sim100$ Myr. Moreover, the clones that survive the longest are often excited to high inclination, thereby producing an orbit akin to that of the  observed object 2015\,BP$_{519}$. This same dynamical trend applies to the long-lived clones of objects that were labeled as unstable in section \ref{anomalous}. More specifically, we find the realizations of the objects 2015\,GT$_{50}$, 2015\,KG$_{163}$, and 2007\,TG$_{422}$ that remain long-lived in the
simulations with Planet Nine derive their dynamical stability by acquiring high inclinations (see sections \ref{sec:analytical} and \ref{sec:numerical} for further discussion).

A pair of outliers also exist within the nominally unstable group on KBOs, namely the objects 2013\,RF$_{98}$ and 2014\,FE$_{72}$. In the absence of Planet Nine, these objects experiences relatively rapid orbital diffusion, primarily due to their low perihelion (which allows for strong interactions with Neptune). In the presence of Planet Nine, however, the orbits of 2013\,RF$_{98}$ and and 2014\,FE$_{72}$ are significantly stabilized, so that its long-term evolution is characterized by apsidal confinement akin to that exhibited by objects like Sedna, 2012\,VP$_{113}$, and others.  

\begin{figure}[tbp]
\centering 
\includegraphics[width=1.0\textwidth,clip]{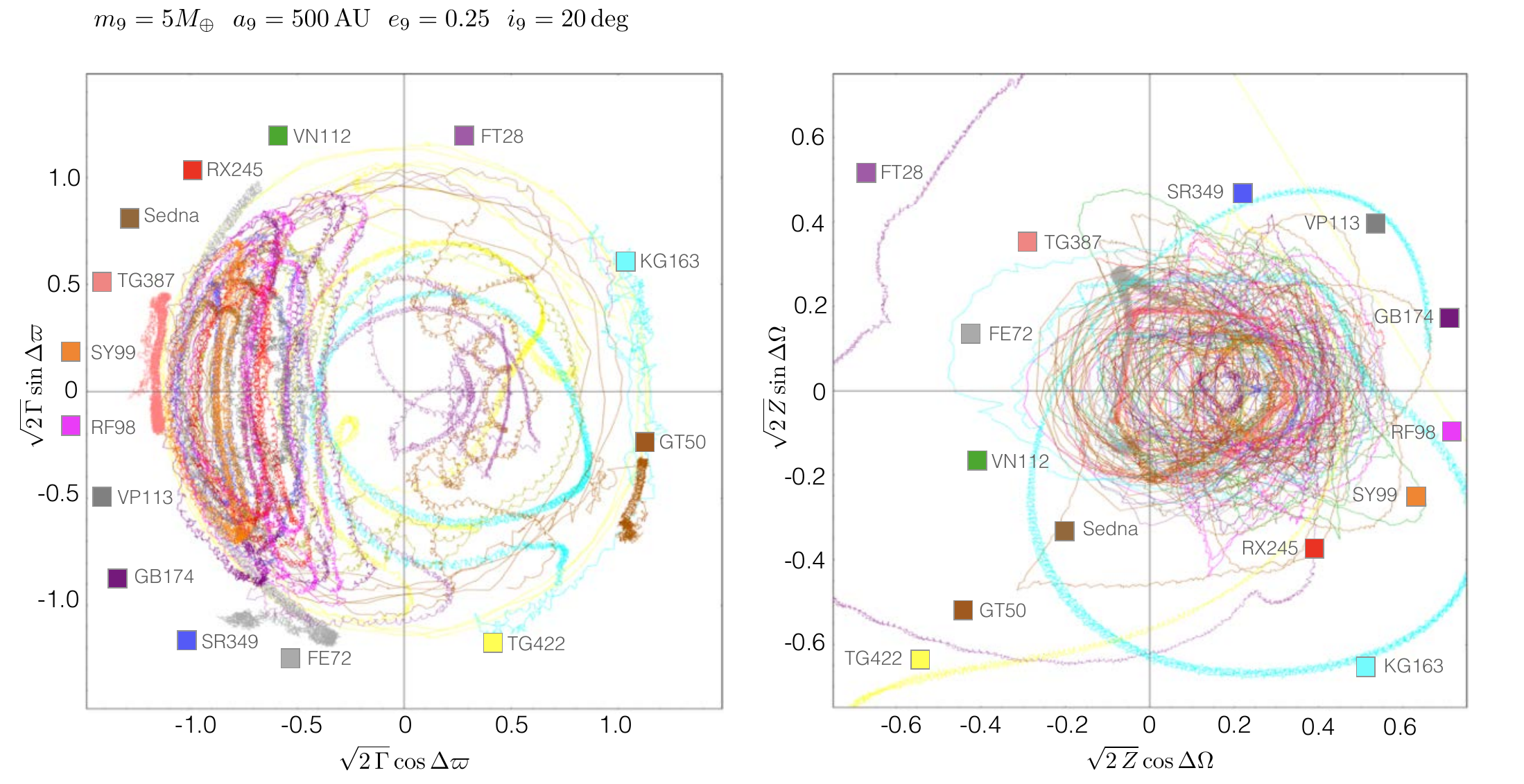}
\caption{Phase space evolution of distant KBOs. This Figure projects the phase-space evolution of the observed long-period KBOs in the same way as in Figure \ref{fig:pspace5Me}. Under the influence of P9's gravitational potential, the KBOs 2014\,SR$_{349}$, 2010\,GB$_{174}$, 2012\,VP$_{113}$, Sedna, 2004\,VN$_{112}$, 2015\,TG$_{387}$, 2013\,SY$_{99}$, 2013\,RF$_{98}$, 2014\,FE$_{72}$ and 2015\,RX$_{245}$ maintain long-term apsidal confinement and clustering of the orbital poles. Conversely, the KBOs 2013\,FT$_{28}$, 2015\,GT$_{50}$, 2015\,KG$_{163}$, and 2007\,TG$_{422}$ are typically unstable and their depicted long-lived clones derive orbital stability primarily for excitation of extreme orbital inclinations.} 
\label{fig:tnophase} 
\end{figure} 

In addition to demonstrating long-term stability of the observed KBO orbits (shown in Figure \ref{fig:tnotime}), the simulations also determine the evolution of the orbits in phase space (shown in Figure \ref{fig:tnophase}). For every long-period KBO in the sample, one clone that survives over the entire integration time of 4 Gyr is shown in the figure, which depicts the evolution in phase-space, as done in Figure \ref{fig:pspace5Me}. The phase-space trajectories corresponding to individual objects are labeled by color, as shown. In the presence of Planet Nine, the observed distant KBOs display the same type of dynamics as that described in sections \ref{sec:analytical} and \ref{sec:numerical}. All of the
objects that are long-term stable (particularly, 2014\,SR$_{349}$, 2010\,GB$_{174}$, 2012\,VP$_{113}$, Sedna, 2004\,VN$_{112}$, 2015\,TG$_{387}$, 2013\,SY$_{99}$, 2013\,RF$_{98}$, 2014\,FE$_{72}$ and 2015\,RX$_{245}$) execute secular apsidal librations (shown in the left panel), while their orbital poles cluster together as expected (shown in the right panel). The objects 2013\,FT$_{28}$, 2015\,GT$_{50}$, 2015\,KG$_{163}$, and 2007\,TG$_{422}$ are typically unstable. However, their stable
clones shown in the figure execute high-inclination dynamics, as shown by the light purple, cyan, yellow, and dark orange paths in the right panel of Figure \ref{fig:tnophase}. 

In summary, this appendix has presented simulations of the observed $a>250\,$AU KBOs in the presence of P9 with revised parameters. The results of these simulations show that this Planet Nine model allows the real object to remain stable over billions of years (Figure \ref{fig:tnotime}) and facilitates sustained clustering of their orbits in a manner that is consistent with observations (Figure \ref{fig:tnophase}).

\newpage

\bigskip

\end{document}